\documentstyle[preprint,aps]{revtex}

\def\Rel{Re_{\lambda}}

\def\beq{\begin{equation}}
\def\be{\begin{equation}}
\def\ee{\end{equation}}
\def\eeq{\end{equation}}
\def\noin{\noindent}

\begin{document}

\draft
\tighten

\title{Velocity Probability Distribution Function in the Crossover
and the Viscous Range}

\author{Jens Eggers$^{1}$ and  Z.~Jane Wang$^{2\dagger}$}
\address{(1)Universit\"{a}t Gesamthochschule Essen, Fachbereich Physik, 
        45117 Essen, Germany \\ 
        (2) Department of Theoretical Physics,University of Oxford,
            Oxford, OX1 3NP, UK }

\date{\today}

\maketitle

\begin{abstract} 

We derive a multifractal model for the velocity probability density 
distribution function (PDF), which is valid from the inertial range
to the viscous range. The model gives a continuous 
evolution of velocity PDFs from large to small scales. 
It also predicts the asymptotic form of the PDF of 
velocity gradients. More importantly, the model captures 
the crossover range behavior and thus allows us to address
the transition observed in the Helium turbulence 
experiment[{\it Tabeling et. al., Phys. Rev. E53, 1613 (1996)}]
quantitatively. We compare both the PDFs and the structure functions 
predicted by this model with the experiments from the
inertial range to the smallest scale resolved by the
experiment. The model also predicts that the 
flatness measured in the crossover region can
decrease with the Reynolds number.
\end{abstract}
\vspace{0.3cm}
{PACS numbers: 47.27.-i, 47.80.+v  \hspace{0.5cm} { \it submitted to Physical
Review E}}

\section{Introduction}

The statistics of a turbulent flow field can be
characterized by a family of probability 
distribution functions (PDF) of velocity differences over
varying spatial scales. At large scales, where 
the energy is fed into the system, the probability 
distribution is approximately Gaussian. As the scale decreases, the
distribution develops increasingly stretched tails, which
corresponds to large deviations from its average values, a
phenomenon called intermittency. These intermittent 
fluctuations are believed to be a consequence of the 
nonlinear interactions between  different scales. 
Through these nonlinear interactions the energy 
is transported from large to small scales. 
At each stage of the energy cascade, the velocity
fluctuations are amplified. On a crude level, the
cascade can be described by a random multiplicative process, 
which eventually is terminated by viscosity. At
scales smaller than the viscous scale, the velocity 
field becomes smooth and its PDF reaches an asymptotic form.
Previous studies have already examined the PDF of velocity
gradients \cite{ben91,bi93,sre92}.

An equivalent description of the statistics of the velocity field 
is given by the structure functions $D_q(r)$:
\be
D_q(r)=\langle u_r^{q}  \rangle ,
\label{eq-str}
\ee
\noin where $u_r(x)$ is the velocity difference over
$r: [u(x + r) -u(x)]$.

In the inertial range, where the details of the 
large scale flow and the viscosity are unimportant, 
the velocity distribution is self-similar in $r$.
Consequently, $D_q(r)$ is given by a power law:
\be
D_q(r)= C_q (r/L)^{q/3 + \delta \zeta_q}  ,
\label{eq-pl}
\ee
\noin where the $C_q$ are constants independent of $r$ 
and the $\delta\zeta_q$ are the so-called intermittency corrections
to the scaling exponents. In the absence of intermittency,
the scaling is described by the classical Kolmogorov 41 law \cite{Kol:41a}:
\be
\delta \zeta_q = 0  .
\ee
\noin This corresponds to a case where the shape of 
the PDF does not depend on scale, and their variances 
decrease as $r^{1/3}$. 
In the viscous range, $D_q(r)$ is simple because of the
viscous smoothing:   $D_q(r) \sim r^q$. On the other hand,
the crossover region from the inertial range 
to the viscous range is far more complex.
Nelkin \cite{nel90} noted that because
the local velocity is fluctuating in space, the 
viscous cutoff scale $\eta_c$, which depends on the local Reynolds number,
must also be fluctuating in space. The local Reynolds number
is defined as
\be
Re_r = \frac{u_r r}{\nu},
\ee
\noin where $\nu$ is viscosity. The fluctuations in 
$\eta_c$ lead to an additional broadening of 
the velocity distributions. This effect can best be
seen by considering a case in which the velocity
distribution in the inertial range is Gaussian.  
Benzi et. al. \cite{ben91} showed that the fluctuating 
$\eta_c$ alone can lead to a stretched exponential
distribution of the PDF of velocity gradients.  
In the presence of intermittency, these authors further derived a formula
for the PDF of velocity gradients and of PDFs of velocity differences
in the inertial range. We will generalize their 
results to include the crossover regime from the 
inertial range to the dissipation range, and we derive
our results using  continuous evolutions.
A related work on comparing experiment data and multifractal model of
PDFs can be found in reference\cite{sre92}.
Another parallel work by Frisch and Vergassola\cite{fri91}
studied the consequence of the  fluctuating $\eta_c$ in structure functions.

Our motivation to develop a model for 
PDFs of velocity differences over the full range of scales
is threefold:
\begin{itemize}
\item[(1)] The PDFs are the experimentally measured quantities, 
and thus a model for the PDF allows for the most direct comparison 
with experimental data. 
\item[(2)] The crossover from inertial range to  viscous range
can be analyzed quantitatively. 
\item[(3)] The model allows us to go one step beyond
the order of magnitude arguments generally made when 
comparing with the experiments. Once the
adjustable parameters are fixed by comparing with one particular flow,
parameters are known for all flows and predictions 
can be made and compared with the experiments. 
\end{itemize}

Our model is particularly useful for 
analyzing a series of recent Helium turbulence experiments 
by Tabeling et al. \cite{zoc94}, \cite{tab95}, 
\cite{bel96}. We study in detail the quantitative comparison between
the model and the experiments. We are particularly
interested in the flatness because of the reported
transition seen in flatness at a Taylor 
Reynolds number of $Re_{\lambda} \approx 700$.
Within our model, we find that the flatness is rapidly varying
over a spatial scale between $2\eta$ and $20\eta$.
This means that measurements of the flatness 
are extremely sensitive to the experimental
resolution, which lies in this range of scales. 
More interestingly, the ordering of the magnitude of the flatness
as function of Reynolds number
depends on the particular scale in the crossover range.
This further suggests that the asymptotic value of the flatness can only be
measured below $\eta$, where we find
the PDF to reach its asymptotic form.
We emphasize the important difference
between the scale $1\eta$ and $10\eta$, which is often indistinguishable
in order of magnitude arguments like those made in reference\cite{ems96}.

In the next section we briefly describe the multifractal model of turbulence,
and in Section 3 we generalize this idea to include viscous effects.
The resulting model for PDFs is the central starting point of the paper. 
In Section 4 we determine the adjustable parameters of the model using 
the experimental data  by Tabeling's group. In the subsequent 
section we discuss the evolution of the PDFs as a function of the 
separation evolving from  the inertial to the viscous range.
In Section 6 we compute the flatness as  a function of scale and
focus on its crossover behavior.
In the final discussion, we review all points of agreement and the 
discrepancies  between the experimental data and the multifractal model
and relate our observation to the recently reported  transition in 
the flatness.

\section{Basic Idea of Multiplicative Processes in Turbulence}

Experimental studies have shown that the turbulence
velocity field in the inertial range exhibit
power law scaling as described in Eq. (\ref{eq-pl}).
All information on two-point correlations of the velocity field is contained 
in the spectrum of exponents $\zeta_q$. 
The scaling (\ref{eq-pl}) can also be viewed as a consequence of
a underlying multiplicative process, by which the energy 
is cascaded progressively from eddies of size $r$ into 
eddies of size $r/\lambda$. The ratio of the velocities
in successive steps is a stochastic variable $s_i$,
whose statistics is determined by its probability
distribution $p_s(s_i)$. The index $i$ denotes 
the $ith$-step in the cascade from a given large scale $L$ to $r$: 

\begin{equation}
u_r = u_L \prod_{i=1}^n s_i .
\label{eq:1-2}
\end{equation}

\noin Assuming that the $s_i$ are identically distributed 
and uncorrelated,
we then have for the velocity moments:

\[
\langle u_r^q \rangle = \langle s^q \rangle^n \langle u_L^q \rangle ,
\quad n = -\log_\lambda \frac{r}{L}
\]

\noin and by comparison with (\ref{eq-pl}) 

\begin{equation}
\zeta_q = \frac{q}{3} + \delta\zeta_q = -\log_\lambda\langle s^q \rangle .
\label{eq:1-3}
\end{equation}
Since  $\zeta_3 = 1$ by the Kolmogorov structure equation, 
we have the constraint
\begin{equation}
\langle s^3 \rangle = 1/\lambda .
\label{eq:1-4}
\end{equation}
In the following we will be making the conventional choice 
$\lambda = 2$. In the context of our phenomenological description,
this is of a little consequence. Using the known exponent spectrum 
$\zeta_q$, information on the distribution $p_s(s)$ can be extracted 
through its moments. There have been attempts to measure $p_s$
directly \cite{chh92}, but little is known about its shape theoretically. 
Technically it is convenient to approximate $p_s$
by a bimodal distribution 

\begin{equation}
p_s(s) = p\delta(s-s_1) + (1-p)\delta(s-s_2).
\label{eq-ps}
\end{equation}

Imposing (\ref{eq:1-4}) as a constraint, the two remaining parameters 
can be used to reproduce the first 18 moments to within experimental 
error \cite{EG:92}. One can imagine that many 
other parameterizations of $p_s$ would work equally well. 

Here we base our fit on the experimental exponents 
by Tabeling's group\cite{bel96}, which were obtained
by using the method of extended self-similarity. We find

\begin{equation}
p = 0.688, \hspace{0.2cm} s1 = 0.699, \hspace{0.2cm}  s2 = 0.947 .
\label{eq-para}
\end{equation}

\noindent The predicted scaling exponents using the above
parameters are compared with the experiments 
in the table below:
\noindent The exponents agree with the established values 
obtained in various turbulence experiments \cite{Ans:84,Cas:90}
to within the error.

Given this simple model of a multiplier distribution, the PDF 
$P_j(u)$ of velocity differences on level 
\[
j = -\log_2(r/L)
\]
can be readily calculated: it is the distribution $P_0(u)$ on 
the outer scale, convoluted with $p_s$ $j$ times, giving 
\begin{equation}
P_j(u) = \sum^{j}_{\kappa=0} {j \choose \kappa} 
\left(\frac{p_1}{s_1}\right)^{j-\kappa} \left(\frac{p_2}{s_2}\right)^{\kappa}
P_0\left(\frac{u}{s_1^{j-\kappa}s_2^{\kappa}}\right) .
\label{eq:2-5}
\end{equation}
An expression equivalent to (\ref{eq:2-5}), but using the random 
$\beta$ model, was already given in \cite{ben91}.
Since $r$ is a continuous variable, it is useful to generalize 
(\ref{eq:2-5}) to continuous values of $j$. This can be done 
using the Euler-McLaurin sum formula, which gives to lowest order 
\begin{equation}
P_j(u) = \int^{j}_{\kappa} {j \choose \kappa} 
\left(\frac{p_1}{s_1}\right)^{j-\kappa} \left(\frac{p_2}{s_2}\right)^{\kappa}
P_0\left(\frac{u}{s_1^{j-\kappa}s_2^{\kappa}}\right) d\kappa +
\frac{1}{2} \left[\left(\frac{p_1}{s_1}\right)^j 
P_0\left(\frac{u}{s_1^j}\right) +
\left(\frac{p_2}{s_2}\right)^j P_0\left(\frac{u}{s_2^j}\right)\right] .
\label{eq:2-6}
\end{equation}
This proved to be an adequate approximation of (\ref{eq:2-5}) for 
integer values of $j$ and smoothly interpolates in between. 
In the next section, we follow 
the above basic idea to model the inertial range 
scaling, and in addition we introduce
a fluctuating cutoff to 
model the viscous effects.

\section{Model for the Probability Distributions of Velocity Differences}

Our model of PDFs of velocity differences includes 
three basic elements: 1) the experimentally
measured PDF of velocities at a large scale $P_0$, 2) the multiplier
distribution $p_s$, and 3) the viscous cutoff mechanism. 
Once they are determined, we can compute 
the PDF of velocities for any given scale. Among the
three elements, $P_0$ is experimentally given,  and
$p_s(s)$ is assumed to be of the form of Eq. (\ref{eq-ps}), with
three parameters given by  Eq. (\ref{eq-para}).
We then follow the Nelkin's idea\cite{nel90}
and introduce a viscous cutoff scale $\eta_c$ at which
the local Reynolds number is equal to a fixed value. 
The physical idea of this cutoff is that whenever 
the {\it local} Reynolds number of 
an eddy reaches a critical value $Re_{cr}$, it
is smoothed out by viscosity and the cascade stops. 
Since one expects the structure of small-scale velocity 
fluctuations to be universal, the details of the cutoff 
mechanism will not depend on the specific flows, thus
$Re_{cr}$  should be universal.  Furthermore, 
because the local Reynolds number 
$|u_r| r /\nu$ is a fluctuating quantity, so is $\eta_c$.
These fluctuations further modify the scaling in the dissipation
range in addition to the inertial range fluctuations.

More specifically, we can construct the following ensemble 
for velocity differences $u_r$. For convenience, all quantities
are assumed to be non-dimensionalized by the large
scale $L$ and the root-mean-squared velocity fluctuations on that
scale

\[
U_{rms} = [\langle u_L^2 \rangle ]^{1/2} .
\]

In computing the PDFs we treat the the positive and negative
velocities independently. 
Here we only look at positive values of
$u_r$, the negative values are treated separately, replacing $u_r$
by its modulus. Letting $r = 2^{-n}$, each ensemble member is then 
constructed by the following procedure:
\begin{itemize}
\item[(1)] 
Choose a realization of the large scale velocity field $u_0\equiv u_L$. 
\item[(2)]
Multiply $u_0$ with random multipliers $s$ to obtain a realization
on step $j$ until either
(a) level $n$ is reached or
(b) the condition $u_j 2^{-j} \geq R$ is no longer satisfied. 
The constant $R$ is simply $R = Re_{cr}/ Re$. This cutoff level 
in a particular realization will be denoted by $j_c$.
\item[(3)]
In case (a) $u_r = u_n$ is the desired value of $u_r$. 
If the cascade has stopped due to the action of viscosity
on level $j_c < n$, the velocity difference over the distance 
$2^{-j_c}$ is $u_{j_c}$. Therefore, on scale $r$ we have 
$u_r = u_{j_c} 2^{j_c-n}$. 
\end{itemize}
To derive probability distributions from this procedure, it is 
more convenient to treat the level number as a continuous
variable. Accordingly, we will assume that 
the velocity distributions $P_j$ on level $j$
have been continued to all real values $j$, as we did in (\ref{eq:2-6}). 

To find an explicit expression for the probability distribution 
of $u_r$, we need to find the relation between $j_c$ and $u_r$.
If $u_0$ is already smaller than $R$, $j_c$ is equal to zero, 
and the corresponding $u_r$ is $u_0 2^{-n}$. Thus $j_c=0$ for
$u_r \leq R 2^{-n}$. If the cascade reaches step $n$, $u_n$ 
must have been larger than $R 2^n$, and thus $j_c = n$ for 
$u_r \geq R 2^n$. In between, the cascade is terminated by 
viscosity. This means the cutoff condition $u_{j_c}2^{-j_c}=R$
must be satisfied, and thus $u_r = R 2^{2j_c-n}$. To summarize,
we have 

\begin{equation}
j_c = \left \{ \begin{array}{l@{\quad}l}
             0
               & \log_2 \frac{u_r}{R} \leq -n \\
             n
                & \log_2 \frac{u_r}{R} \geq n \\
             \frac{1}{2} (n + \log_2 \frac{u_r}{R})
                & \mbox{elsewhere} \quad\; .
                 \end{array}\right.
\label{eq:2-2}
\end{equation}

Physically, we expect the relation between $j_c$ and $u_r$ 
to be smooth, so we will be using a smoothed version of 
(\ref{eq:2-2}), given in Appendix A, which is one-to-one 
between $j_c$ and $u_r$ and has the property $j_c(u_r) \rightarrow 0$
for $u_r \rightarrow 0$ and $j_c(u_r) \rightarrow n$ for 
$u_r \rightarrow \infty$. 
The smoothing introduces an additional parameter 
$\Delta$, cf. (\ref{eq:A-1}), which is the number of levels over 
which the cascade is gradually cut off. Thus the description of the 
viscous range introduces two parameters: $Re_{cr}$, 
which measures the relative importance of the viscous term,
and $\Delta$, which measures the effectiveness of the viscous 
smoothing. In a small-scale regime of universal isotropic 
fluctuations, we expect both to be independent of the large-scale
flow. Thus once they are adjusted, they can be applied to all 
other flows at different Reynolds numbers. 

To compute the probability distribution $P_r$ of $u_r$, we 
express the total probability of finding
$u_r$ between $u_r(j_c)$ and $u_r(j_c+\delta)$ through the densities 
$P_j$. Namely 
\begin{equation}
\int^{u_r(j_c+\delta)}_{u_r(j_c)} P_r(u) du  =
\int^{\infty}_{u_{j_c}} P_{j_c}(u) du - 
\int^{\infty}_{u_{j_c+\delta}} P_{j_c+\delta}(u) du ,
\label{eq:2-3}
\end{equation}
where $u_{j_c}$ is the velocity on level $j_c$, which contributed 
to $u_r$. Hence $u_{j_c} = u_r 2^{n-j_c}$. 
The normalization condition (\ref{eq:2-3}) simply expresses the
fact that contributions from levels between $j_c$ and $j_c+\delta$
correspond to the probability of cascading down at least to level
$j_c$, minus the probability of even making it to level $j_c+\delta$. 
Letting $\delta$ go to zero, we obtain
\begin{equation}
P_r(u_r) = \frac{\partial j_c}{\partial u_r} \left\{
\frac{\partial u_{j_c}}{\partial j_c} P_{j_c}(u_{j_c}) - 
\int^{\infty}_{u_{j_c}} \left.\frac{\partial P}{\partial j}\right|
_{j = j_c}(u) du
\right\} ,
\label{eq:2-4}
\end{equation}
where $j_c$ on the right hand side of (\ref{eq:2-4}) can be expressed 
through $u_r$. If $r$ tends to zero, (\ref{eq:2-4}) gives the 
distribution of velocity gradients, for which the distribution
was also derived in \cite{ben91}. However, the approximation 
of \cite{ben91} just corresponds to the first term in the curly brackets,
while the second term was neglected. Note that 
(\ref{eq:2-3}) automatically ensures that $P_r$ is normalized, since
\[
\int^{\infty}_{0} P_r(u) du = \int^{\infty}_{0} P_0(u) du ,
\]
which means that $P_r$ inherits its normalization from the top level 
distribution $P_0$. 
Thus (\ref{eq:2-4}), together with $P_j(u)$ (cf. (\ref{eq:2-6}))
and $j_c(u_r)$ gives an explicit 
formula for $P_r(u_r)$, which we implemented numerically.
This is the central result of the paper, which will be explored in
the sections below. 

\section{Comparison with Experiment}
We apply this model to a set of 
experimental data measured by  Tabeling et. al. \cite{zoc94},
\cite{tab95}, \cite{bel96}. 
The experiment measures the longitudinal component of the turbulent 
velocity field  inside a closed cylinder filled with Helium.
The viscosity is varied by a technique similar to that 
used in the Rayleigh-Bernard
system\cite{thr,hes87} to achieve a variation 
over 3 decades in Reynolds number.
The measurements of various scaling laws are described in
the above series of papers \cite{zoc94},\cite{tab95},\cite{bel96}. 
The Reynolds number dependence of the flatness is reported
to have an abrupt change at the Taylor Reynolds number $\approx 700$:
the flatness increases with the Reynolds number up to $\Rel \approx 700$
and then it decreases with the Reynolds number. 
However the origin of this transition has been subjected 
to much debate \cite{bel95a},\cite{sre95a},\cite{ems96}.

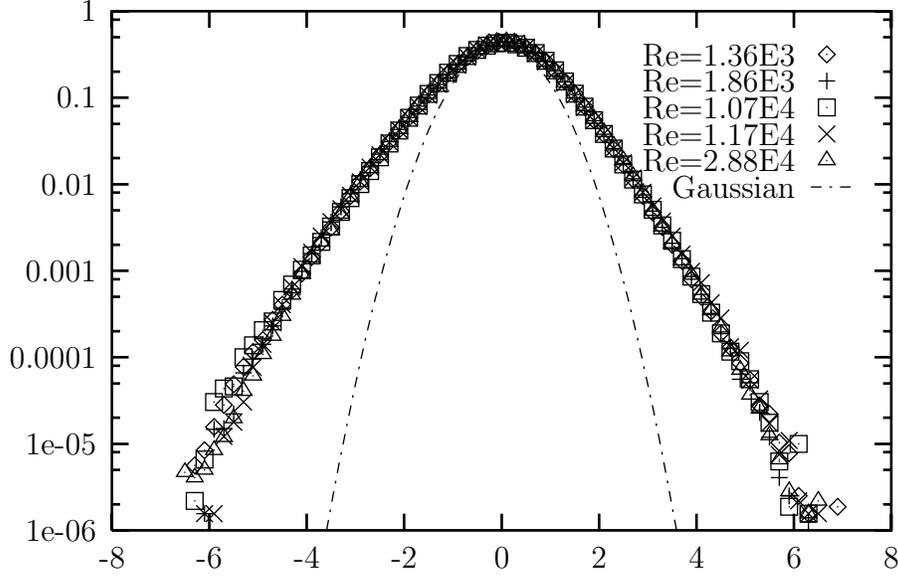
\begin{figure}[htb]
\caption{PDF of the velocity difference at the outer scale $L$ are
plotted on a log-linear scale for all 
experimental runs under consideration. The velocity differences
is normalized by its variance, and the PDF is normalized to unity.
 }
\setlength{\unitlength}{0.1bp}
\special{!
/gnudict 40 dict def
gnudict begin
/Color false def
/Solid false def
/gnulinewidth 5.000 def
/vshift -33 def
/dl {10 mul} def
/hpt 31.5 def
/vpt 31.5 def
/M {moveto} bind def
/L {lineto} bind def
/R {rmoveto} bind def
/V {rlineto} bind def
/vpt2 vpt 2 mul def
/hpt2 hpt 2 mul def
/Lshow { currentpoint stroke M
  0 vshift R show } def
/Rshow { currentpoint stroke M
  dup stringwidth pop neg vshift R show } def
/Cshow { currentpoint stroke M
  dup stringwidth pop -2 div vshift R show } def
/DL { Color {setrgbcolor Solid {pop []} if 0 setdash }
 {pop pop pop Solid {pop []} if 0 setdash} ifelse } def
/BL { stroke gnulinewidth 2 mul setlinewidth } def
/AL { stroke gnulinewidth 2 div setlinewidth } def
/PL { stroke gnulinewidth setlinewidth } def
/LTb { BL [] 0 0 0 DL } def
/LTa { AL [1 dl 2 dl] 0 setdash 0 0 0 setrgbcolor } def
/LT0 { PL [] 0 1 0 DL } def
/LT1 { PL [4 dl 2 dl] 0 0 1 DL } def
/LT2 { PL [2 dl 3 dl] 1 0 0 DL } def
/LT3 { PL [1 dl 1.5 dl] 1 0 1 DL } def
/LT4 { PL [5 dl 2 dl 1 dl 2 dl] 0 1 1 DL } def
/LT5 { PL [4 dl 3 dl 1 dl 3 dl] 1 1 0 DL } def
/LT6 { PL [2 dl 2 dl 2 dl 4 dl] 0 0 0 DL } def
/LT7 { PL [2 dl 2 dl 2 dl 2 dl 2 dl 4 dl] 1 0.3 0 DL } def
/LT8 { PL [2 dl 2 dl 2 dl 2 dl 2 dl 2 dl 2 dl 4 dl] 0.5 0.5 0.5 DL } def
/P { stroke [] 0 setdash
  currentlinewidth 2 div sub M
  0 currentlinewidth V stroke } def
/D { stroke [] 0 setdash 2 copy vpt add M
  hpt neg vpt neg V hpt vpt neg V
  hpt vpt V hpt neg vpt V closepath stroke
  P } def
/A { stroke [] 0 setdash vpt sub M 0 vpt2 V
  currentpoint stroke M
  hpt neg vpt neg R hpt2 0 V stroke
  } def
/B { stroke [] 0 setdash 2 copy exch hpt sub exch vpt add M
  0 vpt2 neg V hpt2 0 V 0 vpt2 V
  hpt2 neg 0 V closepath stroke
  P } def
/C { stroke [] 0 setdash exch hpt sub exch vpt add M
  hpt2 vpt2 neg V currentpoint stroke M
  hpt2 neg 0 R hpt2 vpt2 V stroke } def
/T { stroke [] 0 setdash 2 copy vpt 1.12 mul add M
  hpt neg vpt -1.62 mul V
  hpt 2 mul 0 V
  hpt neg vpt 1.62 mul V closepath stroke
  P  } def
/S { 2 copy A C} def
end
}
\begin{picture}(3600,2160)(0,0)
\special{"
gnudict begin
gsave
50 50 translate
0.100 0.100 scale
0 setgray
/Helvetica findfont 100 scalefont setfont
newpath
-500.000000 -500.000000 translate
LTa
LTb
480 151 M
63 0 V
2874 0 R
-63 0 V
480 249 M
31 0 V
2906 0 R
-31 0 V
480 379 M
31 0 V
2906 0 R
-31 0 V
480 446 M
31 0 V
2906 0 R
-31 0 V
480 477 M
63 0 V
2874 0 R
-63 0 V
480 576 M
31 0 V
2906 0 R
-31 0 V
480 705 M
31 0 V
2906 0 R
-31 0 V
480 772 M
31 0 V
2906 0 R
-31 0 V
480 804 M
63 0 V
2874 0 R
-63 0 V
480 902 M
31 0 V
2906 0 R
-31 0 V
480 1032 M
31 0 V
2906 0 R
-31 0 V
480 1098 M
31 0 V
2906 0 R
-31 0 V
480 1130 M
63 0 V
2874 0 R
-63 0 V
480 1228 M
31 0 V
2906 0 R
-31 0 V
480 1358 M
31 0 V
2906 0 R
-31 0 V
480 1425 M
31 0 V
2906 0 R
-31 0 V
480 1456 M
63 0 V
2874 0 R
-63 0 V
480 1555 M
31 0 V
2906 0 R
-31 0 V
480 1684 M
31 0 V
2906 0 R
-31 0 V
480 1751 M
31 0 V
2906 0 R
-31 0 V
480 1783 M
63 0 V
2874 0 R
-63 0 V
480 1881 M
31 0 V
2906 0 R
-31 0 V
480 2011 M
31 0 V
2906 0 R
-31 0 V
480 2077 M
31 0 V
2906 0 R
-31 0 V
480 2109 M
63 0 V
2874 0 R
-63 0 V
480 151 M
0 63 V
0 1895 R
0 -63 V
847 151 M
0 63 V
0 1895 R
0 -63 V
1214 151 M
0 63 V
0 1895 R
0 -63 V
1581 151 M
0 63 V
0 1895 R
0 -63 V
1949 151 M
0 63 V
0 1895 R
0 -63 V
2316 151 M
0 63 V
0 1895 R
0 -63 V
2683 151 M
0 63 V
0 1895 R
0 -63 V
3050 151 M
0 63 V
0 1895 R
0 -63 V
3417 151 M
0 63 V
0 1895 R
0 -63 V
480 151 M
2937 0 V
0 1958 V
-2937 0 V
480 151 L
LT0
3174 1946 D
792 396 D
829 453 D
865 541 D
902 624 D
939 702 D
976 768 D
1012 824 D
1049 869 D
1086 945 D
1122 1002 D
1159 1057 D
1196 1120 D
1233 1182 D
1269 1237 D
1306 1293 D
1343 1354 D
1379 1412 D
1416 1467 D
1453 1517 D
1490 1567 D
1526 1616 D
1563 1663 D
1600 1708 D
1636 1753 D
1673 1796 D
1710 1837 D
1747 1875 D
1783 1908 D
1820 1936 D
1857 1960 D
1893 1978 D
1930 1989 D
1967 1992 D
2004 1987 D
2040 1972 D
2077 1951 D
2114 1921 D
2150 1885 D
2187 1844 D
2224 1799 D
2261 1750 D
2297 1698 D
2334 1644 D
2371 1588 D
2407 1530 D
2444 1474 D
2481 1412 D
2518 1350 D
2554 1295 D
2591 1228 D
2628 1167 D
2664 1098 D
2701 1032 D
2738 977 D
2775 894 D
2811 841 D
2848 782 D
2885 718 D
2921 622 D
2958 590 D
2995 482 D
3032 442 D
3068 281 D
3105 214 D
3215 240 D
LT1
3174 1846 A
829 214 A
865 532 A
902 535 A
939 590 A
976 745 A
1012 797 A
1049 853 A
1086 922 A
1122 985 A
1159 1061 A
1196 1128 A
1233 1195 A
1269 1255 A
1306 1314 A
1343 1371 A
1379 1421 A
1416 1475 A
1453 1522 A
1490 1570 A
1526 1617 A
1563 1663 A
1600 1708 A
1636 1751 A
1673 1793 A
1710 1833 A
1747 1871 A
1783 1905 A
1820 1936 A
1857 1960 A
1893 1980 A
1930 1991 A
1967 1994 A
2004 1988 A
2040 1973 A
2077 1950 A
2114 1921 A
2150 1884 A
2187 1842 A
2224 1795 A
2261 1747 A
2297 1696 A
2334 1645 A
2371 1591 A
2407 1535 A
2444 1477 A
2481 1416 A
2518 1355 A
2554 1294 A
2591 1234 A
2628 1176 A
2664 1108 A
2701 1041 A
2738 969 A
2775 893 A
2811 838 A
2848 721 A
2885 708 A
2921 596 A
2958 502 A
2995 350 A
3032 281 A
3105 183 A
LT2
3174 1746 B
792 262 B
829 418 B
865 635 B
902 686 B
939 694 B
976 804 B
1012 849 B
1049 906 B
1086 939 B
1122 1020 B
1159 1078 B
1196 1134 B
1233 1188 B
1269 1239 B
1306 1297 B
1343 1352 B
1379 1404 B
1416 1457 B
1453 1506 B
1490 1558 B
1526 1610 B
1563 1659 B
1600 1707 B
1636 1753 B
1673 1797 B
1710 1838 B
1747 1877 B
1783 1910 B
1820 1940 B
1857 1963 B
1893 1979 B
1930 1988 B
1967 1990 B
2004 1984 B
2040 1971 B
2077 1951 B
2114 1922 B
2150 1886 B
2187 1844 B
2224 1798 B
2261 1749 B
2297 1698 B
2334 1646 B
2371 1591 B
2407 1531 B
2444 1474 B
2481 1418 B
2518 1360 B
2554 1301 B
2591 1243 B
2628 1174 B
2664 1108 B
2701 1042 B
2738 972 B
2775 894 B
2811 825 B
2848 789 B
2885 721 B
2921 635 B
2958 557 B
2995 411 B
3032 240 B
3068 477 B
3105 214 B
LT3
3174 1646 C
829 214 C
865 214 C
902 505 C
939 559 C
976 635 C
1012 766 C
1049 853 C
1086 934 C
1122 1002 C
1159 1069 C
1196 1146 C
1233 1200 C
1269 1261 C
1306 1315 C
1343 1364 C
1379 1415 C
1416 1466 C
1453 1521 C
1490 1568 C
1526 1614 C
1563 1661 C
1600 1707 C
1636 1750 C
1673 1792 C
1710 1833 C
1747 1871 C
1783 1905 C
1820 1936 C
1857 1963 C
1893 1982 C
1930 1993 C
1967 1996 C
2004 1988 C
2040 1972 C
2077 1949 C
2114 1918 C
2150 1881 C
2187 1838 C
2224 1794 C
2261 1746 C
2297 1696 C
2334 1647 C
2371 1592 C
2407 1537 C
2444 1480 C
2481 1426 C
2518 1374 C
2554 1317 C
2591 1261 C
2628 1193 C
2664 1128 C
2701 1085 C
2738 1009 C
2775 952 C
2811 844 C
2848 830 C
2885 717 C
2921 647 C
2958 554 C
2995 437 C
3032 490 C
3068 262 C
3105 214 C
3142 214 C
LT4
3174 1546 T
755 370 T
792 350 T
829 379 T
865 453 T
902 502 T
939 573 T
976 676 T
1012 734 T
1049 815 T
1086 884 T
1122 959 T
1159 1039 T
1196 1117 T
1233 1179 T
1269 1239 T
1306 1282 T
1343 1344 T
1379 1408 T
1416 1464 T
1453 1516 T
1490 1565 T
1526 1615 T
1563 1664 T
1600 1709 T
1636 1754 T
1673 1796 T
1710 1835 T
1747 1873 T
1783 1906 T
1820 1937 T
1857 1962 T
1893 1981 T
1930 1992 T
1967 1994 T
2004 1987 T
2040 1971 T
2077 1947 T
2114 1918 T
2150 1881 T
2187 1839 T
2224 1797 T
2261 1750 T
2297 1702 T
2334 1652 T
2371 1601 T
2407 1546 T
2444 1486 T
2481 1423 T
2518 1355 T
2554 1296 T
2591 1233 T
2628 1168 T
2664 1114 T
2701 1039 T
2738 980 T
2775 915 T
2811 825 T
2848 756 T
2885 662 T
2921 609 T
2958 509 T
2995 418 T
3032 298 T
3105 214 T
3142 262 T
LT5
3114 1446 M
180 0 V
1289 151 M
21 114 V
25 132 V
26 128 V
25 122 V
25 116 V
25 112 V
25 106 V
26 100 V
25 96 V
25 90 V
25 84 V
25 80 V
26 74 V
25 68 V
25 63 V
25 58 V
26 53 V
25 47 V
25 42 V
25 36 V
25 31 V
26 26 V
25 20 V
25 15 V
25 10 V
26 4 V
25 0 V
25 -7 V
25 -12 V
25 -17 V
26 -22 V
25 -28 V
25 -33 V
25 -38 V
25 -44 V
26 -49 V
25 -55 V
25 -59 V
25 -66 V
26 -70 V
25 -76 V
25 -81 V
25 -87 V
25 -92 V
26 -97 V
25 -103 V
25 -107 V
25 -114 V
26 -118 V
25 -124 V
25 -130 V
25 -135 V
12 -63 V
stroke
grestore
end
showpage
}
\put(3054,1446){\makebox(0,0)[r]{Gaussian}}
\put(3054,1546){\makebox(0,0)[r]{Re=2.88E4}}
\put(3054,1646){\makebox(0,0)[r]{Re=1.17E4}}
\put(3054,1746){\makebox(0,0)[r]{Re=1.07E4}}
\put(3054,1846){\makebox(0,0)[r]{Re=1.86E3}}
\put(3054,1946){\makebox(0,0)[r]{Re=1.36E3}}
\put(3417,51){\makebox(0,0){8}}
\put(3050,51){\makebox(0,0){6}}
\put(2683,51){\makebox(0,0){4}}
\put(2316,51){\makebox(0,0){2}}
\put(1949,51){\makebox(0,0){0}}
\put(1581,51){\makebox(0,0){-2}}
\put(1214,51){\makebox(0,0){-4}}
\put(847,51){\makebox(0,0){-6}}
\put(480,51){\makebox(0,0){-8}}
\put(420,2109){\makebox(0,0)[r]{1}}
\put(420,1783){\makebox(0,0)[r]{0.1}}
\put(420,1456){\makebox(0,0)[r]{0.01}}
\put(420,1130){\makebox(0,0)[r]{0.001}}
\put(420,804){\makebox(0,0)[r]{0.0001}}
\put(420,477){\makebox(0,0)[r]{1e-05}}
\put(420,151){\makebox(0,0)[r]{1e-06}}
\end{picture}
\label{fig-Lpdf}
\end{figure}

Because our model offers a comparison for the
PDF, or equivalently for all structure functions, rather
than just the scaling exponents, we can do a careful
study to compare the model with the experiments and 
have a better understanding of the viscous scaling
where the transition is seen.
The PDF at the beginning of the cascade can be determined experimentally
once we choose a sensible outer scale. Since
the experiments were carried out in a fixed geometry,
we expect that at a fixed outer scale, where the energy is fed in, 
the PDF of velocity differences is independent of the 
Reynolds number. We confirm this by analyzing the 
experimental data. By inspecting the scaling of the second order
structure functions, we choose the outer scale to be $L=0.73cm$ 
for all flows. The scale is roughly 1/5 of the integral scale
quoted in reference \cite{zoc94}. The reason we chose $L$ to be 
somewhat smaller is that below $L$ the scaling 
follows a power law, and boundary effects seem to be negligible. We 
then examine the PDFs of the normalized velocity differences over
the separation $L$, and find that they collapse for all the
flows as shown in Fig.\ref{fig-Lpdf}. 
The velocities are normalized by their variance $U_{rms}$.
In the same plot, a Gaussian distribution 
is shown as a dot-dashed line for comparison.
The PDF at the outer scale is already non-Gaussian because 
we have chosen $L$ to be smaller than the integral scale 
of \cite{zoc94}, for the reasons given above. 
Thus some growth of intermittent fluctuations has already taken place. 
Also the dynamics of the boundary layer may have an important 
effect, because the experiment is in a closed geometry.
The small asymmetry of the PDF  is consistent 
with the  Kolmogorov structure equation, which gives
a non-vanishing  skewness of the velocity.
By the construction of the model, the asymmetry will 
propagate down to smaller scales.
The large scale velocity variances, which differ because of different 
driving, set the velocity scales at each cascade level.
The ratio of the velocity variance  and the 
rotational velocity  at a fixed scale will be shown later. 
We emphasize that the collapse of the PDFs holds for 
Taylor Reynolds number ($\Rel$) ranging from $300$ to $2000$, 
which include both the flows below and above the reported 
transition at $\Rel \approx 700$.

\begin{figure}[htb]
\caption{PDF of the velocity difference at the scale $L/4$ 
(still in the inertial range) for the same 
experimental runs as shown in Fig. 1. Again,
the velocity difference is normalized by its variance, 
and the PDF is normalized to unity.  }
\setlength{\unitlength}{0.1bp}
\special{!
/gnudict 40 dict def
gnudict begin
/Color false def
/Solid false def
/gnulinewidth 5.000 def
/vshift -33 def
/dl {10 mul} def
/hpt 31.5 def
/vpt 31.5 def
/M {moveto} bind def
/L {lineto} bind def
/R {rmoveto} bind def
/V {rlineto} bind def
/vpt2 vpt 2 mul def
/hpt2 hpt 2 mul def
/Lshow { currentpoint stroke M
  0 vshift R show } def
/Rshow { currentpoint stroke M
  dup stringwidth pop neg vshift R show } def
/Cshow { currentpoint stroke M
  dup stringwidth pop -2 div vshift R show } def
/DL { Color {setrgbcolor Solid {pop []} if 0 setdash }
 {pop pop pop Solid {pop []} if 0 setdash} ifelse } def
/BL { stroke gnulinewidth 2 mul setlinewidth } def
/AL { stroke gnulinewidth 2 div setlinewidth } def
/PL { stroke gnulinewidth setlinewidth } def
/LTb { BL [] 0 0 0 DL } def
/LTa { AL [1 dl 2 dl] 0 setdash 0 0 0 setrgbcolor } def
/LT0 { PL [] 0 1 0 DL } def
/LT1 { PL [4 dl 2 dl] 0 0 1 DL } def
/LT2 { PL [2 dl 3 dl] 1 0 0 DL } def
/LT3 { PL [1 dl 1.5 dl] 1 0 1 DL } def
/LT4 { PL [5 dl 2 dl 1 dl 2 dl] 0 1 1 DL } def
/LT5 { PL [4 dl 3 dl 1 dl 3 dl] 1 1 0 DL } def
/LT6 { PL [2 dl 2 dl 2 dl 4 dl] 0 0 0 DL } def
/LT7 { PL [2 dl 2 dl 2 dl 2 dl 2 dl 4 dl] 1 0.3 0 DL } def
/LT8 { PL [2 dl 2 dl 2 dl 2 dl 2 dl 2 dl 2 dl 4 dl] 0.5 0.5 0.5 DL } def
/P { stroke [] 0 setdash
  currentlinewidth 2 div sub M
  0 currentlinewidth V stroke } def
/D { stroke [] 0 setdash 2 copy vpt add M
  hpt neg vpt neg V hpt vpt neg V
  hpt vpt V hpt neg vpt V closepath stroke
  P } def
/A { stroke [] 0 setdash vpt sub M 0 vpt2 V
  currentpoint stroke M
  hpt neg vpt neg R hpt2 0 V stroke
  } def
/B { stroke [] 0 setdash 2 copy exch hpt sub exch vpt add M
  0 vpt2 neg V hpt2 0 V 0 vpt2 V
  hpt2 neg 0 V closepath stroke
  P } def
/C { stroke [] 0 setdash exch hpt sub exch vpt add M
  hpt2 vpt2 neg V currentpoint stroke M
  hpt2 neg 0 R hpt2 vpt2 V stroke } def
/T { stroke [] 0 setdash 2 copy vpt 1.12 mul add M
  hpt neg vpt -1.62 mul V
  hpt 2 mul 0 V
  hpt neg vpt 1.62 mul V closepath stroke
  P  } def
/S { 2 copy A C} def
end
}
\begin{picture}(3600,2160)(0,0)
\special{"
gnudict begin
gsave
50 50 translate
0.100 0.100 scale
0 setgray
/Helvetica findfont 100 scalefont setfont
newpath
-500.000000 -500.000000 translate
LTa
LTb
480 151 M
63 0 V
2874 0 R
-63 0 V
480 249 M
31 0 V
2906 0 R
-31 0 V
480 379 M
31 0 V
2906 0 R
-31 0 V
480 446 M
31 0 V
2906 0 R
-31 0 V
480 477 M
63 0 V
2874 0 R
-63 0 V
480 576 M
31 0 V
2906 0 R
-31 0 V
480 705 M
31 0 V
2906 0 R
-31 0 V
480 772 M
31 0 V
2906 0 R
-31 0 V
480 804 M
63 0 V
2874 0 R
-63 0 V
480 902 M
31 0 V
2906 0 R
-31 0 V
480 1032 M
31 0 V
2906 0 R
-31 0 V
480 1098 M
31 0 V
2906 0 R
-31 0 V
480 1130 M
63 0 V
2874 0 R
-63 0 V
480 1228 M
31 0 V
2906 0 R
-31 0 V
480 1358 M
31 0 V
2906 0 R
-31 0 V
480 1425 M
31 0 V
2906 0 R
-31 0 V
480 1456 M
63 0 V
2874 0 R
-63 0 V
480 1555 M
31 0 V
2906 0 R
-31 0 V
480 1684 M
31 0 V
2906 0 R
-31 0 V
480 1751 M
31 0 V
2906 0 R
-31 0 V
480 1783 M
63 0 V
2874 0 R
-63 0 V
480 1881 M
31 0 V
2906 0 R
-31 0 V
480 2011 M
31 0 V
2906 0 R
-31 0 V
480 2077 M
31 0 V
2906 0 R
-31 0 V
480 2109 M
63 0 V
2874 0 R
-63 0 V
480 151 M
0 63 V
0 1895 R
0 -63 V
970 151 M
0 63 V
0 1895 R
0 -63 V
1459 151 M
0 63 V
0 1895 R
0 -63 V
1949 151 M
0 63 V
0 1895 R
0 -63 V
2438 151 M
0 63 V
0 1895 R
0 -63 V
2928 151 M
0 63 V
0 1895 R
0 -63 V
3417 151 M
0 63 V
0 1895 R
0 -63 V
480 151 M
2937 0 V
0 1958 V
-2937 0 V
480 151 L
LT0
3174 1946 D
906 204 D
1023 240 D
1082 240 D
1111 183 D
1141 322 D
1170 353 D
1200 373 D
1229 507 D
1258 557 D
1288 616 D
1317 716 D
1346 755 D
1376 822 D
1405 893 D
1435 962 D
1464 1026 D
1493 1100 D
1523 1164 D
1552 1231 D
1581 1293 D
1611 1361 D
1640 1428 D
1669 1495 D
1699 1559 D
1728 1624 D
1758 1687 D
1787 1750 D
1816 1810 D
1846 1868 D
1875 1920 D
1904 1966 D
1934 1998 D
1963 2007 D
1993 1985 D
2022 1941 D
2051 1883 D
2081 1816 D
2110 1745 D
2139 1670 D
2169 1595 D
2198 1519 D
2228 1443 D
2257 1371 D
2286 1292 D
2316 1219 D
2345 1144 D
2374 1063 D
2404 993 D
2433 924 D
2462 862 D
2492 792 D
2521 722 D
2551 653 D
2580 581 D
2609 555 D
2639 424 D
2668 322 D
2697 269 D
2727 223 D
LT1
3174 1846 A
1200 338 A
1229 353 A
1258 428 A
1288 551 A
1317 617 A
1346 731 A
1376 778 A
1405 870 A
1435 945 A
1464 1016 A
1493 1091 A
1523 1149 A
1552 1217 A
1581 1284 A
1611 1350 A
1640 1420 A
1669 1487 A
1699 1555 A
1728 1624 A
1758 1689 A
1787 1753 A
1816 1814 A
1846 1872 A
1875 1925 A
1904 1967 A
1934 1995 A
1963 2001 A
1993 1983 A
2022 1942 A
2051 1887 A
2081 1821 A
2110 1749 A
2139 1673 A
2169 1598 A
2198 1520 A
2228 1442 A
2257 1365 A
2286 1285 A
2316 1204 A
2345 1125 A
2374 1052 A
2404 970 A
2433 889 A
2462 826 A
2492 746 A
2521 674 A
2551 647 A
2580 557 A
2609 497 A
2639 385 A
2668 240 A
LT2
3174 1746 B
1170 240 B
1200 204 B
1229 240 B
1258 497 B
1288 548 B
1317 705 B
1346 718 B
1376 748 B
1405 852 B
1435 922 B
1464 1001 B
1493 1077 B
1523 1145 B
1552 1208 B
1581 1276 B
1611 1349 B
1640 1417 B
1669 1484 B
1699 1553 B
1728 1620 B
1758 1688 B
1787 1755 B
1816 1817 B
1846 1875 B
1875 1926 B
1904 1968 B
1934 1994 B
1963 1999 B
1993 1981 B
2022 1943 B
2051 1888 B
2081 1822 B
2110 1750 B
2139 1675 B
2169 1599 B
2198 1519 B
2228 1440 B
2257 1359 B
2286 1288 B
2316 1210 B
2345 1133 B
2374 1051 B
2404 974 B
2433 890 B
2462 841 B
2492 774 B
2521 676 B
2551 581 B
2580 499 B
2609 511 B
2639 312 B
LT3
3174 1646 C
1200 353 C
1229 255 C
1258 424 C
1288 514 C
1317 665 C
1346 735 C
1376 831 C
1405 903 C
1435 962 C
1464 1025 C
1493 1086 C
1523 1151 C
1552 1220 C
1581 1290 C
1611 1357 C
1640 1420 C
1669 1487 C
1699 1554 C
1728 1621 C
1758 1686 C
1787 1749 C
1816 1813 C
1846 1872 C
1875 1925 C
1904 1969 C
1934 1997 C
1963 2003 C
1993 1983 C
2022 1940 C
2051 1884 C
2081 1819 C
2110 1747 C
2139 1672 C
2169 1596 C
2198 1520 C
2228 1450 C
2257 1372 C
2286 1296 C
2316 1224 C
2345 1145 C
2374 1074 C
2404 1000 C
2433 931 C
2462 867 C
2492 751 C
2521 625 C
2551 520 C
2580 483 C
2609 390 C
2639 428 C
2668 281 C
LT4
3174 1546 T
1170 183 T
1200 424 T
1229 428 T
1258 562 T
1288 558 T
1317 631 T
1346 651 T
1376 764 T
1405 871 T
1435 931 T
1464 991 T
1493 1066 T
1523 1148 T
1552 1214 T
1581 1280 T
1611 1348 T
1640 1419 T
1669 1486 T
1699 1554 T
1728 1621 T
1758 1687 T
1787 1752 T
1816 1814 T
1846 1874 T
1875 1926 T
1904 1968 T
1934 1996 T
1963 2002 T
1993 1982 T
2022 1940 T
2051 1884 T
2081 1819 T
2110 1749 T
2139 1677 T
2169 1601 T
2198 1523 T
2228 1446 T
2257 1375 T
2286 1293 T
2316 1211 T
2345 1140 T
2374 1073 T
2404 983 T
2433 934 T
2462 866 T
2492 756 T
2521 648 T
2551 600 T
2580 507 T
2609 477 T
2639 373 T
2668 240 T
2697 183 T
2727 204 T
2756 157 T
LT5
3114 1446 M
180 0 V
1597 151 M
14 147 V
22 209 V
22 195 V
21 181 V
22 168 V
22 153 V
21 140 V
22 126 V
22 111 V
21 99 V
22 84 V
22 70 V
21 56 V
22 43 V
22 29 V
21 15 V
22 0 V
22 -12 V
21 -27 V
22 -41 V
21 -54 V
22 -69 V
22 -82 V
21 -96 V
22 -110 V
22 -124 V
21 -138 V
22 -152 V
22 -165 V
21 -179 V
22 -194 V
22 -207 V
17 -176 V
stroke
grestore
end
showpage
}
\put(3054,1446){\makebox(0,0)[r]{Gaussian}}
\put(3054,1546){\makebox(0,0)[r]{Re=2.88E4}}
\put(3054,1646){\makebox(0,0)[r]{Re=1.17E4}}
\put(3054,1746){\makebox(0,0)[r]{Re=1.07E4}}
\put(3054,1846){\makebox(0,0)[r]{Re=1.86E3}}
\put(3054,1946){\makebox(0,0)[r]{Re=1.36E3}}
\put(3417,51){\makebox(0,0){15}}
\put(2928,51){\makebox(0,0){10}}
\put(2438,51){\makebox(0,0){5}}
\put(1949,51){\makebox(0,0){0}}
\put(1459,51){\makebox(0,0){-5}}
\put(970,51){\makebox(0,0){-10}}
\put(480,51){\makebox(0,0){-15}}
\put(420,2109){\makebox(0,0)[r]{1}}
\put(420,1783){\makebox(0,0)[r]{0.1}}
\put(420,1456){\makebox(0,0)[r]{0.01}}
\put(420,1130){\makebox(0,0)[r]{0.001}}
\put(420,804){\makebox(0,0)[r]{0.0001}}
\put(420,477){\makebox(0,0)[r]{1e-05}}
\put(420,151){\makebox(0,0)[r]{1e-06}}
\end{picture}
\label{fig-Lpdf2}
\end{figure}
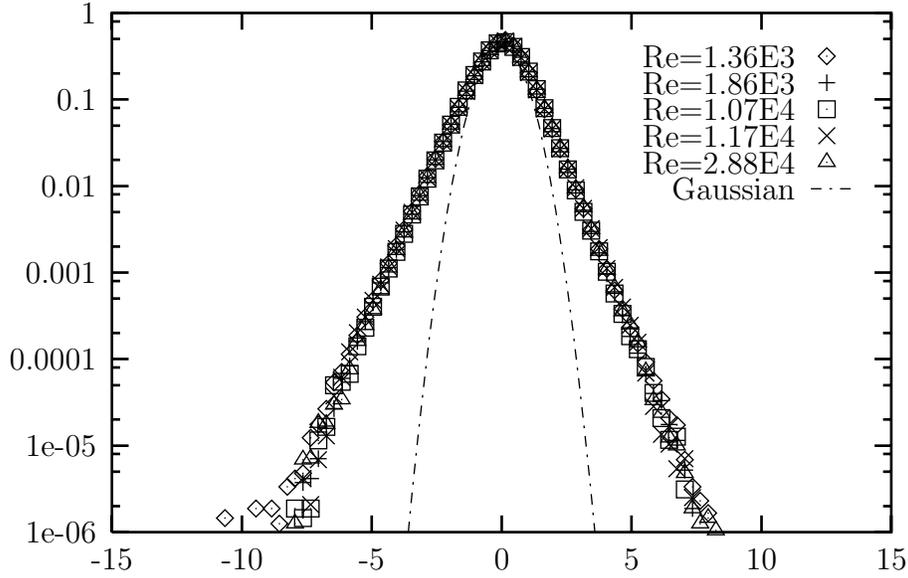

\noindent The collapse of the outer scale PDF $P_0(u)$ 
implies the collapse of PDF 
in the inertial range at a fixed
separation. To check this, we calculate the  experimental
PDF at a separation scale of $L/4$, and again we observe the 
collapse of the PDFs as shown in Fig.\ref{fig-Lpdf2}.

We recall that Tabeling et. al. observed the collapse 
of the PDF in the inertial range only for sufficiently
large Reynolds numbers \cite{tab95}.  However, 
as we show here, the collapse works equally well
for small Reynolds number ($\Rel < 700$, and correspondingly,
$Re < 10E4$) flows in the inertial
range.  We suspect the reason that the previous authors 
did not observe the same collapse for small
Reynolds number flows is because they fixed
a scale too small to be in the inertial range of the
small Reynolds number flows.

Because the shape of the normalized
outer scale PDF is constant at the energy-input scale,
the corresponding velocity scale is uniquely 
determined by its variance $U_{rms}(L)$. 
It allows us to define a Reynolds number 

\be
Re = \frac{U_{rms}(L) L}{\nu}. \label{eq-Re}
\ee

\noindent We remark that this definition is different from 
the one used previously\cite{zoc94}:

\be
Re = \frac{\Omega R^2}{\nu}, \label{eq-exRe}
\ee

\noindent where $\Omega$ is the rotation frequency and $R$ is 
the radius of the apparatus. Equation (\ref{eq-exRe})
assumes that the velocity at the energy-input scale is 
proportional to $\Omega R$. We tabulate the ratio of $U_{rms}$ 
and $\Omega R$ for different Reynolds numbers in Table $2$ 
and find that they vary considerably.  Many factors
can contribute to these differences. Apart from the systematic
errors in the measurements, the shear velocity profile depends
on the Reynolds number. A better understanding
of the instabilities of the driving flow  and  
the dependence of the large-scale quantity $\Omega R/U_{rms}(L)$ 
on viscosity is desirable. For our purposes, the Reynolds 
number defined in Equation (\ref{eq-Re}) is adequate. 

Another frequently used dimensionless quantity is the Taylor
Reynolds number:

\begin{eqnarray}
&&R_\lambda = \frac{U_{rms} \lambda}{\nu}, \nonumber\\
&&\lambda = \frac{U_{rms}}{\sqrt{(du/dx)^2}} = 
U_{rms} \sqrt{\frac{15\nu}{\epsilon}},
\label{rl}
\end{eqnarray}

\noin where $\epsilon$ is the average energy dissipation. The last
identity assumes isotropy of the flows. Here $\epsilon$ can
be estimated by the large scale flow:

\be
\epsilon = -\frac{4}{5} \frac{U_{rms}^3(L)}{L}.
\ee

\noin In the subsequent discussions, we shall use the $Re$ defined in
Eq. (\ref{eq-Re}), and the corresponding $\Rel$ 
can be found in Table (\ref{tab3}) below.

The remaining two parameters are 
$Re_{cr}$ and $\Delta$ defined in the previous section.
The critical Reynolds number 
$Re_{cr}$ defines the threshold to be compared with the local Reynolds number 
$Re_r$. If $Re_r < Re_{cr}$, viscous 
diffusion dominates and the cascade stops. 
In real turbulence, the termination of the cascade 
is a gradual process. The gradual crossover is parameterized
by $\Delta$ in our model. One expects  
$Re_{cr}$ and $\Delta$ to characterize the 
viscous cutoff mechanism, independent of large scale flow. 
Ideally, one would like to choose one particular flow to fix 
$Re_{cr}$ and $\Delta$, and use the same values 
for the rest of the flows studied. 
We remark that the constancy of $Re_{cr}$ is an assumption 
of the multifractal theory of turbulence, which has not been checked 
explicitly before.

We now determine the values of $Re_{cr}$ and $\Delta$ using 
the flow with $Re = 1.36E3$. Starting from the PDF $P_0$ of
the velocity at the outer scale, we compute the 
evolution of the PDFs, and consequently the structure functions
at smaller scales. The values of $Re_{cr}$ and $\Delta$ are
adjusted such  that the resulting $D_2(r)$ agrees best
with experiment. Because
our algorithm 
allows for continuous evolution steps, we can compute a
sufficient number of points along the $D_2(r)$ curve
necessary for a good comparison. We find 

\be
Re_{cr} = 85 \pm 3, \hspace{0.2cm} \mbox{and} 
\hspace{0.2cm} \Delta = 0.4 \pm 0.1.
\ee

\begin{figure}[htb]
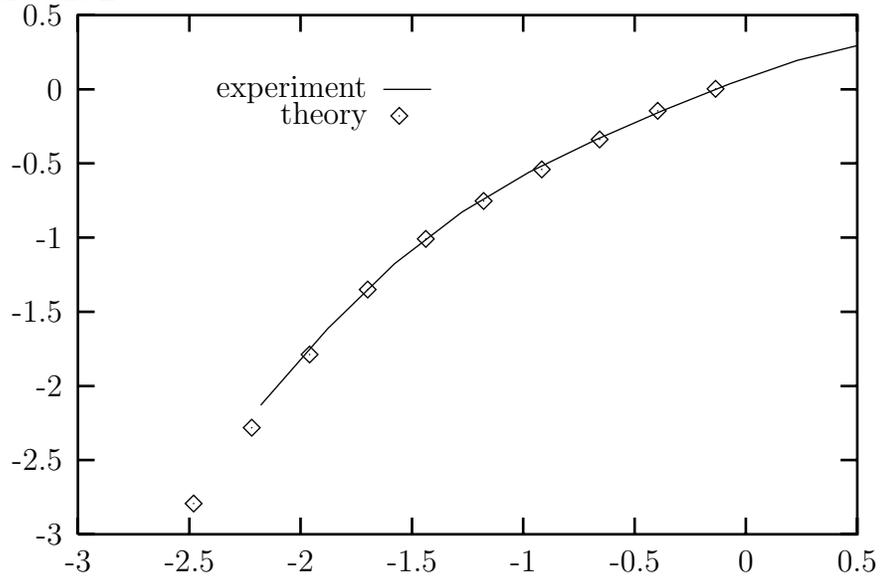

\caption{$\log D_2(r)$ plotted against $\log(r)$ on a log-log scale. 
The separation $r$ is normalized by the outer scale $L$.
  }
\input pictures/fig-d2.tex
\label{fig-d2}
\end{figure}

\noindent Figure \ref{fig-d2} shows the comparison 
of the predicted $D_2(r)$ and the experimental measurement.
Because the crossover in $D_2(r)$ is a sensitive
function of $Re_{cr}$, this comparison gives a relative
small fitting error, which is about $3\%$, as opposed to
inspecting the overlap of the PDFs, which gives errors of
about 15\%.

We find that the predicted PDFs are insensitive to the
values of  $\Delta$, which is fixed to be $0.4$ for all flows.
To check whether $Re_{cr}$ is indeed independent of Reynolds number, 
we repeat the above procedure for all the flows. 
The value of $Re_{cr}$ is shown in the fourth column
of Table \ref{tab3}. It is evident that there are considerable 
fluctuations in $Re_{cr}$, which correspond to about a factor 
of two in the value of the crossover scale between inertial and
viscous range. In what follows we will use $Re_{cr}$ as quoted 
in Table \ref{tab3} for the individual runs, since otherwise the 
fits of the PDFs in the viscous range would be poor. Unfortunately,
this means that there is an adjustable parameter for each run. 
We will bear this in mind when we compare the theory and the experiments.
From the experimental data available to us, we are not able
to pinpoint the reason for the deviations of $Re_{cr}$ from a 
constant value. However, these fluctuations are at least
consistent with the fluctuations seen in the energy
dissipation\cite{zoc94,ems96}.

Having determined all the adjustable parameters in the model, we 
can proceed to make predictions for various quantities of interest.
Table \ref{tab3} is a summary of the experimental and model 
parameters for the flows studied in this paper.

\section{Evolution of PDFs and their asymptotics}

In this section we show three results: 1) the 
model gives a correct description of the inertial range
behavior, 2) it also captures the crossover from the inertial
range to the dissipation range, and 3) the model gives
an asymptotic probability distribution for the velocity
derivatives, based upon which we compute the flatness.
1) and 2) have  partly been seen in our previous discussion
on $D_2(r)$. Here we  look at them again in the PDF of the 
velocity differences. 

We start from the PDF of velocity differences at the outer scale,
and compute the subsequent PDFs as the scale decreases.
We show the evolution of the PDFs for two
typical flows: one below and one above the transition at the Reynolds number
$\Rel \approx 700$.
Figure \ref{fig-pdfevol-35V} shows the evolution of the PDFs
for $Re = 1.36E3 $, which corresponds to the Taylor Reynolds
number $\Rel= 344$, and Fig. \ref{fig-pdfevol-15V}
for $Re= 1.17E4 $ at $Re_{\lambda}=1626$.
The solid line represents the experimental
measurement, while the diamonds show the theoretical prediction.
Each series covers scales ranging from 
the outer scale to the smallest scale measured by the
experiments.

Similar pictures are obtained for all the
flows we study.  As expected, the shape of the PDF evolves
toward stretched-exponential curves. The stretched
tails describe the increasingly frequent occurrences
of large intermittent events.
The distribution of 
the large events make significant contributions
to higher moments. A typical measure of 
intermittency is the flatness, which will be the focus
of the next section.

In the case of the smaller $Re$ flow, theory and experiment
agree down to the smallest scale resolved by the experiment.
Both the experimental and the theoretical PDFs show no 
sign of reaching their asymptotics even at the smallest scales.
In the case of the large $Re$ flow, theory and 
experiment agree for  a number of cascade steps, but eventually
deviate as the scale decreases toward the limit of experimental
resolution.  While the theoretical PDF
continues to evolve, the experimental PDF saturates.
This saturation corresponds to a saturation in the flatness,
as shown below. 

We recall that the second order structure function 
$D_2(r)$ agrees very well 
between theory and experiment at all scales and for both the smaller
and the large Reynolds number flows.
This however does not contradict the deviation we see
in the PDF, which occur only in the tail of the distribution,
because $D_2(r)$ does not sensitively depend on the tail distribution.

To better understand the deviations in the PDF,
we tabulate the scale $L_c$, where theory and experiment
starts to deviate for all experimental runs in Table \ref{tab4}.
At the smallest Reynolds number there is agreement on all available scales, 
so $L_c$ can only be bounded from above. For the other Reynolds numbers
we have given the ratios $L_c/L$ and $L_c/\eta$. 
The fact that $L_c/L$ is constant for the 
higher Reynolds numbers indicates that there is a constant scale in the 
system. The other ratio $L_c/\eta$, on the 
other hand, does not offer a simple pattern. 

We next use our model to follow up on the evolution of the PDFs below 
the smallest spatial resolution of the experiments. This should allow 
us to estimate whether experiments can reliably approximate the 
distribution of velocity derivatives. Figure \ref{fig-pdfasym} 
shows the theoretical PDFs at four different values of $r/\eta$. 
The largest, $r/\eta = 19.3$, is the separation where theory and 
experiment start to deviate. At half that separation, which is the 
smallest one measured experimentally, the theory predicts a much wider 
distribution. The theoretical distribution continues to evolve 
down to about half that scale, as shown by the distributions 
at $r/\eta = 4.8$ and $r/\eta = 2.4$. 

Thus the theoretical PDF  reaches its
asymptotics close to the Kolmogorov
scale, typically at  $r \sim \eta$.  This then
suggests that in order to understand the scaling of the velocity
derivatives, the experiments need to resolve a smaller
scale than they currently do, which is on the order
of $ 3\eta$ to $10 \eta$, and is $9.6 \eta$ for this particular
flow. On the basis of order of magnitude arguments,
the difference between $ 1\eta$ and $10\eta$ is perhaps insignificant.
However, we will show in the next section that the difference in 
the prefactor is crucial and is significant in interpreting
the experimental measurements of the flatness.

\section{Flatness}

Once we have the PDF $P(x)$ for the velocity difference 
in the asymptotic limit $r\rightarrow0$, it is straightforward 
to compute the flatness: 
\begin{equation}
F_0 = \frac{\int P(x) x^4 dx}{(\int P(x) x^2 dx)^2} .
\label{flat}
\end{equation}
It is also more revealing to plot the flatness $F(r)$ as
function of scale
\be
F(r) = \frac{D_4(r)}{(D_2(r))^2},
\label{flatr}
\ee
which for $r \rightarrow0$ coincides with the definition (\ref{flat}).
This allows us to compare our model with the experiments.
Figure \ref{fig-flat-theo} shows the 
theoretical prediction for $F(r)$ for the Reynolds numbers studied, which
is seen to saturate at a constant value $F_0$. This asymptotic value
is consistent with a power law 
\begin{equation}
F_0 \sim Re_{\lambda}^{0.15}.
\label{f0}
\end{equation}
The exponent agrees with the compilation of flatness by Van Atta et. al.
\cite{van80}.
The most important feature we observe in Fig. \ref{fig-flat-theo}
is the pronounced crossover behavior, which depends
on the Reynolds number. In particular, although in the
true asymptotic limit the flatness increases with the Reynolds
number, at a fixed scale in the crossover region the flatness decreases
with the Reynolds number as clearly shown in this figure. Therefore,
if experiments are limited to the resolution in the crossover region (which
is typically $3 -10 \eta$), it is conceivable that the experiments
can yield a result which shows a decrease of flatness with
Reynolds number.

To understand the difference between theory and experiments
in the crossover range  in more detail, 
we plot $F(r)$ for a single flow and also 
compare it with experimental measurements. 
In the theoretical curve, three distinct regions are seen. At large 
scales, which correspond to the inertial range, $F(r)$ increases like 
$r^{\zeta_4 - 2\zeta_2}$. At about $15\eta$ viscosity becomes effective
and $F(r)$ increases much more sharply, because of the additional fluctuations 
introduced by the fluctuating cutoff. But only at $r \approx 1\eta$ does
$F(r)$ fully saturate, in agreement with our observations of the 
probability distribution. By comparison,
the experimental results shows a relatively small viscous rise.
Thus the experiments show only a small part of
the viscous cutoff fluctuations predicted by theory.

One can further understand the width of the crossover range as
a function of Reynolds number.
The higher the viscosity, the earlier  does the rise of $F(r)$ set in.
At the same time, the rise is sharper compared to the low viscosity
flows, since the viscous cutoff fluctuates
over a narrower range of scales \cite{Egg:91c}. The combined effect 
leads to a {\it reversal} in the magnitude of $F$ at intermediate 
separations, the lowest Reynolds number leading to the largest value 
of the flatness, as we see in Fig. 7.  This suggests strongly 
that one has to be extremely  careful about
comparing $F$ at different Reynolds numbers if the spatial  
resolution is in the crossover range.

\section{Discussion}

We have studied in detail a model of PDFs of turbulent
velocity differences with the aim of understanding 
the crossover below the inertial range for two reasons:
1) the crossover range is far more complex and less understood
than both the inertial range and the far dissipation range, and
2) we hope to gain some quantitative understanding of the behavior
of flatness in the crossover range, which may shed some light
on the recent controversy about the transition seen in the 
flatness \cite{tab95}.

In particular, we compare our model with the recent series of experiments
by Tabeling's group. We find that in the small Reynolds number flows,
the model and the experiments agree on the shape of PDFs down 
to the smallest scale resolved. On the other hand, in  the
large Reynolds number flows, the model shows departure from
the experiment below a scale defined as $L_c$, which is on
the order of $10 \eta$. 

We also observe that in the crossover range between $2 \eta$ 
and $20 \eta$ the model predicts that the flatness
does not necessarily increase with Reynolds number.
The flatness may also decrease as function of the Reynolds number
at an intermediate scale. This results from a rapid growth of the 
flatness in the crossover regime, which is often neglected
in the usual picture of the scale dependence of the flatness.
Furthermore, the width of the crossover region does not
simply scale a la Kolmogorov because of the fluctuations
in the cutoff scale. Both effects are due to the viscous
fluctuation in the crossover range. Thus the observed decrease of the flatness
with Reynolds number in the crossover region can be interpreted
as a signature of multifractality.
We emphasize that it is crucial to take the crossover
into account when interpreting the experiments, since most often
the experiment resolution coincides with the crossover range.

There are some remaining puzzles. For example, 
we do not completely understand the fluctuations seen in the
critical Reynolds number $Re_{cr}$ defined for this model, 
nor do we understand why the flatness does not rise in
the crossover range as sharply as predicted by the model.
A better understanding of these issues requires further
comparisons between this model and other experiments such
as large closed geometry experiments or open flow
experiments.

\acknowledgements
We thank L.~Biferale, L.~Kadanoff, D.~Lohse, and P.~Tabeling 
for useful discussions. We are indebted 
to Tabeling's group at Ecole Normale Superieure for
sharing their data with us. ZJW is grateful to the 
hospitality of Tabeling's group during her visits supported
by the NATO grant under award number CRG-950245. 
She also acknowledges the support by the NSF-MRSEC
Program at the University of Chicago and by the EPSRC
grant through University of Oxford.
JE is supported by the Deutsche Forschungsgemeinschaft 
through Sonderforschungsbereich 237.

\vspace{0.5cm}

\noindent $\dagger$ Current address: 
Courant Institute of Mathematical Sciences,
New York University, 251 Mercer St., New York, NY 10012.
Electronic mail: jwang@cims.nyu.edu.

\appendix
\section{Smoothing}
Here we briefly describe how a smooth function $j_c(u_r)$ 
was obtained from (\ref{eq:2-2}). Putting 
$\ell = \log_2\frac{u_r}{R}$, we first make sure that 
$j_c = \frac{1}{2}(n + \ell)$ smoothly merges into $j=n$ 
for $\ell \approx n$. A convenient parameterization is
\begin{equation}
\bar{j} = n + \frac{1}{2}\left[\frac{\ell-n}{2} - 
(\Delta + \frac{(\ell-n)^2}{4})^{1/2}\right] .
\label{eq:A-1}
\end{equation}
Here we introduced a parameter $\Delta$, which measures the
width of the transition region. In Section 3 we compare with
the viscous crossover of an experimentally measured structure function 
$D_2(r)$ and find $\Delta=1$ to accurately describe experiments. 

But $\bar{j}$, as defined by (\ref{eq:A-1}) still goes to $-\infty$
as $u_r \rightarrow 0$, while the lowest available level is $0$. 
The approach of $j = 0$ has to be fast enough to make 
$\partial j_c/\partial u_r$ go to zero as $u_r \rightarrow 0$.
Namely, in view of (\ref{eq:2-4}) this means that 
\begin{equation}
P_r(u_r) \approx 2^n P_0(2^n u_r) ,
\label{eq:A-2}
\end{equation}
so one simply sees the gradient of the large scale fluctuations,
as expected. This is achieved by putting
\begin{equation}
j_c = \frac{1}{4} \log_2(1 + 2^{4\bar{j}}) .
\label{eq:A-3}
\end{equation}
We found that the details of the smoothing were inconsequential
to the shape of the distribution as long as the basic properties
of $j_c(u_r)$ were satisfied. Thus (\ref{eq:A-1}) and (\ref{eq:A-3})
determine the function $j_c(u_r)$ we will be using throughout 
this paper.

\bibliographystyle{unsrt}

\clearpage

\begin{table}[htb]
  \begin{center}
     \leavevmode
\begin{tabular}{|c|c|c|c|c|c|c|c|c|c|}   \hline
q         &2     &3      &4     &5     &6     &7      &8     &9
&10 \\ \hline
exp.      &0.70  &1.00   &1.26  &1.50  &1.71  &1.90   &2.08  &2.19
&2.30 \\ \hline
model     &0.70  &1.00   &1.26  &1.50  &1.71  &1.89   &2.05  &2.19
&2.31 \\ \hline
\end{tabular}
   \end{center}
\caption{ Inertial range 
scaling exponents $\zeta_q$ of the velocity field. 
Compared are the experimental measurements 
and our fit Eqs. (\ref{eq-ps}), (\ref{eq-para}).} 
  \label{tab1}
\end{table}

\vspace{0.3cm}

\begin{table}[]
  \begin{center}
     \leavevmode
\begin{tabular}{|c|c|c|c|c|c|}   \hline
$Re$                  &1.36E3   &1.86E3  &1.07E4  &1.17E4 &2.88E4 \\ \hline
$\Omega R/U_{rms}(L)$  &2.31     &0.71    &0.89    &0.46   &1.26   \\  \hline
\end{tabular}
   \end{center}
\caption{Comparison of our definition (\protect{\ref{eq-Re}}) of the Reynolds
number, and the definition (\protect{\ref{eq-exRe}}) of the 
experimental group for the different flows under consideration.    }
  \label{tab2}
\end{table}

\vspace{0.3cm}

\vspace{0.3cm}
\begin{table}[]
  \begin{center}
     \leavevmode
\begin{tabular}{|c|c|c|c|c|c|}\hline
   $Re$   & $Urms(L)(cm/s)$  &$\nu(cm^{2}/s)$ &$Re_{cr}$  & $\Rel$ 
&$\eta(\mu m)$ \\ \hline
  1.360E3 & 34.93   &1.88E-2 & 85.7  & 344.0 &49.0 \\ \hline
  1.863E3 & 10.18   &4.E-3   & 41.0  & 600.0 &20.0 \\ \hline
  1.170E4 & 18.54   &1.16E-3 & 58.5  & 1626. &5.9 \\ \hline
  1.073E4 & 20.52   &1.4E-3  & 64.38 & 1802. &6.0  \\ \hline
  2.881E4 & 44.08   &1.12E-3 & 115.2 & 2394. &4.1  \\  \hline
\end{tabular}
   \end{center}
\caption{Experimental parameters and fitted values of $Re_{cr}$ for the 
experimental runs studied in this paper.     }
  \label{tab3}
\end{table}

\vspace{0.3cm}

\vspace{0.3cm}
\begin{table}[]
   \begin{center}
     \leavevmode
\begin{tabular}{|c|c|c|} \hline
$Re$    & $L_c/L$     & $L_c/\eta$ \\ \hline
1.360E3 & $< 3E{-2}$       & $<4.5$  \\ \hline
1.863E3 & $2.2E{-2}$   & 8.1\\ \hline
1.073E4 & $1.1E{-2}$   & 13.5\\ \hline
1.170E4 & $1.1E{-2}$   & 13.8\\ \hline
2.881E4 & $1.1E{-2}$   & 19.6\\ \hline
\end{tabular}
  \end{center}
   \caption{The smallest scale $L_c$ where the multifractal model still works.}
  \label{tab4}
 \end{table}
\vspace{0.3cm}

\clearpage

\vspace{0.3cm}

\newpage

\newpage

\begin{figure}[htb]
\caption{Evolution of the logarithm of PDFs of velocity differences
over different separations at $Re=1.36E3$. 
The velocity difference is normalized by its variance, and
the PDF is normalized to unity.}

\setlength{\unitlength}{0.1bp}
\special{!
/gnudict 40 dict def
gnudict begin
/Color false def
/Solid false def
/gnulinewidth 5.000 def
/vshift -33 def
/dl {10 mul} def
/hpt 31.5 def
/vpt 31.5 def
/M {moveto} bind def
/L {lineto} bind def
/R {rmoveto} bind def
/V {rlineto} bind def
/vpt2 vpt 2 mul def
/hpt2 hpt 2 mul def
/Lshow { currentpoint stroke M
  0 vshift R show } def
/Rshow { currentpoint stroke M
  dup stringwidth pop neg vshift R show } def
/Cshow { currentpoint stroke M
  dup stringwidth pop -2 div vshift R show } def
/DL { Color {setrgbcolor Solid {pop []} if 0 setdash }
 {pop pop pop Solid {pop []} if 0 setdash} ifelse } def
/BL { stroke gnulinewidth 2 mul setlinewidth } def
/AL { stroke gnulinewidth 2 div setlinewidth } def
/PL { stroke gnulinewidth setlinewidth } def
/LTb { BL [] 0 0 0 DL } def
/LTa { AL [1 dl 2 dl] 0 setdash 0 0 0 setrgbcolor } def
/LT0 { PL [] 0 1 0 DL } def
/LT1 { PL [4 dl 2 dl] 0 0 1 DL } def
/LT2 { PL [2 dl 3 dl] 1 0 0 DL } def
/LT3 { PL [1 dl 1.5 dl] 1 0 1 DL } def
/LT4 { PL [5 dl 2 dl 1 dl 2 dl] 0 1 1 DL } def
/LT5 { PL [4 dl 3 dl 1 dl 3 dl] 1 1 0 DL } def
/LT6 { PL [2 dl 2 dl 2 dl 4 dl] 0 0 0 DL } def
/LT7 { PL [2 dl 2 dl 2 dl 2 dl 2 dl 4 dl] 1 0.3 0 DL } def
/LT8 { PL [2 dl 2 dl 2 dl 2 dl 2 dl 2 dl 2 dl 4 dl] 0.5 0.5 0.5 DL } def
/P { stroke [] 0 setdash
  currentlinewidth 2 div sub M
  0 currentlinewidth V stroke } def
/D { stroke [] 0 setdash 2 copy vpt add M
  hpt neg vpt neg V hpt vpt neg V
  hpt vpt V hpt neg vpt V closepath stroke
  P } def
/A { stroke [] 0 setdash vpt sub M 0 vpt2 V
  currentpoint stroke M
  hpt neg vpt neg R hpt2 0 V stroke
  } def
/B { stroke [] 0 setdash 2 copy exch hpt sub exch vpt add M
  0 vpt2 neg V hpt2 0 V 0 vpt2 V
  hpt2 neg 0 V closepath stroke
  P } def
/C { stroke [] 0 setdash exch hpt sub exch vpt add M
  hpt2 vpt2 neg V currentpoint stroke M
  hpt2 neg 0 R hpt2 vpt2 V stroke } def
/T { stroke [] 0 setdash 2 copy vpt 1.12 mul add M
  hpt neg vpt -1.62 mul V
  hpt 2 mul 0 V
  hpt neg vpt 1.62 mul V closepath stroke
  P  } def
/S { 2 copy A C} def
end
}
\begin{picture}(3600,2160)(0,0)
\special{"
gnudict begin
gsave
50 50 translate
0.100 0.100 scale
0 setgray
/Helvetica findfont 100 scalefont setfont
newpath
-500.000000 -500.000000 translate
LTa
LTb
480 151 M
63 0 V
2874 0 R
-63 0 V
480 477 M
63 0 V
2874 0 R
-63 0 V
480 804 M
63 0 V
2874 0 R
-63 0 V
480 1130 M
63 0 V
2874 0 R
-63 0 V
480 1456 M
63 0 V
2874 0 R
-63 0 V
480 1783 M
63 0 V
2874 0 R
-63 0 V
480 2109 M
63 0 V
2874 0 R
-63 0 V
480 151 M
0 63 V
0 1895 R
0 -63 V
774 151 M
0 63 V
0 1895 R
0 -63 V
1067 151 M
0 63 V
0 1895 R
0 -63 V
1361 151 M
0 63 V
0 1895 R
0 -63 V
1655 151 M
0 63 V
0 1895 R
0 -63 V
1949 151 M
0 63 V
0 1895 R
0 -63 V
2242 151 M
0 63 V
0 1895 R
0 -63 V
2536 151 M
0 63 V
0 1895 R
0 -63 V
2830 151 M
0 63 V
0 1895 R
0 -63 V
3123 151 M
0 63 V
0 1895 R
0 -63 V
3417 151 M
0 63 V
0 1895 R
0 -63 V
480 151 M
2937 0 V
0 1958 V
-2937 0 V
480 151 L
LT0
617 151 M
9 19 V
15 12 V
14 3 V
15 -2 V
15 -4 V
14 3 V
15 16 V
14 16 V
15 4 V
15 19 V
14 43 V
15 31 V
14 3 V
15 9 V
15 41 V
14 43 V
15 28 V
15 32 V
14 44 V
15 30 V
14 4 V
15 -9 V
15 -3 V
14 12 V
15 29 V
14 37 V
15 36 V
15 26 V
14 16 V
15 14 V
14 24 V
15 29 V
15 32 V
14 28 V
15 23 V
15 21 V
14 22 V
15 25 V
14 29 V
15 26 V
15 21 V
14 21 V
15 26 V
14 28 V
15 25 V
15 23 V
14 24 V
15 22 V
14 22 V
15 22 V
15 25 V
14 25 V
15 23 V
15 21 V
14 21 V
15 22 V
14 23 V
15 23 V
15 24 V
14 23 V
15 22 V
14 22 V
15 23 V
15 23 V
14 24 V
15 23 V
15 21 V
14 21 V
15 21 V
14 22 V
15 21 V
15 22 V
14 21 V
15 21 V
14 21 V
15 21 V
15 20 V
14 20 V
15 20 V
14 19 V
15 18 V
15 17 V
14 17 V
15 17 V
15 15 V
14 15 V
15 13 V
14 12 V
15 10 V
15 8 V
14 6 V
15 3 V
14 -1 V
15 -3 V
15 -7 V
14 -10 V
15 -12 V
14 -14 V
15 -15 V
15 -18 V
14 -19 V
15 -20 V
15 -21 V
14 -22 V
15 -22 V
14 -24 V
15 -24 V
15 -23 V
14 -24 V
15 -25 V
14 -25 V
15 -26 V
15 -26 V
14 -26 V
15 -25 V
14 -26 V
15 -26 V
15 -26 V
14 -26 V
15 -27 V
15 -26 V
14 -26 V
15 -24 V
14 -25 V
15 -25 V
15 -27 V
14 -27 V
15 -26 V
14 -25 V
15 -24 V
15 -25 V
14 -28 V
15 -28 V
15 -26 V
14 -25 V
15 -26 V
14 -23 V
15 -20 V
15 -21 V
14 -28 V
15 -28 V
14 -22 V
15 -21 V
15 -26 V
14 -28 V
15 -22 V
14 -19 V
15 -20 V
15 -24 V
14 -32 V
15 -26 V
15 -9 V
14 -9 V
15 -27 V
14 -31 V
15 -21 V
15 -31 V
14 -64 V
15 -75 V
14 0 V
15 24 V
15 -5 V
14 -11 V
15 1 V
14 -11 V
15 -51 V
15 -77 V
14 -46 V
15 -14 V
15 -28 V
10 -33 V
262 0 R
5 32 V
7 -32 V
LT1
2509 804 D
843 153 D
857 193 D
872 232 D
887 269 D
901 305 D
916 341 D
931 376 D
945 408 D
960 439 D
974 470 D
989 500 D
1004 528 D
1018 553 D
1033 577 D
1048 602 D
1062 630 D
1077 661 D
1092 690 D
1106 717 D
1121 743 D
1136 769 D
1150 794 D
1165 821 D
1180 848 D
1194 876 D
1209 903 D
1223 930 D
1238 957 D
1253 983 D
1267 1009 D
1282 1035 D
1297 1061 D
1311 1087 D
1326 1114 D
1341 1140 D
1355 1167 D
1370 1194 D
1385 1220 D
1399 1247 D
1414 1273 D
1429 1299 D
1443 1324 D
1458 1350 D
1472 1376 D
1487 1401 D
1502 1427 D
1516 1452 D
1531 1477 D
1546 1502 D
1560 1527 D
1575 1551 D
1590 1575 D
1604 1599 D
1619 1623 D
1634 1646 D
1648 1669 D
1663 1692 D
1678 1715 D
1692 1738 D
1707 1760 D
1721 1782 D
1736 1803 D
1751 1824 D
1765 1845 D
1780 1865 D
1795 1884 D
1809 1902 D
1824 1918 D
1839 1934 D
1853 1948 D
1868 1961 D
1883 1972 D
1897 1981 D
1912 1989 D
1927 2001 D
1941 1976 D
1956 1975 D
1970 2007 D
1985 1997 D
2000 1991 D
2014 1984 D
2029 1974 D
2044 1962 D
2058 1948 D
2073 1931 D
2088 1913 D
2102 1894 D
2117 1873 D
2132 1852 D
2146 1829 D
2161 1806 D
2176 1781 D
2190 1756 D
2205 1731 D
2219 1705 D
2234 1679 D
2249 1652 D
2263 1625 D
2278 1598 D
2293 1571 D
2307 1543 D
2322 1516 D
2337 1488 D
2351 1461 D
2366 1433 D
2381 1405 D
2395 1377 D
2410 1349 D
2425 1321 D
2439 1293 D
2454 1265 D
2468 1237 D
2483 1208 D
2498 1179 D
2512 1151 D
2527 1123 D
2542 1096 D
2556 1068 D
2571 1039 D
2586 1009 D
2600 980 D
2615 953 D
2630 925 D
2644 895 D
2659 866 D
2674 837 D
2688 808 D
2703 780 D
2717 754 D
2732 726 D
2747 696 D
2761 665 D
2776 635 D
2791 609 D
2805 585 D
2820 558 D
2835 529 D
2849 500 D
2864 472 D
2879 443 D
2893 409 D
2908 373 D
2923 341 D
2937 320 D
2952 300 D
2966 270 D
2981 229 D
2996 188 D
3010 164 D
stroke
grestore
end
showpage
}
\put(2389,804){\makebox(0,0)[r]{$r=L/2, Re = 1.36 E3$}}
\put(3417,51){\makebox(0,0){10}}
\put(3123,51){\makebox(0,0){8}}
\put(2830,51){\makebox(0,0){6}}
\put(2536,51){\makebox(0,0){4}}
\put(2242,51){\makebox(0,0){2}}
\put(1949,51){\makebox(0,0){0}}
\put(1655,51){\makebox(0,0){-2}}
\put(1361,51){\makebox(0,0){-4}}
\put(1067,51){\makebox(0,0){-6}}
\put(774,51){\makebox(0,0){-8}}
\put(480,51){\makebox(0,0){-10}}
\put(420,2109){\makebox(0,0)[r]{0}}
\put(420,1783){\makebox(0,0)[r]{-1}}
\put(420,1456){\makebox(0,0)[r]{-2}}
\put(420,1130){\makebox(0,0)[r]{-3}}
\put(420,804){\makebox(0,0)[r]{-4}}
\put(420,477){\makebox(0,0)[r]{-5}}
\put(420,151){\makebox(0,0)[r]{-6}}
\end{picture}
\setlength{\unitlength}{0.1bp}
\special{!
/gnudict 40 dict def
gnudict begin
/Color false def
/Solid false def
/gnulinewidth 5.000 def
/vshift -33 def
/dl {10 mul} def
/hpt 31.5 def
/vpt 31.5 def
/M {moveto} bind def
/L {lineto} bind def
/R {rmoveto} bind def
/V {rlineto} bind def
/vpt2 vpt 2 mul def
/hpt2 hpt 2 mul def
/Lshow { currentpoint stroke M
  0 vshift R show } def
/Rshow { currentpoint stroke M
  dup stringwidth pop neg vshift R show } def
/Cshow { currentpoint stroke M
  dup stringwidth pop -2 div vshift R show } def
/DL { Color {setrgbcolor Solid {pop []} if 0 setdash }
 {pop pop pop Solid {pop []} if 0 setdash} ifelse } def
/BL { stroke gnulinewidth 2 mul setlinewidth } def
/AL { stroke gnulinewidth 2 div setlinewidth } def
/PL { stroke gnulinewidth setlinewidth } def
/LTb { BL [] 0 0 0 DL } def
/LTa { AL [1 dl 2 dl] 0 setdash 0 0 0 setrgbcolor } def
/LT0 { PL [] 0 1 0 DL } def
/LT1 { PL [4 dl 2 dl] 0 0 1 DL } def
/LT2 { PL [2 dl 3 dl] 1 0 0 DL } def
/LT3 { PL [1 dl 1.5 dl] 1 0 1 DL } def
/LT4 { PL [5 dl 2 dl 1 dl 2 dl] 0 1 1 DL } def
/LT5 { PL [4 dl 3 dl 1 dl 3 dl] 1 1 0 DL } def
/LT6 { PL [2 dl 2 dl 2 dl 4 dl] 0 0 0 DL } def
/LT7 { PL [2 dl 2 dl 2 dl 2 dl 2 dl 4 dl] 1 0.3 0 DL } def
/LT8 { PL [2 dl 2 dl 2 dl 2 dl 2 dl 2 dl 2 dl 4 dl] 0.5 0.5 0.5 DL } def
/P { stroke [] 0 setdash
  currentlinewidth 2 div sub M
  0 currentlinewidth V stroke } def
/D { stroke [] 0 setdash 2 copy vpt add M
  hpt neg vpt neg V hpt vpt neg V
  hpt vpt V hpt neg vpt V closepath stroke
  P } def
/A { stroke [] 0 setdash vpt sub M 0 vpt2 V
  currentpoint stroke M
  hpt neg vpt neg R hpt2 0 V stroke
  } def
/B { stroke [] 0 setdash 2 copy exch hpt sub exch vpt add M
  0 vpt2 neg V hpt2 0 V 0 vpt2 V
  hpt2 neg 0 V closepath stroke
  P } def
/C { stroke [] 0 setdash exch hpt sub exch vpt add M
  hpt2 vpt2 neg V currentpoint stroke M
  hpt2 neg 0 R hpt2 vpt2 V stroke } def
/T { stroke [] 0 setdash 2 copy vpt 1.12 mul add M
  hpt neg vpt -1.62 mul V
  hpt 2 mul 0 V
  hpt neg vpt 1.62 mul V closepath stroke
  P  } def
/S { 2 copy A C} def
end
}
\begin{picture}(3600,2160)(0,0)
\special{"
gnudict begin
gsave
50 50 translate
0.100 0.100 scale
0 setgray
/Helvetica findfont 100 scalefont setfont
newpath
-500.000000 -500.000000 translate
LTa
LTb
480 151 M
63 0 V
2874 0 R
-63 0 V
480 477 M
63 0 V
2874 0 R
-63 0 V
480 804 M
63 0 V
2874 0 R
-63 0 V
480 1130 M
63 0 V
2874 0 R
-63 0 V
480 1456 M
63 0 V
2874 0 R
-63 0 V
480 1783 M
63 0 V
2874 0 R
-63 0 V
480 2109 M
63 0 V
2874 0 R
-63 0 V
480 151 M
0 63 V
0 1895 R
0 -63 V
774 151 M
0 63 V
0 1895 R
0 -63 V
1067 151 M
0 63 V
0 1895 R
0 -63 V
1361 151 M
0 63 V
0 1895 R
0 -63 V
1655 151 M
0 63 V
0 1895 R
0 -63 V
1949 151 M
0 63 V
0 1895 R
0 -63 V
2242 151 M
0 63 V
0 1895 R
0 -63 V
2536 151 M
0 63 V
0 1895 R
0 -63 V
2830 151 M
0 63 V
0 1895 R
0 -63 V
3123 151 M
0 63 V
0 1895 R
0 -63 V
3417 151 M
0 63 V
0 1895 R
0 -63 V
480 151 M
2937 0 V
0 1958 V
-2937 0 V
480 151 L
LT0
538 272 M
15 53 V
15 17 V
14 -3 V
15 -16 V
15 -24 V
14 -18 V
15 41 V
14 81 V
15 39 V
15 -21 V
14 -33 V
15 37 V
14 57 V
15 36 V
15 16 V
14 0 V
15 5 V
14 27 V
15 28 V
15 13 V
14 10 V
15 19 V
15 20 V
14 17 V
15 18 V
14 24 V
15 21 V
15 14 V
14 12 V
15 17 V
14 20 V
15 21 V
15 17 V
14 13 V
15 12 V
14 14 V
15 20 V
15 24 V
14 23 V
15 19 V
15 18 V
14 19 V
15 19 V
14 15 V
15 15 V
15 18 V
14 19 V
15 18 V
14 18 V
15 20 V
15 20 V
14 19 V
15 19 V
14 20 V
15 20 V
15 22 V
14 21 V
15 19 V
15 19 V
14 21 V
15 20 V
14 20 V
15 19 V
15 20 V
14 20 V
15 21 V
14 21 V
15 20 V
15 20 V
14 19 V
15 19 V
15 21 V
14 21 V
15 19 V
14 20 V
15 20 V
15 20 V
14 20 V
15 21 V
14 20 V
15 19 V
15 20 V
14 20 V
15 20 V
14 19 V
15 19 V
15 19 V
14 19 V
15 19 V
15 17 V
14 17 V
15 16 V
14 15 V
15 14 V
15 11 V
14 8 V
15 4 V
14 -1 V
15 -5 V
15 -10 V
14 -12 V
15 -16 V
14 -17 V
15 -19 V
15 -21 V
14 -21 V
15 -22 V
15 -22 V
14 -23 V
15 -23 V
14 -23 V
15 -24 V
15 -24 V
14 -24 V
15 -23 V
14 -24 V
15 -24 V
15 -23 V
14 -23 V
15 -24 V
14 -24 V
15 -23 V
15 -22 V
14 -22 V
15 -23 V
15 -24 V
14 -24 V
15 -24 V
14 -23 V
15 -23 V
15 -23 V
14 -24 V
15 -22 V
14 -21 V
15 -21 V
15 -23 V
14 -24 V
15 -24 V
15 -22 V
14 -21 V
15 -22 V
14 -23 V
15 -25 V
15 -22 V
14 -19 V
15 -19 V
14 -23 V
15 -22 V
15 -19 V
14 -22 V
15 -28 V
14 -21 V
15 -3 V
15 -9 V
14 -30 V
15 -31 V
15 -10 V
14 -11 V
15 -32 V
14 -41 V
15 -23 V
15 -6 V
14 -6 V
15 -13 V
14 -28 V
15 -28 V
15 -12 V
14 -18 V
15 -47 V
14 -51 V
15 -4 V
15 -6 V
14 -64 V
15 -63 V
15 59 V
14 40 V
15 -32 V
14 -73 V
15 -41 V
15 -3 V
14 -12 V
15 -10 V
14 10 V
15 12 V
15 0 V
14 -24 V
15 -63 V
4 -24 V
LT1
2509 804 D
685 157 D
701 188 D
716 218 D
732 247 D
748 275 D
763 302 D
779 329 D
795 355 D
810 379 D
826 401 D
842 423 D
857 447 D
873 473 D
889 500 D
905 527 D
920 552 D
936 576 D
952 599 D
967 623 D
983 646 D
999 670 D
1014 695 D
1030 719 D
1046 744 D
1062 769 D
1077 793 D
1093 817 D
1109 841 D
1124 865 D
1140 888 D
1156 912 D
1171 936 D
1187 961 D
1203 985 D
1219 1010 D
1234 1034 D
1250 1059 D
1266 1083 D
1281 1108 D
1297 1132 D
1313 1156 D
1328 1181 D
1344 1205 D
1360 1229 D
1376 1253 D
1391 1277 D
1407 1302 D
1423 1326 D
1438 1350 D
1454 1374 D
1470 1399 D
1485 1423 D
1501 1447 D
1517 1471 D
1533 1495 D
1548 1519 D
1564 1542 D
1580 1566 D
1595 1590 D
1611 1613 D
1627 1636 D
1642 1659 D
1658 1682 D
1674 1705 D
1689 1727 D
1705 1749 D
1721 1771 D
1737 1792 D
1752 1813 D
1768 1833 D
1784 1853 D
1799 1871 D
1815 1889 D
1831 1905 D
1846 1920 D
1862 1936 D
1878 1952 D
1894 1970 D
1909 1992 D
1925 2023 D
1941 2061 D
1956 2064 D
1972 2030 D
1988 2002 D
2003 1981 D
2019 1965 D
2035 1949 D
2051 1934 D
2066 1918 D
2082 1900 D
2098 1881 D
2113 1861 D
2129 1840 D
2145 1818 D
2160 1795 D
2176 1771 D
2192 1746 D
2208 1721 D
2223 1696 D
2239 1670 D
2255 1644 D
2270 1618 D
2286 1591 D
2302 1565 D
2317 1538 D
2333 1512 D
2349 1485 D
2364 1458 D
2380 1432 D
2396 1405 D
2412 1379 D
2427 1352 D
2443 1326 D
2459 1299 D
2474 1272 D
2490 1246 D
2506 1220 D
2521 1194 D
2537 1168 D
2553 1141 D
2569 1115 D
2584 1089 D
2600 1062 D
2616 1036 D
2631 1010 D
2647 985 D
2663 959 D
2678 934 D
2694 908 D
2710 881 D
2726 855 D
2741 829 D
2757 804 D
2773 778 D
2788 752 D
2804 726 D
2820 699 D
2835 673 D
2851 648 D
2867 623 D
2883 599 D
2898 575 D
2914 549 D
2930 521 D
2945 492 D
2961 466 D
2977 443 D
2992 421 D
3008 397 D
3024 372 D
3040 347 D
3055 321 D
3071 296 D
3087 269 D
3102 238 D
3118 207 D
3134 179 D
3149 159 D
stroke
grestore
end
showpage
}
\put(2389,804){\makebox(0,0)[r]{$r=L/2^2, Re = 1.36 E3$}}
\put(3417,51){\makebox(0,0){10}}
\put(3123,51){\makebox(0,0){8}}
\put(2830,51){\makebox(0,0){6}}
\put(2536,51){\makebox(0,0){4}}
\put(2242,51){\makebox(0,0){2}}
\put(1949,51){\makebox(0,0){0}}
\put(1655,51){\makebox(0,0){-2}}
\put(1361,51){\makebox(0,0){-4}}
\put(1067,51){\makebox(0,0){-6}}
\put(774,51){\makebox(0,0){-8}}
\put(480,51){\makebox(0,0){-10}}
\put(420,2109){\makebox(0,0)[r]{0}}
\put(420,1783){\makebox(0,0)[r]{-1}}
\put(420,1456){\makebox(0,0)[r]{-2}}
\put(420,1130){\makebox(0,0)[r]{-3}}
\put(420,804){\makebox(0,0)[r]{-4}}
\put(420,477){\makebox(0,0)[r]{-5}}
\put(420,151){\makebox(0,0)[r]{-6}}
\end{picture}
\newpage
\setlength{\unitlength}{0.1bp}
\special{!
/gnudict 40 dict def
gnudict begin
/Color false def
/Solid false def
/gnulinewidth 5.000 def
/vshift -33 def
/dl {10 mul} def
/hpt 31.5 def
/vpt 31.5 def
/M {moveto} bind def
/L {lineto} bind def
/R {rmoveto} bind def
/V {rlineto} bind def
/vpt2 vpt 2 mul def
/hpt2 hpt 2 mul def
/Lshow { currentpoint stroke M
  0 vshift R show } def
/Rshow { currentpoint stroke M
  dup stringwidth pop neg vshift R show } def
/Cshow { currentpoint stroke M
  dup stringwidth pop -2 div vshift R show } def
/DL { Color {setrgbcolor Solid {pop []} if 0 setdash }
 {pop pop pop Solid {pop []} if 0 setdash} ifelse } def
/BL { stroke gnulinewidth 2 mul setlinewidth } def
/AL { stroke gnulinewidth 2 div setlinewidth } def
/PL { stroke gnulinewidth setlinewidth } def
/LTb { BL [] 0 0 0 DL } def
/LTa { AL [1 dl 2 dl] 0 setdash 0 0 0 setrgbcolor } def
/LT0 { PL [] 0 1 0 DL } def
/LT1 { PL [4 dl 2 dl] 0 0 1 DL } def
/LT2 { PL [2 dl 3 dl] 1 0 0 DL } def
/LT3 { PL [1 dl 1.5 dl] 1 0 1 DL } def
/LT4 { PL [5 dl 2 dl 1 dl 2 dl] 0 1 1 DL } def
/LT5 { PL [4 dl 3 dl 1 dl 3 dl] 1 1 0 DL } def
/LT6 { PL [2 dl 2 dl 2 dl 4 dl] 0 0 0 DL } def
/LT7 { PL [2 dl 2 dl 2 dl 2 dl 2 dl 4 dl] 1 0.3 0 DL } def
/LT8 { PL [2 dl 2 dl 2 dl 2 dl 2 dl 2 dl 2 dl 4 dl] 0.5 0.5 0.5 DL } def
/P { stroke [] 0 setdash
  currentlinewidth 2 div sub M
  0 currentlinewidth V stroke } def
/D { stroke [] 0 setdash 2 copy vpt add M
  hpt neg vpt neg V hpt vpt neg V
  hpt vpt V hpt neg vpt V closepath stroke
  P } def
/A { stroke [] 0 setdash vpt sub M 0 vpt2 V
  currentpoint stroke M
  hpt neg vpt neg R hpt2 0 V stroke
  } def
/B { stroke [] 0 setdash 2 copy exch hpt sub exch vpt add M
  0 vpt2 neg V hpt2 0 V 0 vpt2 V
  hpt2 neg 0 V closepath stroke
  P } def
/C { stroke [] 0 setdash exch hpt sub exch vpt add M
  hpt2 vpt2 neg V currentpoint stroke M
  hpt2 neg 0 R hpt2 vpt2 V stroke } def
/T { stroke [] 0 setdash 2 copy vpt 1.12 mul add M
  hpt neg vpt -1.62 mul V
  hpt 2 mul 0 V
  hpt neg vpt 1.62 mul V closepath stroke
  P  } def
/S { 2 copy A C} def
end
}
\begin{picture}(3600,2160)(0,0)
\special{"
gnudict begin
gsave
50 50 translate
0.100 0.100 scale
0 setgray
/Helvetica findfont 100 scalefont setfont
newpath
-500.000000 -500.000000 translate
LTa
LTb
480 151 M
63 0 V
2874 0 R
-63 0 V
480 477 M
63 0 V
2874 0 R
-63 0 V
480 804 M
63 0 V
2874 0 R
-63 0 V
480 1130 M
63 0 V
2874 0 R
-63 0 V
480 1456 M
63 0 V
2874 0 R
-63 0 V
480 1783 M
63 0 V
2874 0 R
-63 0 V
480 2109 M
63 0 V
2874 0 R
-63 0 V
480 151 M
0 63 V
0 1895 R
0 -63 V
774 151 M
0 63 V
0 1895 R
0 -63 V
1067 151 M
0 63 V
0 1895 R
0 -63 V
1361 151 M
0 63 V
0 1895 R
0 -63 V
1655 151 M
0 63 V
0 1895 R
0 -63 V
1949 151 M
0 63 V
0 1895 R
0 -63 V
2242 151 M
0 63 V
0 1895 R
0 -63 V
2536 151 M
0 63 V
0 1895 R
0 -63 V
2830 151 M
0 63 V
0 1895 R
0 -63 V
3123 151 M
0 63 V
0 1895 R
0 -63 V
3417 151 M
0 63 V
0 1895 R
0 -63 V
480 151 M
2937 0 V
0 1958 V
-2937 0 V
480 151 L
LT0
538 473 M
15 39 V
15 33 V
14 17 V
15 -9 V
15 -12 V
14 9 V
15 24 V
14 24 V
15 23 V
15 21 V
14 9 V
15 -9 V
14 -9 V
15 13 V
15 27 V
14 27 V
15 21 V
14 11 V
15 11 V
15 14 V
14 15 V
15 11 V
15 7 V
14 5 V
15 6 V
14 11 V
15 20 V
15 26 V
14 21 V
15 12 V
14 10 V
15 17 V
15 18 V
14 12 V
15 11 V
14 15 V
15 19 V
15 18 V
14 16 V
15 14 V
15 13 V
14 15 V
15 15 V
14 16 V
15 17 V
15 17 V
14 16 V
15 12 V
14 12 V
15 17 V
15 19 V
14 18 V
15 16 V
14 17 V
15 16 V
15 17 V
14 16 V
15 18 V
15 18 V
14 18 V
15 17 V
14 15 V
15 16 V
15 18 V
14 18 V
15 18 V
14 18 V
15 17 V
15 18 V
14 18 V
15 18 V
15 19 V
14 19 V
15 18 V
14 19 V
15 18 V
15 19 V
14 19 V
15 19 V
14 20 V
15 19 V
15 19 V
14 20 V
15 21 V
14 20 V
15 21 V
15 21 V
14 21 V
15 21 V
15 20 V
14 20 V
15 19 V
14 19 V
15 17 V
15 14 V
14 10 V
15 5 V
14 -1 V
15 -8 V
15 -12 V
14 -17 V
15 -19 V
14 -21 V
15 -22 V
15 -24 V
14 -24 V
15 -24 V
15 -25 V
14 -24 V
15 -25 V
14 -23 V
15 -24 V
15 -22 V
14 -22 V
15 -22 V
14 -22 V
15 -23 V
15 -22 V
14 -22 V
15 -21 V
14 -21 V
15 -21 V
15 -21 V
14 -20 V
15 -21 V
15 -20 V
14 -21 V
15 -21 V
14 -20 V
15 -20 V
15 -20 V
14 -20 V
15 -20 V
14 -19 V
15 -19 V
15 -18 V
14 -18 V
15 -19 V
15 -19 V
14 -18 V
15 -17 V
14 -19 V
15 -21 V
15 -21 V
14 -17 V
15 -16 V
14 -22 V
15 -20 V
15 -15 V
14 -14 V
15 -18 V
14 -18 V
15 -15 V
15 -16 V
14 -19 V
15 -19 V
15 -14 V
14 -13 V
15 -15 V
14 -18 V
15 -20 V
15 -22 V
14 -20 V
15 -11 V
14 -3 V
15 -8 V
15 -24 V
14 -32 V
15 -28 V
14 -21 V
15 -16 V
15 -11 V
14 -8 V
15 -11 V
15 -18 V
14 -17 V
15 -9 V
14 -11 V
15 -20 V
15 -17 V
14 0 V
15 -4 V
14 -27 V
15 -40 V
15 -26 V
14 0 V
15 16 V
14 -2 V
15 -48 V
15 -66 V
14 -1 V
15 17 V
15 -25 V
14 -25 V
15 36 V
14 28 V
15 -25 V
LT1
2509 804 D
483 243 D
501 264 D
519 288 D
538 312 D
556 336 D
575 359 D
593 382 D
612 403 D
630 425 D
648 446 D
667 467 D
685 489 D
704 511 D
722 533 D
741 555 D
759 577 D
778 600 D
796 622 D
814 644 D
833 666 D
851 687 D
870 709 D
888 730 D
907 752 D
925 774 D
944 796 D
962 818 D
980 841 D
999 863 D
1017 885 D
1036 907 D
1054 930 D
1073 952 D
1091 975 D
1109 997 D
1128 1019 D
1146 1041 D
1165 1064 D
1183 1086 D
1202 1109 D
1220 1131 D
1239 1154 D
1257 1176 D
1275 1199 D
1294 1221 D
1312 1244 D
1331 1267 D
1349 1289 D
1368 1312 D
1386 1334 D
1405 1357 D
1423 1380 D
1441 1402 D
1460 1425 D
1478 1447 D
1497 1470 D
1515 1492 D
1534 1514 D
1552 1536 D
1570 1558 D
1589 1580 D
1607 1602 D
1626 1624 D
1644 1646 D
1663 1668 D
1681 1690 D
1700 1712 D
1718 1734 D
1736 1757 D
1755 1779 D
1773 1803 D
1792 1827 D
1810 1851 D
1829 1876 D
1847 1902 D
1866 1929 D
1884 1958 D
1902 1991 D
1921 2031 D
1939 2094 D
1958 2099 D
1976 2041 D
1995 2003 D
2013 1972 D
2031 1943 D
2050 1915 D
2068 1887 D
2087 1861 D
2105 1834 D
2124 1808 D
2142 1783 D
2161 1758 D
2179 1732 D
2197 1708 D
2216 1683 D
2234 1658 D
2253 1634 D
2271 1609 D
2290 1584 D
2308 1560 D
2327 1535 D
2345 1510 D
2363 1486 D
2382 1461 D
2400 1436 D
2419 1411 D
2437 1387 D
2456 1362 D
2474 1337 D
2492 1313 D
2511 1288 D
2529 1263 D
2548 1239 D
2566 1214 D
2585 1190 D
2603 1166 D
2622 1141 D
2640 1117 D
2658 1093 D
2677 1069 D
2695 1045 D
2714 1020 D
2732 997 D
2751 973 D
2769 949 D
2788 925 D
2806 901 D
2824 877 D
2843 853 D
2861 830 D
2880 807 D
2898 784 D
2917 761 D
2935 737 D
2953 713 D
2972 689 D
2990 666 D
3009 643 D
3027 620 D
3046 597 D
3064 573 D
3083 550 D
3101 526 D
3119 503 D
3138 480 D
3156 458 D
3175 436 D
3193 415 D
3212 392 D
3230 368 D
3249 343 D
3267 318 D
3285 295 D
3304 274 D
3322 254 D
3341 233 D
3359 211 D
3378 188 D
3396 166 D
stroke
grestore
end
showpage
}
\put(2389,804){\makebox(0,0)[r]{$r=L/2^3, Re = 1.36 E3$}}
\put(3417,51){\makebox(0,0){10}}
\put(3123,51){\makebox(0,0){8}}
\put(2830,51){\makebox(0,0){6}}
\put(2536,51){\makebox(0,0){4}}
\put(2242,51){\makebox(0,0){2}}
\put(1949,51){\makebox(0,0){0}}
\put(1655,51){\makebox(0,0){-2}}
\put(1361,51){\makebox(0,0){-4}}
\put(1067,51){\makebox(0,0){-6}}
\put(774,51){\makebox(0,0){-8}}
\put(480,51){\makebox(0,0){-10}}
\put(420,2109){\makebox(0,0)[r]{0}}
\put(420,1783){\makebox(0,0)[r]{-1}}
\put(420,1456){\makebox(0,0)[r]{-2}}
\put(420,1130){\makebox(0,0)[r]{-3}}
\put(420,804){\makebox(0,0)[r]{-4}}
\put(420,477){\makebox(0,0)[r]{-5}}
\put(420,151){\makebox(0,0)[r]{-6}}
\end{picture}
\setlength{\unitlength}{0.1bp}
\special{!
/gnudict 40 dict def
gnudict begin
/Color false def
/Solid false def
/gnulinewidth 5.000 def
/vshift -33 def
/dl {10 mul} def
/hpt 31.5 def
/vpt 31.5 def
/M {moveto} bind def
/L {lineto} bind def
/R {rmoveto} bind def
/V {rlineto} bind def
/vpt2 vpt 2 mul def
/hpt2 hpt 2 mul def
/Lshow { currentpoint stroke M
  0 vshift R show } def
/Rshow { currentpoint stroke M
  dup stringwidth pop neg vshift R show } def
/Cshow { currentpoint stroke M
  dup stringwidth pop -2 div vshift R show } def
/DL { Color {setrgbcolor Solid {pop []} if 0 setdash }
 {pop pop pop Solid {pop []} if 0 setdash} ifelse } def
/BL { stroke gnulinewidth 2 mul setlinewidth } def
/AL { stroke gnulinewidth 2 div setlinewidth } def
/PL { stroke gnulinewidth setlinewidth } def
/LTb { BL [] 0 0 0 DL } def
/LTa { AL [1 dl 2 dl] 0 setdash 0 0 0 setrgbcolor } def
/LT0 { PL [] 0 1 0 DL } def
/LT1 { PL [4 dl 2 dl] 0 0 1 DL } def
/LT2 { PL [2 dl 3 dl] 1 0 0 DL } def
/LT3 { PL [1 dl 1.5 dl] 1 0 1 DL } def
/LT4 { PL [5 dl 2 dl 1 dl 2 dl] 0 1 1 DL } def
/LT5 { PL [4 dl 3 dl 1 dl 3 dl] 1 1 0 DL } def
/LT6 { PL [2 dl 2 dl 2 dl 4 dl] 0 0 0 DL } def
/LT7 { PL [2 dl 2 dl 2 dl 2 dl 2 dl 4 dl] 1 0.3 0 DL } def
/LT8 { PL [2 dl 2 dl 2 dl 2 dl 2 dl 2 dl 2 dl 4 dl] 0.5 0.5 0.5 DL } def
/P { stroke [] 0 setdash
  currentlinewidth 2 div sub M
  0 currentlinewidth V stroke } def
/D { stroke [] 0 setdash 2 copy vpt add M
  hpt neg vpt neg V hpt vpt neg V
  hpt vpt V hpt neg vpt V closepath stroke
  P } def
/A { stroke [] 0 setdash vpt sub M 0 vpt2 V
  currentpoint stroke M
  hpt neg vpt neg R hpt2 0 V stroke
  } def
/B { stroke [] 0 setdash 2 copy exch hpt sub exch vpt add M
  0 vpt2 neg V hpt2 0 V 0 vpt2 V
  hpt2 neg 0 V closepath stroke
  P } def
/C { stroke [] 0 setdash exch hpt sub exch vpt add M
  hpt2 vpt2 neg V currentpoint stroke M
  hpt2 neg 0 R hpt2 vpt2 V stroke } def
/T { stroke [] 0 setdash 2 copy vpt 1.12 mul add M
  hpt neg vpt -1.62 mul V
  hpt 2 mul 0 V
  hpt neg vpt 1.62 mul V closepath stroke
  P  } def
/S { 2 copy A C} def
end
}
\begin{picture}(3600,2160)(0,0)
\special{"
gnudict begin
gsave
50 50 translate
0.100 0.100 scale
0 setgray
/Helvetica findfont 100 scalefont setfont
newpath
-500.000000 -500.000000 translate
LTa
LTb
480 151 M
63 0 V
2874 0 R
-63 0 V
480 477 M
63 0 V
2874 0 R
-63 0 V
480 804 M
63 0 V
2874 0 R
-63 0 V
480 1130 M
63 0 V
2874 0 R
-63 0 V
480 1456 M
63 0 V
2874 0 R
-63 0 V
480 1783 M
63 0 V
2874 0 R
-63 0 V
480 2109 M
63 0 V
2874 0 R
-63 0 V
480 151 M
0 63 V
0 1895 R
0 -63 V
774 151 M
0 63 V
0 1895 R
0 -63 V
1067 151 M
0 63 V
0 1895 R
0 -63 V
1361 151 M
0 63 V
0 1895 R
0 -63 V
1655 151 M
0 63 V
0 1895 R
0 -63 V
1949 151 M
0 63 V
0 1895 R
0 -63 V
2242 151 M
0 63 V
0 1895 R
0 -63 V
2536 151 M
0 63 V
0 1895 R
0 -63 V
2830 151 M
0 63 V
0 1895 R
0 -63 V
3123 151 M
0 63 V
0 1895 R
0 -63 V
3417 151 M
0 63 V
0 1895 R
0 -63 V
480 151 M
2937 0 V
0 1958 V
-2937 0 V
480 151 L
LT0
538 645 M
15 11 V
15 -3 V
14 -7 V
15 0 V
15 7 V
14 14 V
15 23 V
14 30 V
15 23 V
15 7 V
14 -5 V
15 -7 V
14 5 V
15 22 V
15 21 V
14 6 V
15 2 V
14 11 V
15 14 V
15 9 V
14 11 V
15 20 V
15 20 V
14 11 V
15 8 V
14 11 V
15 11 V
15 9 V
14 7 V
15 10 V
14 15 V
15 21 V
15 17 V
14 6 V
15 5 V
14 15 V
15 20 V
15 17 V
14 13 V
15 10 V
15 11 V
14 13 V
15 15 V
14 14 V
15 14 V
15 13 V
14 13 V
15 15 V
14 15 V
15 14 V
15 13 V
14 14 V
15 15 V
14 16 V
15 15 V
15 13 V
14 13 V
15 17 V
15 18 V
14 15 V
15 15 V
14 15 V
15 17 V
15 14 V
14 14 V
15 16 V
14 15 V
15 15 V
15 15 V
14 15 V
15 16 V
15 17 V
14 19 V
15 19 V
14 20 V
15 19 V
15 19 V
14 19 V
15 20 V
14 21 V
15 20 V
15 20 V
14 21 V
15 21 V
14 22 V
15 22 V
15 23 V
14 22 V
15 23 V
15 23 V
14 22 V
15 22 V
14 20 V
15 18 V
15 14 V
14 10 V
15 4 V
14 -2 V
15 -8 V
15 -13 V
14 -18 V
15 -20 V
14 -23 V
15 -25 V
15 -25 V
14 -26 V
15 -26 V
15 -26 V
14 -26 V
15 -25 V
14 -25 V
15 -24 V
15 -23 V
14 -24 V
15 -22 V
14 -23 V
15 -23 V
15 -23 V
14 -22 V
15 -23 V
14 -21 V
15 -20 V
15 -17 V
14 -17 V
15 -17 V
15 -16 V
14 -16 V
15 -15 V
14 -15 V
15 -17 V
15 -19 V
14 -19 V
15 -18 V
14 -16 V
15 -15 V
15 -17 V
14 -18 V
15 -17 V
15 -16 V
14 -16 V
15 -16 V
14 -17 V
15 -16 V
15 -15 V
14 -17 V
15 -15 V
14 -16 V
15 -16 V
15 -17 V
14 -17 V
15 -14 V
14 -14 V
15 -15 V
15 -16 V
14 -14 V
15 -12 V
15 -10 V
14 -14 V
15 -19 V
14 -17 V
15 -8 V
15 -9 V
14 -19 V
15 -19 V
14 -10 V
15 -11 V
15 -24 V
14 -22 V
15 -6 V
14 -2 V
15 -13 V
15 -16 V
14 -10 V
15 -9 V
15 -17 V
14 -25 V
15 -26 V
14 -15 V
15 5 V
15 3 V
14 -15 V
15 -20 V
14 -9 V
15 -3 V
15 -11 V
14 -18 V
15 -23 V
14 -24 V
15 -15 V
15 -5 V
14 3 V
15 1 V
15 -8 V
14 -8 V
15 0 V
14 -4 V
15 -22 V
LT1
2509 804 D
496 621 D
521 640 D
546 659 D
572 679 D
597 698 D
622 718 D
648 737 D
673 757 D
698 777 D
723 796 D
749 816 D
774 835 D
799 855 D
824 875 D
850 894 D
875 914 D
900 934 D
925 954 D
951 974 D
976 995 D
1001 1015 D
1026 1035 D
1052 1056 D
1077 1077 D
1102 1097 D
1128 1119 D
1153 1140 D
1178 1161 D
1203 1183 D
1229 1205 D
1254 1227 D
1279 1249 D
1304 1272 D
1330 1295 D
1355 1318 D
1380 1341 D
1405 1365 D
1431 1390 D
1456 1415 D
1481 1440 D
1506 1466 D
1532 1492 D
1557 1519 D
1582 1547 D
1607 1575 D
1633 1604 D
1658 1634 D
1683 1665 D
1709 1697 D
1734 1730 D
1759 1764 D
1784 1800 D
1810 1837 D
1835 1876 D
1860 1918 D
1885 1962 D
1911 2014 D
1936 2092 D
1961 2098 D
1986 2026 D
2012 1976 D
2037 1930 D
2062 1887 D
2087 1845 D
2113 1805 D
2138 1766 D
2163 1728 D
2188 1692 D
2214 1656 D
2239 1622 D
2264 1589 D
2290 1557 D
2315 1526 D
2340 1495 D
2365 1466 D
2391 1437 D
2416 1408 D
2441 1381 D
2466 1354 D
2492 1327 D
2517 1301 D
2542 1276 D
2567 1250 D
2593 1226 D
2618 1201 D
2643 1177 D
2668 1154 D
2694 1130 D
2719 1107 D
2744 1084 D
2769 1061 D
2795 1039 D
2820 1017 D
2845 995 D
2871 973 D
2896 951 D
2921 929 D
2946 908 D
2972 887 D
2997 865 D
3022 844 D
3047 823 D
3073 802 D
3098 781 D
3123 760 D
3148 739 D
3174 719 D
3199 698 D
3224 678 D
3249 657 D
3275 637 D
3300 617 D
3325 597 D
3351 576 D
3376 555 D
3401 535 D
stroke
grestore
end
showpage
}
\put(2389,804){\makebox(0,0)[r]{$r=L/2^4, Re = 1.36 E3$}}
\put(3417,51){\makebox(0,0){10}}
\put(3123,51){\makebox(0,0){8}}
\put(2830,51){\makebox(0,0){6}}
\put(2536,51){\makebox(0,0){4}}
\put(2242,51){\makebox(0,0){2}}
\put(1949,51){\makebox(0,0){0}}
\put(1655,51){\makebox(0,0){-2}}
\put(1361,51){\makebox(0,0){-4}}
\put(1067,51){\makebox(0,0){-6}}
\put(774,51){\makebox(0,0){-8}}
\put(480,51){\makebox(0,0){-10}}
\put(420,2109){\makebox(0,0)[r]{0}}
\put(420,1783){\makebox(0,0)[r]{-1}}
\put(420,1456){\makebox(0,0)[r]{-2}}
\put(420,1130){\makebox(0,0)[r]{-3}}
\put(420,804){\makebox(0,0)[r]{-4}}
\put(420,477){\makebox(0,0)[r]{-5}}
\put(420,151){\makebox(0,0)[r]{-6}}
\end{picture}
\setlength{\unitlength}{0.1bp}
\special{!
/gnudict 40 dict def
gnudict begin
/Color false def
/Solid false def
/gnulinewidth 5.000 def
/vshift -33 def
/dl {10 mul} def
/hpt 31.5 def
/vpt 31.5 def
/M {moveto} bind def
/L {lineto} bind def
/R {rmoveto} bind def
/V {rlineto} bind def
/vpt2 vpt 2 mul def
/hpt2 hpt 2 mul def
/Lshow { currentpoint stroke M
  0 vshift R show } def
/Rshow { currentpoint stroke M
  dup stringwidth pop neg vshift R show } def
/Cshow { currentpoint stroke M
  dup stringwidth pop -2 div vshift R show } def
/DL { Color {setrgbcolor Solid {pop []} if 0 setdash }
 {pop pop pop Solid {pop []} if 0 setdash} ifelse } def
/BL { stroke gnulinewidth 2 mul setlinewidth } def
/AL { stroke gnulinewidth 2 div setlinewidth } def
/PL { stroke gnulinewidth setlinewidth } def
/LTb { BL [] 0 0 0 DL } def
/LTa { AL [1 dl 2 dl] 0 setdash 0 0 0 setrgbcolor } def
/LT0 { PL [] 0 1 0 DL } def
/LT1 { PL [4 dl 2 dl] 0 0 1 DL } def
/LT2 { PL [2 dl 3 dl] 1 0 0 DL } def
/LT3 { PL [1 dl 1.5 dl] 1 0 1 DL } def
/LT4 { PL [5 dl 2 dl 1 dl 2 dl] 0 1 1 DL } def
/LT5 { PL [4 dl 3 dl 1 dl 3 dl] 1 1 0 DL } def
/LT6 { PL [2 dl 2 dl 2 dl 4 dl] 0 0 0 DL } def
/LT7 { PL [2 dl 2 dl 2 dl 2 dl 2 dl 4 dl] 1 0.3 0 DL } def
/LT8 { PL [2 dl 2 dl 2 dl 2 dl 2 dl 2 dl 2 dl 4 dl] 0.5 0.5 0.5 DL } def
/P { stroke [] 0 setdash
  currentlinewidth 2 div sub M
  0 currentlinewidth V stroke } def
/D { stroke [] 0 setdash 2 copy vpt add M
  hpt neg vpt neg V hpt vpt neg V
  hpt vpt V hpt neg vpt V closepath stroke
  P } def
/A { stroke [] 0 setdash vpt sub M 0 vpt2 V
  currentpoint stroke M
  hpt neg vpt neg R hpt2 0 V stroke
  } def
/B { stroke [] 0 setdash 2 copy exch hpt sub exch vpt add M
  0 vpt2 neg V hpt2 0 V 0 vpt2 V
  hpt2 neg 0 V closepath stroke
  P } def
/C { stroke [] 0 setdash exch hpt sub exch vpt add M
  hpt2 vpt2 neg V currentpoint stroke M
  hpt2 neg 0 R hpt2 vpt2 V stroke } def
/T { stroke [] 0 setdash 2 copy vpt 1.12 mul add M
  hpt neg vpt -1.62 mul V
  hpt 2 mul 0 V
  hpt neg vpt 1.62 mul V closepath stroke
  P  } def
/S { 2 copy A C} def
end
}
\begin{picture}(3600,2160)(0,0)
\special{"
gnudict begin
gsave
50 50 translate
0.100 0.100 scale
0 setgray
/Helvetica findfont 100 scalefont setfont
newpath
-500.000000 -500.000000 translate
LTa
LTb
480 151 M
63 0 V
2874 0 R
-63 0 V
480 477 M
63 0 V
2874 0 R
-63 0 V
480 804 M
63 0 V
2874 0 R
-63 0 V
480 1130 M
63 0 V
2874 0 R
-63 0 V
480 1456 M
63 0 V
2874 0 R
-63 0 V
480 1783 M
63 0 V
2874 0 R
-63 0 V
480 2109 M
63 0 V
2874 0 R
-63 0 V
480 151 M
0 63 V
0 1895 R
0 -63 V
774 151 M
0 63 V
0 1895 R
0 -63 V
1067 151 M
0 63 V
0 1895 R
0 -63 V
1361 151 M
0 63 V
0 1895 R
0 -63 V
1655 151 M
0 63 V
0 1895 R
0 -63 V
1949 151 M
0 63 V
0 1895 R
0 -63 V
2242 151 M
0 63 V
0 1895 R
0 -63 V
2536 151 M
0 63 V
0 1895 R
0 -63 V
2830 151 M
0 63 V
0 1895 R
0 -63 V
3123 151 M
0 63 V
0 1895 R
0 -63 V
3417 151 M
0 63 V
0 1895 R
0 -63 V
480 151 M
2937 0 V
0 1958 V
-2937 0 V
480 151 L
LT0
538 682 M
15 7 V
15 6 V
14 5 V
15 5 V
15 12 V
14 20 V
15 14 V
14 -2 V
15 -2 V
15 17 V
14 21 V
15 12 V
14 9 V
15 11 V
15 14 V
14 15 V
15 14 V
14 9 V
15 10 V
15 13 V
14 14 V
15 13 V
15 11 V
14 10 V
15 9 V
14 9 V
15 11 V
15 16 V
14 16 V
15 13 V
14 11 V
15 14 V
15 15 V
14 13 V
15 11 V
14 11 V
15 13 V
15 14 V
14 14 V
15 12 V
15 10 V
14 10 V
15 9 V
14 8 V
15 7 V
15 6 V
14 5 V
15 6 V
14 5 V
15 5 V
15 6 V
14 6 V
15 8 V
14 11 V
15 15 V
15 18 V
14 21 V
15 19 V
15 17 V
14 17 V
15 15 V
14 14 V
15 15 V
15 16 V
14 17 V
15 16 V
14 14 V
15 15 V
15 16 V
14 16 V
15 18 V
15 18 V
14 19 V
15 18 V
14 19 V
15 20 V
15 20 V
14 21 V
15 20 V
14 22 V
15 22 V
15 23 V
14 23 V
15 23 V
14 25 V
15 24 V
15 24 V
14 24 V
15 24 V
15 24 V
14 22 V
15 20 V
14 19 V
15 15 V
15 12 V
14 8 V
15 3 V
14 -3 V
15 -6 V
15 -12 V
14 -16 V
15 -19 V
14 -21 V
15 -24 V
15 -26 V
14 -26 V
15 -27 V
15 -28 V
14 -27 V
15 -28 V
14 -27 V
15 -26 V
15 -26 V
14 -25 V
15 -25 V
14 -24 V
15 -24 V
15 -23 V
14 -21 V
15 -22 V
14 -22 V
15 -21 V
15 -18 V
14 -18 V
15 -20 V
15 -18 V
14 -17 V
15 -17 V
14 -18 V
15 -18 V
15 -15 V
14 -16 V
15 -16 V
14 -18 V
15 -21 V
15 -22 V
14 -24 V
15 -22 V
15 -17 V
14 -10 V
15 -2 V
14 -1 V
15 -2 V
15 -1 V
14 0 V
15 -1 V
14 -4 V
15 -5 V
15 -5 V
14 -4 V
15 -4 V
14 -7 V
15 -11 V
15 -14 V
14 -12 V
15 -14 V
15 -18 V
14 -18 V
15 -13 V
14 -15 V
15 -20 V
15 -22 V
14 -17 V
15 -13 V
14 -11 V
15 -11 V
15 -11 V
14 -14 V
15 -20 V
14 -15 V
15 -3 V
15 -7 V
14 -24 V
15 -27 V
15 -10 V
14 -1 V
15 -10 V
14 -16 V
15 -17 V
15 -14 V
14 -8 V
15 -10 V
14 -18 V
15 -19 V
15 -12 V
14 -5 V
15 -7 V
14 -8 V
15 -12 V
15 -18 V
14 -27 V
15 -22 V
15 0 V
14 0 V
15 -22 V
14 -23 V
15 10 V
LT1
2509 804 D
510 717 D
548 740 D
587 762 D
625 785 D
663 809 D
702 832 D
740 856 D
779 881 D
817 906 D
855 931 D
894 957 D
932 983 D
970 1010 D
1009 1037 D
1047 1065 D
1085 1093 D
1124 1122 D
1162 1152 D
1201 1183 D
1239 1214 D
1277 1246 D
1316 1279 D
1354 1313 D
1392 1348 D
1431 1384 D
1469 1421 D
1507 1459 D
1546 1499 D
1584 1541 D
1622 1585 D
1661 1630 D
1699 1678 D
1738 1729 D
1776 1783 D
1814 1841 D
1853 1903 D
1891 1973 D
1929 2068 D
1968 2077 D
2006 1987 D
2044 1915 D
2083 1849 D
2121 1786 D
2159 1727 D
2198 1671 D
2236 1618 D
2275 1568 D
2313 1520 D
2351 1474 D
2390 1430 D
2428 1388 D
2466 1348 D
2505 1309 D
2543 1271 D
2581 1235 D
2620 1199 D
2658 1165 D
2696 1132 D
2735 1099 D
2773 1067 D
2812 1037 D
2850 1006 D
2888 977 D
2927 948 D
2965 920 D
3003 892 D
3042 865 D
3080 838 D
3118 812 D
3157 786 D
3195 761 D
3234 736 D
3272 711 D
3310 687 D
3349 663 D
3387 639 D
stroke
grestore
end
showpage
}
\put(2389,804){\makebox(0,0)[r]{$r=L/2^5, Re = 1.36 E3$}}
\put(3417,51){\makebox(0,0){10}}
\put(3123,51){\makebox(0,0){8}}
\put(2830,51){\makebox(0,0){6}}
\put(2536,51){\makebox(0,0){4}}
\put(2242,51){\makebox(0,0){2}}
\put(1949,51){\makebox(0,0){0}}
\put(1655,51){\makebox(0,0){-2}}
\put(1361,51){\makebox(0,0){-4}}
\put(1067,51){\makebox(0,0){-6}}
\put(774,51){\makebox(0,0){-8}}
\put(480,51){\makebox(0,0){-10}}
\put(420,2109){\makebox(0,0)[r]{0}}
\put(420,1783){\makebox(0,0)[r]{-1}}
\put(420,1456){\makebox(0,0)[r]{-2}}
\put(420,1130){\makebox(0,0)[r]{-3}}
\put(420,804){\makebox(0,0)[r]{-4}}
\put(420,477){\makebox(0,0)[r]{-5}}
\put(420,151){\makebox(0,0)[r]{-6}}
\end{picture}
\label{fig-pdfevol-35V}
\end{figure}

\newpage

\begin{figure}[htb]
\caption{Evolution of the logarithm of PDFs of velocity differences
over different separations at $Re=1.17E4$. 
The velocity difference is normalized by its variance, and
the PDF is normalized to unity.
  }
\setlength{\unitlength}{0.1bp}
\special{!
/gnudict 40 dict def
gnudict begin
/Color false def
/Solid false def
/gnulinewidth 5.000 def
/vshift -33 def
/dl {10 mul} def
/hpt 31.5 def
/vpt 31.5 def
/M {moveto} bind def
/L {lineto} bind def
/R {rmoveto} bind def
/V {rlineto} bind def
/vpt2 vpt 2 mul def
/hpt2 hpt 2 mul def
/Lshow { currentpoint stroke M
  0 vshift R show } def
/Rshow { currentpoint stroke M
  dup stringwidth pop neg vshift R show } def
/Cshow { currentpoint stroke M
  dup stringwidth pop -2 div vshift R show } def
/DL { Color {setrgbcolor Solid {pop []} if 0 setdash }
 {pop pop pop Solid {pop []} if 0 setdash} ifelse } def
/BL { stroke gnulinewidth 2 mul setlinewidth } def
/AL { stroke gnulinewidth 2 div setlinewidth } def
/PL { stroke gnulinewidth setlinewidth } def
/LTb { BL [] 0 0 0 DL } def
/LTa { AL [1 dl 2 dl] 0 setdash 0 0 0 setrgbcolor } def
/LT0 { PL [] 0 1 0 DL } def
/LT1 { PL [4 dl 2 dl] 0 0 1 DL } def
/LT2 { PL [2 dl 3 dl] 1 0 0 DL } def
/LT3 { PL [1 dl 1.5 dl] 1 0 1 DL } def
/LT4 { PL [5 dl 2 dl 1 dl 2 dl] 0 1 1 DL } def
/LT5 { PL [4 dl 3 dl 1 dl 3 dl] 1 1 0 DL } def
/LT6 { PL [2 dl 2 dl 2 dl 4 dl] 0 0 0 DL } def
/LT7 { PL [2 dl 2 dl 2 dl 2 dl 2 dl 4 dl] 1 0.3 0 DL } def
/LT8 { PL [2 dl 2 dl 2 dl 2 dl 2 dl 2 dl 2 dl 4 dl] 0.5 0.5 0.5 DL } def
/P { stroke [] 0 setdash
  currentlinewidth 2 div sub M
  0 currentlinewidth V stroke } def
/D { stroke [] 0 setdash 2 copy vpt add M
  hpt neg vpt neg V hpt vpt neg V
  hpt vpt V hpt neg vpt V closepath stroke
  P } def
/A { stroke [] 0 setdash vpt sub M 0 vpt2 V
  currentpoint stroke M
  hpt neg vpt neg R hpt2 0 V stroke
  } def
/B { stroke [] 0 setdash 2 copy exch hpt sub exch vpt add M
  0 vpt2 neg V hpt2 0 V 0 vpt2 V
  hpt2 neg 0 V closepath stroke
  P } def
/C { stroke [] 0 setdash exch hpt sub exch vpt add M
  hpt2 vpt2 neg V currentpoint stroke M
  hpt2 neg 0 R hpt2 vpt2 V stroke } def
/T { stroke [] 0 setdash 2 copy vpt 1.12 mul add M
  hpt neg vpt -1.62 mul V
  hpt 2 mul 0 V
  hpt neg vpt 1.62 mul V closepath stroke
  P  } def
/S { 2 copy A C} def
end
}
\begin{picture}(3600,2160)(0,0)
\special{"
gnudict begin
gsave
50 50 translate
0.100 0.100 scale
0 setgray
/Helvetica findfont 100 scalefont setfont
newpath
-500.000000 -500.000000 translate
LTa
LTb
480 151 M
63 0 V
2874 0 R
-63 0 V
480 477 M
63 0 V
2874 0 R
-63 0 V
480 804 M
63 0 V
2874 0 R
-63 0 V
480 1130 M
63 0 V
2874 0 R
-63 0 V
480 1456 M
63 0 V
2874 0 R
-63 0 V
480 1783 M
63 0 V
2874 0 R
-63 0 V
480 2109 M
63 0 V
2874 0 R
-63 0 V
480 151 M
0 63 V
0 1895 R
0 -63 V
774 151 M
0 63 V
0 1895 R
0 -63 V
1067 151 M
0 63 V
0 1895 R
0 -63 V
1361 151 M
0 63 V
0 1895 R
0 -63 V
1655 151 M
0 63 V
0 1895 R
0 -63 V
1949 151 M
0 63 V
0 1895 R
0 -63 V
2242 151 M
0 63 V
0 1895 R
0 -63 V
2536 151 M
0 63 V
0 1895 R
0 -63 V
2830 151 M
0 63 V
0 1895 R
0 -63 V
3123 151 M
0 63 V
0 1895 R
0 -63 V
3417 151 M
0 63 V
0 1895 R
0 -63 V
480 151 M
2937 0 V
0 1958 V
-2937 0 V
480 151 L
LT0
938 151 M
10 246 V
14 104 V
15 21 V
14 3 V
15 33 V
15 39 V
14 23 V
15 36 V
14 57 V
15 43 V
15 13 V
14 8 V
15 31 V
15 36 V
14 26 V
15 18 V
14 15 V
15 23 V
15 37 V
14 32 V
15 13 V
14 11 V
15 25 V
15 31 V
14 25 V
15 22 V
14 23 V
15 26 V
15 31 V
14 31 V
15 27 V
15 24 V
14 24 V
15 23 V
14 22 V
15 22 V
15 22 V
14 23 V
15 22 V
14 22 V
15 25 V
15 25 V
14 23 V
15 22 V
15 22 V
14 21 V
15 22 V
14 23 V
15 21 V
15 22 V
14 22 V
15 22 V
14 21 V
15 21 V
15 21 V
14 20 V
15 20 V
14 20 V
15 18 V
15 18 V
14 18 V
15 16 V
15 15 V
14 14 V
15 13 V
14 11 V
15 9 V
15 6 V
14 4 V
15 2 V
14 -2 V
15 -3 V
15 -6 V
14 -9 V
15 -11 V
14 -13 V
15 -15 V
15 -16 V
14 -19 V
15 -19 V
15 -21 V
14 -21 V
15 -23 V
14 -23 V
15 -24 V
15 -24 V
14 -24 V
15 -25 V
14 -25 V
15 -25 V
15 -25 V
14 -25 V
15 -25 V
14 -25 V
15 -25 V
15 -26 V
14 -27 V
15 -28 V
15 -28 V
14 -24 V
15 -25 V
14 -26 V
15 -28 V
15 -27 V
14 -24 V
15 -21 V
14 -23 V
15 -28 V
15 -34 V
14 -36 V
15 -35 V
15 -29 V
14 -21 V
15 -15 V
14 -17 V
15 -26 V
15 -32 V
14 -33 V
15 -30 V
14 -21 V
15 -22 V
15 -34 V
14 -35 V
15 -23 V
14 -27 V
15 -48 V
15 -66 V
14 -63 V
15 -50 V
15 -38 V
14 -35 V
15 -47 V
14 -43 V
15 -10 V
15 3 V
14 -11 V
15 -46 V
13 -118 V
LT1
2509 804 D
974 169 D
993 256 D
1012 299 D
1031 334 D
1050 385 D
1069 458 D
1088 525 D
1107 581 D
1126 633 D
1144 684 D
1163 733 D
1182 779 D
1201 820 D
1220 865 D
1239 913 D
1258 957 D
1277 995 D
1296 1035 D
1315 1075 D
1334 1115 D
1353 1152 D
1371 1188 D
1390 1223 D
1409 1258 D
1428 1293 D
1447 1327 D
1466 1361 D
1485 1395 D
1504 1429 D
1523 1460 D
1542 1492 D
1561 1524 D
1580 1556 D
1598 1588 D
1617 1619 D
1636 1649 D
1655 1679 D
1674 1709 D
1693 1738 D
1712 1767 D
1731 1795 D
1750 1823 D
1769 1850 D
1788 1876 D
1807 1900 D
1826 1923 D
1844 1944 D
1863 1964 D
1882 1982 D
1901 1995 D
1920 2005 D
1939 2012 D
1958 2014 D
1977 2010 D
1996 2002 D
2015 1991 D
2034 1976 D
2053 1958 D
2071 1936 D
2090 1911 D
2109 1885 D
2128 1856 D
2147 1826 D
2166 1796 D
2185 1764 D
2204 1732 D
2223 1699 D
2242 1665 D
2261 1631 D
2280 1596 D
2299 1561 D
2317 1526 D
2336 1491 D
2355 1455 D
2374 1420 D
2393 1384 D
2412 1348 D
2431 1311 D
2450 1277 D
2469 1242 D
2488 1204 D
2507 1167 D
2526 1128 D
2544 1090 D
2563 1053 D
2582 1010 D
2601 971 D
2620 936 D
2639 900 D
2658 856 D
2677 814 D
2696 775 D
2715 730 D
2734 676 D
2753 636 D
2771 613 D
2790 568 D
2809 499 D
2828 449 D
2847 416 D
2866 367 D
2885 303 D
2904 245 D
2923 233 D
2942 233 D
2961 166 D
stroke
grestore
end
showpage
}
\put(2389,804){\makebox(0,0)[r]{$r=L/2^1$, Re=1.17E4}}
\put(3417,51){\makebox(0,0){10}}
\put(3123,51){\makebox(0,0){8}}
\put(2830,51){\makebox(0,0){6}}
\put(2536,51){\makebox(0,0){4}}
\put(2242,51){\makebox(0,0){2}}
\put(1949,51){\makebox(0,0){0}}
\put(1655,51){\makebox(0,0){-2}}
\put(1361,51){\makebox(0,0){-4}}
\put(1067,51){\makebox(0,0){-6}}
\put(774,51){\makebox(0,0){-8}}
\put(480,51){\makebox(0,0){-10}}
\put(420,2109){\makebox(0,0)[r]{0}}
\put(420,1783){\makebox(0,0)[r]{-1}}
\put(420,1456){\makebox(0,0)[r]{-2}}
\put(420,1130){\makebox(0,0)[r]{-3}}
\put(420,804){\makebox(0,0)[r]{-4}}
\put(420,477){\makebox(0,0)[r]{-5}}
\put(420,151){\makebox(0,0)[r]{-6}}
\end{picture}
\setlength{\unitlength}{0.1bp}
\special{!
/gnudict 40 dict def
gnudict begin
/Color false def
/Solid false def
/gnulinewidth 5.000 def
/vshift -33 def
/dl {10 mul} def
/hpt 31.5 def
/vpt 31.5 def
/M {moveto} bind def
/L {lineto} bind def
/R {rmoveto} bind def
/V {rlineto} bind def
/vpt2 vpt 2 mul def
/hpt2 hpt 2 mul def
/Lshow { currentpoint stroke M
  0 vshift R show } def
/Rshow { currentpoint stroke M
  dup stringwidth pop neg vshift R show } def
/Cshow { currentpoint stroke M
  dup stringwidth pop -2 div vshift R show } def
/DL { Color {setrgbcolor Solid {pop []} if 0 setdash }
 {pop pop pop Solid {pop []} if 0 setdash} ifelse } def
/BL { stroke gnulinewidth 2 mul setlinewidth } def
/AL { stroke gnulinewidth 2 div setlinewidth } def
/PL { stroke gnulinewidth setlinewidth } def
/LTb { BL [] 0 0 0 DL } def
/LTa { AL [1 dl 2 dl] 0 setdash 0 0 0 setrgbcolor } def
/LT0 { PL [] 0 1 0 DL } def
/LT1 { PL [4 dl 2 dl] 0 0 1 DL } def
/LT2 { PL [2 dl 3 dl] 1 0 0 DL } def
/LT3 { PL [1 dl 1.5 dl] 1 0 1 DL } def
/LT4 { PL [5 dl 2 dl 1 dl 2 dl] 0 1 1 DL } def
/LT5 { PL [4 dl 3 dl 1 dl 3 dl] 1 1 0 DL } def
/LT6 { PL [2 dl 2 dl 2 dl 4 dl] 0 0 0 DL } def
/LT7 { PL [2 dl 2 dl 2 dl 2 dl 2 dl 4 dl] 1 0.3 0 DL } def
/LT8 { PL [2 dl 2 dl 2 dl 2 dl 2 dl 2 dl 2 dl 4 dl] 0.5 0.5 0.5 DL } def
/P { stroke [] 0 setdash
  currentlinewidth 2 div sub M
  0 currentlinewidth V stroke } def
/D { stroke [] 0 setdash 2 copy vpt add M
  hpt neg vpt neg V hpt vpt neg V
  hpt vpt V hpt neg vpt V closepath stroke
  P } def
/A { stroke [] 0 setdash vpt sub M 0 vpt2 V
  currentpoint stroke M
  hpt neg vpt neg R hpt2 0 V stroke
  } def
/B { stroke [] 0 setdash 2 copy exch hpt sub exch vpt add M
  0 vpt2 neg V hpt2 0 V 0 vpt2 V
  hpt2 neg 0 V closepath stroke
  P } def
/C { stroke [] 0 setdash exch hpt sub exch vpt add M
  hpt2 vpt2 neg V currentpoint stroke M
  hpt2 neg 0 R hpt2 vpt2 V stroke } def
/T { stroke [] 0 setdash 2 copy vpt 1.12 mul add M
  hpt neg vpt -1.62 mul V
  hpt 2 mul 0 V
  hpt neg vpt 1.62 mul V closepath stroke
  P  } def
/S { 2 copy A C} def
end
}
\begin{picture}(3600,2160)(0,0)
\special{"
gnudict begin
gsave
50 50 translate
0.100 0.100 scale
0 setgray
/Helvetica findfont 100 scalefont setfont
newpath
-500.000000 -500.000000 translate
LTa
LTb
480 151 M
63 0 V
2874 0 R
-63 0 V
480 477 M
63 0 V
2874 0 R
-63 0 V
480 804 M
63 0 V
2874 0 R
-63 0 V
480 1130 M
63 0 V
2874 0 R
-63 0 V
480 1456 M
63 0 V
2874 0 R
-63 0 V
480 1783 M
63 0 V
2874 0 R
-63 0 V
480 2109 M
63 0 V
2874 0 R
-63 0 V
480 151 M
0 63 V
0 1895 R
0 -63 V
774 151 M
0 63 V
0 1895 R
0 -63 V
1067 151 M
0 63 V
0 1895 R
0 -63 V
1361 151 M
0 63 V
0 1895 R
0 -63 V
1655 151 M
0 63 V
0 1895 R
0 -63 V
1949 151 M
0 63 V
0 1895 R
0 -63 V
2242 151 M
0 63 V
0 1895 R
0 -63 V
2536 151 M
0 63 V
0 1895 R
0 -63 V
2830 151 M
0 63 V
0 1895 R
0 -63 V
3123 151 M
0 63 V
0 1895 R
0 -63 V
3417 151 M
0 63 V
0 1895 R
0 -63 V
480 151 M
2937 0 V
0 1958 V
-2937 0 V
480 151 L
LT0
802 151 M
14 102 V
15 57 V
14 27 V
15 14 V
15 61 V
14 92 V
15 52 V
14 1 V
15 -6 V
15 38 V
14 49 V
15 32 V
14 16 V
15 3 V
15 13 V
14 40 V
15 41 V
14 27 V
15 17 V
15 16 V
14 19 V
15 25 V
15 24 V
14 16 V
15 18 V
14 24 V
15 25 V
15 18 V
14 16 V
15 21 V
14 24 V
15 24 V
15 23 V
14 24 V
15 23 V
14 23 V
15 20 V
15 20 V
14 20 V
15 23 V
15 24 V
14 23 V
15 22 V
14 21 V
15 21 V
15 22 V
14 21 V
15 21 V
14 21 V
15 22 V
15 22 V
14 21 V
15 21 V
15 22 V
14 21 V
15 22 V
14 22 V
15 22 V
15 21 V
14 22 V
15 21 V
14 22 V
15 20 V
15 21 V
14 21 V
15 20 V
14 20 V
15 20 V
15 19 V
14 18 V
15 17 V
15 15 V
14 15 V
15 13 V
14 12 V
15 10 V
15 7 V
14 5 V
15 2 V
14 -1 V
15 -4 V
15 -6 V
14 -10 V
15 -12 V
14 -15 V
15 -16 V
15 -18 V
14 -19 V
15 -20 V
15 -21 V
14 -22 V
15 -23 V
14 -23 V
15 -23 V
15 -25 V
14 -25 V
15 -25 V
14 -24 V
15 -24 V
15 -25 V
14 -26 V
15 -26 V
14 -25 V
15 -25 V
15 -25 V
14 -25 V
15 -23 V
15 -22 V
14 -21 V
15 -25 V
14 -30 V
15 -28 V
15 -23 V
14 -20 V
15 -23 V
14 -24 V
15 -25 V
15 -25 V
14 -26 V
15 -24 V
15 -20 V
14 -23 V
15 -30 V
14 -29 V
15 -19 V
15 -20 V
14 -32 V
15 -35 V
14 -24 V
15 -22 V
15 -30 V
14 -32 V
15 -24 V
14 -13 V
15 -5 V
15 -3 V
14 -7 V
15 -24 V
15 -49 V
14 -66 V
15 -47 V
14 -23 V
15 -15 V
15 -24 V
14 -45 V
15 -47 V
14 -12 V
15 -4 V
15 -33 V
14 -60 V
15 -81 V
14 -54 V
15 31 V
15 29 V
14 -20 V
15 -43 V
15 -25 V
2 -2 V
LT1
2509 804 D
883 194 D
908 267 D
932 345 D
957 409 D
981 468 D
1006 526 D
1030 580 D
1055 630 D
1079 679 D
1104 730 D
1128 783 D
1153 830 D
1177 874 D
1202 920 D
1226 965 D
1251 1009 D
1275 1051 D
1300 1093 D
1324 1135 D
1349 1177 D
1373 1218 D
1398 1260 D
1422 1301 D
1447 1341 D
1471 1382 D
1496 1421 D
1520 1461 D
1545 1501 D
1569 1539 D
1593 1578 D
1618 1617 D
1642 1656 D
1667 1693 D
1691 1731 D
1716 1767 D
1740 1804 D
1765 1839 D
1789 1873 D
1814 1904 D
1838 1934 D
1863 1961 D
1887 1984 D
1912 2000 D
1936 2012 D
1961 2014 D
1985 2006 D
2010 1992 D
2034 1972 D
2059 1947 D
2083 1916 D
2108 1881 D
2132 1844 D
2157 1805 D
2181 1765 D
2206 1724 D
2230 1682 D
2255 1640 D
2279 1597 D
2304 1554 D
2328 1511 D
2352 1469 D
2377 1425 D
2401 1382 D
2426 1339 D
2450 1296 D
2475 1253 D
2499 1210 D
2524 1167 D
2548 1124 D
2573 1081 D
2597 1037 D
2622 994 D
2646 950 D
2671 904 D
2695 858 D
2720 814 D
2744 773 D
2769 730 D
2793 679 D
2818 632 D
2842 588 D
2867 529 D
2891 479 D
2916 452 D
2940 400 D
2965 325 D
2989 278 D
3014 235 D
3038 171 D
stroke
grestore
end
showpage
}
\put(2389,804){\makebox(0,0)[r]{$r=L/2^2$, Re=1.17E4}}
\put(3417,51){\makebox(0,0){10}}
\put(3123,51){\makebox(0,0){8}}
\put(2830,51){\makebox(0,0){6}}
\put(2536,51){\makebox(0,0){4}}
\put(2242,51){\makebox(0,0){2}}
\put(1949,51){\makebox(0,0){0}}
\put(1655,51){\makebox(0,0){-2}}
\put(1361,51){\makebox(0,0){-4}}
\put(1067,51){\makebox(0,0){-6}}
\put(774,51){\makebox(0,0){-8}}
\put(480,51){\makebox(0,0){-10}}
\put(420,2109){\makebox(0,0)[r]{0}}
\put(420,1783){\makebox(0,0)[r]{-1}}
\put(420,1456){\makebox(0,0)[r]{-2}}
\put(420,1130){\makebox(0,0)[r]{-3}}
\put(420,804){\makebox(0,0)[r]{-4}}
\put(420,477){\makebox(0,0)[r]{-5}}
\put(420,151){\makebox(0,0)[r]{-6}}
\end{picture}
\newpage
\setlength{\unitlength}{0.1bp}
\special{!
/gnudict 40 dict def
gnudict begin
/Color false def
/Solid false def
/gnulinewidth 5.000 def
/vshift -33 def
/dl {10 mul} def
/hpt 31.5 def
/vpt 31.5 def
/M {moveto} bind def
/L {lineto} bind def
/R {rmoveto} bind def
/V {rlineto} bind def
/vpt2 vpt 2 mul def
/hpt2 hpt 2 mul def
/Lshow { currentpoint stroke M
  0 vshift R show } def
/Rshow { currentpoint stroke M
  dup stringwidth pop neg vshift R show } def
/Cshow { currentpoint stroke M
  dup stringwidth pop -2 div vshift R show } def
/DL { Color {setrgbcolor Solid {pop []} if 0 setdash }
 {pop pop pop Solid {pop []} if 0 setdash} ifelse } def
/BL { stroke gnulinewidth 2 mul setlinewidth } def
/AL { stroke gnulinewidth 2 div setlinewidth } def
/PL { stroke gnulinewidth setlinewidth } def
/LTb { BL [] 0 0 0 DL } def
/LTa { AL [1 dl 2 dl] 0 setdash 0 0 0 setrgbcolor } def
/LT0 { PL [] 0 1 0 DL } def
/LT1 { PL [4 dl 2 dl] 0 0 1 DL } def
/LT2 { PL [2 dl 3 dl] 1 0 0 DL } def
/LT3 { PL [1 dl 1.5 dl] 1 0 1 DL } def
/LT4 { PL [5 dl 2 dl 1 dl 2 dl] 0 1 1 DL } def
/LT5 { PL [4 dl 3 dl 1 dl 3 dl] 1 1 0 DL } def
/LT6 { PL [2 dl 2 dl 2 dl 4 dl] 0 0 0 DL } def
/LT7 { PL [2 dl 2 dl 2 dl 2 dl 2 dl 4 dl] 1 0.3 0 DL } def
/LT8 { PL [2 dl 2 dl 2 dl 2 dl 2 dl 2 dl 2 dl 4 dl] 0.5 0.5 0.5 DL } def
/P { stroke [] 0 setdash
  currentlinewidth 2 div sub M
  0 currentlinewidth V stroke } def
/D { stroke [] 0 setdash 2 copy vpt add M
  hpt neg vpt neg V hpt vpt neg V
  hpt vpt V hpt neg vpt V closepath stroke
  P } def
/A { stroke [] 0 setdash vpt sub M 0 vpt2 V
  currentpoint stroke M
  hpt neg vpt neg R hpt2 0 V stroke
  } def
/B { stroke [] 0 setdash 2 copy exch hpt sub exch vpt add M
  0 vpt2 neg V hpt2 0 V 0 vpt2 V
  hpt2 neg 0 V closepath stroke
  P } def
/C { stroke [] 0 setdash exch hpt sub exch vpt add M
  hpt2 vpt2 neg V currentpoint stroke M
  hpt2 neg 0 R hpt2 vpt2 V stroke } def
/T { stroke [] 0 setdash 2 copy vpt 1.12 mul add M
  hpt neg vpt -1.62 mul V
  hpt 2 mul 0 V
  hpt neg vpt 1.62 mul V closepath stroke
  P  } def
/S { 2 copy A C} def
end
}
\begin{picture}(3600,2160)(0,0)
\special{"
gnudict begin
gsave
50 50 translate
0.100 0.100 scale
0 setgray
/Helvetica findfont 100 scalefont setfont
newpath
-500.000000 -500.000000 translate
LTa
LTb
480 151 M
63 0 V
2874 0 R
-63 0 V
480 477 M
63 0 V
2874 0 R
-63 0 V
480 804 M
63 0 V
2874 0 R
-63 0 V
480 1130 M
63 0 V
2874 0 R
-63 0 V
480 1456 M
63 0 V
2874 0 R
-63 0 V
480 1783 M
63 0 V
2874 0 R
-63 0 V
480 2109 M
63 0 V
2874 0 R
-63 0 V
480 151 M
0 63 V
0 1895 R
0 -63 V
774 151 M
0 63 V
0 1895 R
0 -63 V
1067 151 M
0 63 V
0 1895 R
0 -63 V
1361 151 M
0 63 V
0 1895 R
0 -63 V
1655 151 M
0 63 V
0 1895 R
0 -63 V
1949 151 M
0 63 V
0 1895 R
0 -63 V
2242 151 M
0 63 V
0 1895 R
0 -63 V
2536 151 M
0 63 V
0 1895 R
0 -63 V
2830 151 M
0 63 V
0 1895 R
0 -63 V
3123 151 M
0 63 V
0 1895 R
0 -63 V
3417 151 M
0 63 V
0 1895 R
0 -63 V
480 151 M
2937 0 V
0 1958 V
-2937 0 V
480 151 L
LT0
569 151 M
13 62 V
15 -11 V
13 -51 V
94 0 R
10 18 V
14 143 V
15 132 V
15 56 V
14 -12 V
15 -34 V
14 22 V
15 40 V
15 17 V
14 9 V
15 17 V
15 24 V
14 26 V
15 23 V
14 19 V
15 23 V
15 31 V
14 27 V
15 17 V
14 16 V
15 25 V
15 23 V
14 13 V
15 10 V
14 16 V
15 20 V
15 18 V
14 18 V
15 21 V
15 18 V
14 10 V
15 10 V
14 18 V
15 24 V
15 25 V
14 23 V
15 21 V
14 21 V
15 22 V
15 23 V
14 21 V
15 22 V
14 23 V
15 22 V
15 20 V
14 19 V
15 19 V
15 21 V
14 20 V
15 19 V
14 20 V
15 20 V
15 19 V
14 20 V
15 22 V
14 23 V
15 20 V
15 19 V
14 20 V
15 21 V
15 21 V
14 21 V
15 21 V
14 21 V
15 22 V
15 22 V
14 21 V
15 21 V
14 21 V
15 21 V
15 21 V
14 21 V
15 21 V
14 19 V
15 19 V
15 19 V
14 18 V
15 19 V
15 17 V
14 16 V
15 14 V
14 12 V
15 11 V
15 8 V
14 6 V
15 2 V
14 -1 V
15 -4 V
15 -7 V
14 -11 V
15 -13 V
14 -15 V
15 -17 V
15 -20 V
14 -20 V
15 -21 V
15 -22 V
14 -22 V
15 -23 V
14 -23 V
15 -24 V
15 -24 V
14 -24 V
15 -25 V
14 -25 V
15 -25 V
15 -25 V
14 -24 V
15 -24 V
14 -23 V
15 -23 V
15 -24 V
14 -24 V
15 -25 V
15 -25 V
14 -23 V
15 -24 V
14 -23 V
15 -22 V
15 -20 V
14 -21 V
15 -24 V
14 -25 V
15 -24 V
15 -23 V
14 -23 V
15 -23 V
15 -24 V
14 -22 V
15 -21 V
14 -20 V
15 -23 V
15 -23 V
14 -20 V
15 -22 V
14 -27 V
15 -28 V
15 -23 V
14 -17 V
15 -11 V
14 -16 V
15 -31 V
15 -35 V
14 -26 V
15 -21 V
15 -27 V
14 -23 V
15 -12 V
14 -17 V
15 -35 V
15 -32 V
14 2 V
15 11 V
14 -4 V
15 -14 V
15 -19 V
14 -34 V
15 -62 V
14 -72 V
15 -12 V
15 22 V
14 2 V
15 -34 V
15 -105 V
13 -125 V
3 0 R
13 98 V
14 65 V
15 -44 V
12 -119 V
LT1
2509 804 D
751 169 D
767 208 D
783 242 D
799 275 D
814 307 D
830 340 D
846 371 D
862 402 D
878 432 D
894 460 D
910 488 D
925 515 D
941 543 D
957 573 D
973 603 D
989 632 D
1005 659 D
1021 685 D
1036 710 D
1052 737 D
1068 763 D
1084 789 D
1100 815 D
1116 841 D
1132 866 D
1148 891 D
1163 916 D
1179 941 D
1195 965 D
1211 990 D
1227 1015 D
1243 1039 D
1259 1064 D
1274 1088 D
1290 1113 D
1306 1138 D
1322 1162 D
1338 1187 D
1354 1211 D
1370 1235 D
1385 1259 D
1401 1283 D
1417 1308 D
1433 1332 D
1449 1356 D
1465 1381 D
1481 1405 D
1496 1429 D
1512 1454 D
1528 1478 D
1544 1502 D
1560 1526 D
1576 1550 D
1592 1575 D
1607 1599 D
1623 1623 D
1639 1647 D
1655 1671 D
1671 1695 D
1687 1719 D
1703 1742 D
1719 1766 D
1734 1789 D
1750 1812 D
1766 1835 D
1782 1857 D
1798 1878 D
1814 1899 D
1830 1919 D
1845 1938 D
1861 1955 D
1877 1972 D
1893 1986 D
1909 1997 D
1925 2004 D
1941 2027 D
1956 2030 D
1972 2009 D
1988 2003 D
2004 1994 D
2020 1982 D
2036 1968 D
2052 1950 D
2067 1931 D
2083 1910 D
2099 1888 D
2115 1864 D
2131 1840 D
2147 1815 D
2163 1790 D
2178 1764 D
2194 1738 D
2210 1712 D
2226 1685 D
2242 1659 D
2258 1632 D
2274 1606 D
2290 1579 D
2305 1553 D
2321 1526 D
2337 1500 D
2353 1473 D
2369 1447 D
2385 1421 D
2401 1394 D
2416 1369 D
2432 1342 D
2448 1316 D
2464 1291 D
2480 1265 D
2496 1239 D
2512 1213 D
2527 1188 D
2543 1162 D
2559 1137 D
2575 1112 D
2591 1086 D
2607 1061 D
2623 1036 D
2638 1011 D
2654 986 D
2670 961 D
2686 935 D
2702 910 D
2718 885 D
2734 859 D
2749 834 D
2765 809 D
2781 783 D
2797 757 D
2813 730 D
2829 704 D
2845 678 D
2861 654 D
2876 630 D
2892 607 D
2908 582 D
2924 554 D
2940 525 D
2956 497 D
2972 472 D
2987 448 D
3003 420 D
3019 388 D
3035 354 D
3051 326 D
3067 308 D
3083 292 D
3098 269 D
3114 235 D
3130 193 D
3146 155 D
stroke
grestore
end
showpage
}
\put(2389,804){\makebox(0,0)[r]{$r=L/2^3$, Re=1.17E4}}
\put(3417,51){\makebox(0,0){10}}
\put(3123,51){\makebox(0,0){8}}
\put(2830,51){\makebox(0,0){6}}
\put(2536,51){\makebox(0,0){4}}
\put(2242,51){\makebox(0,0){2}}
\put(1949,51){\makebox(0,0){0}}
\put(1655,51){\makebox(0,0){-2}}
\put(1361,51){\makebox(0,0){-4}}
\put(1067,51){\makebox(0,0){-6}}
\put(774,51){\makebox(0,0){-8}}
\put(480,51){\makebox(0,0){-10}}
\put(420,2109){\makebox(0,0)[r]{0}}
\put(420,1783){\makebox(0,0)[r]{-1}}
\put(420,1456){\makebox(0,0)[r]{-2}}
\put(420,1130){\makebox(0,0)[r]{-3}}
\put(420,804){\makebox(0,0)[r]{-4}}
\put(420,477){\makebox(0,0)[r]{-5}}
\put(420,151){\makebox(0,0)[r]{-6}}
\end{picture}
\setlength{\unitlength}{0.1bp}
\special{!
/gnudict 40 dict def
gnudict begin
/Color false def
/Solid false def
/gnulinewidth 5.000 def
/vshift -33 def
/dl {10 mul} def
/hpt 31.5 def
/vpt 31.5 def
/M {moveto} bind def
/L {lineto} bind def
/R {rmoveto} bind def
/V {rlineto} bind def
/vpt2 vpt 2 mul def
/hpt2 hpt 2 mul def
/Lshow { currentpoint stroke M
  0 vshift R show } def
/Rshow { currentpoint stroke M
  dup stringwidth pop neg vshift R show } def
/Cshow { currentpoint stroke M
  dup stringwidth pop -2 div vshift R show } def
/DL { Color {setrgbcolor Solid {pop []} if 0 setdash }
 {pop pop pop Solid {pop []} if 0 setdash} ifelse } def
/BL { stroke gnulinewidth 2 mul setlinewidth } def
/AL { stroke gnulinewidth 2 div setlinewidth } def
/PL { stroke gnulinewidth setlinewidth } def
/LTb { BL [] 0 0 0 DL } def
/LTa { AL [1 dl 2 dl] 0 setdash 0 0 0 setrgbcolor } def
/LT0 { PL [] 0 1 0 DL } def
/LT1 { PL [4 dl 2 dl] 0 0 1 DL } def
/LT2 { PL [2 dl 3 dl] 1 0 0 DL } def
/LT3 { PL [1 dl 1.5 dl] 1 0 1 DL } def
/LT4 { PL [5 dl 2 dl 1 dl 2 dl] 0 1 1 DL } def
/LT5 { PL [4 dl 3 dl 1 dl 3 dl] 1 1 0 DL } def
/LT6 { PL [2 dl 2 dl 2 dl 4 dl] 0 0 0 DL } def
/LT7 { PL [2 dl 2 dl 2 dl 2 dl 2 dl 4 dl] 1 0.3 0 DL } def
/LT8 { PL [2 dl 2 dl 2 dl 2 dl 2 dl 2 dl 2 dl 4 dl] 0.5 0.5 0.5 DL } def
/P { stroke [] 0 setdash
  currentlinewidth 2 div sub M
  0 currentlinewidth V stroke } def
/D { stroke [] 0 setdash 2 copy vpt add M
  hpt neg vpt neg V hpt vpt neg V
  hpt vpt V hpt neg vpt V closepath stroke
  P } def
/A { stroke [] 0 setdash vpt sub M 0 vpt2 V
  currentpoint stroke M
  hpt neg vpt neg R hpt2 0 V stroke
  } def
/B { stroke [] 0 setdash 2 copy exch hpt sub exch vpt add M
  0 vpt2 neg V hpt2 0 V 0 vpt2 V
  hpt2 neg 0 V closepath stroke
  P } def
/C { stroke [] 0 setdash exch hpt sub exch vpt add M
  hpt2 vpt2 neg V currentpoint stroke M
  hpt2 neg 0 R hpt2 vpt2 V stroke } def
/T { stroke [] 0 setdash 2 copy vpt 1.12 mul add M
  hpt neg vpt -1.62 mul V
  hpt 2 mul 0 V
  hpt neg vpt 1.62 mul V closepath stroke
  P  } def
/S { 2 copy A C} def
end
}
\begin{picture}(3600,2160)(0,0)
\special{"
gnudict begin
gsave
50 50 translate
0.100 0.100 scale
0 setgray
/Helvetica findfont 100 scalefont setfont
newpath
-500.000000 -500.000000 translate
LTa
LTb
480 151 M
63 0 V
2874 0 R
-63 0 V
480 477 M
63 0 V
2874 0 R
-63 0 V
480 804 M
63 0 V
2874 0 R
-63 0 V
480 1130 M
63 0 V
2874 0 R
-63 0 V
480 1456 M
63 0 V
2874 0 R
-63 0 V
480 1783 M
63 0 V
2874 0 R
-63 0 V
480 2109 M
63 0 V
2874 0 R
-63 0 V
480 151 M
0 63 V
0 1895 R
0 -63 V
774 151 M
0 63 V
0 1895 R
0 -63 V
1067 151 M
0 63 V
0 1895 R
0 -63 V
1361 151 M
0 63 V
0 1895 R
0 -63 V
1655 151 M
0 63 V
0 1895 R
0 -63 V
1949 151 M
0 63 V
0 1895 R
0 -63 V
2242 151 M
0 63 V
0 1895 R
0 -63 V
2536 151 M
0 63 V
0 1895 R
0 -63 V
2830 151 M
0 63 V
0 1895 R
0 -63 V
3123 151 M
0 63 V
0 1895 R
0 -63 V
3417 151 M
0 63 V
0 1895 R
0 -63 V
480 151 M
2937 0 V
0 1958 V
-2937 0 V
480 151 L
LT0
538 284 M
15 -23 V
15 48 V
14 29 V
15 -25 V
15 -3 V
14 82 V
15 59 V
14 15 V
15 -3 V
15 6 V
14 23 V
15 35 V
14 28 V
15 12 V
15 -4 V
14 -19 V
15 -8 V
14 29 V
15 34 V
15 17 V
14 12 V
15 21 V
15 23 V
14 19 V
15 12 V
14 7 V
15 14 V
15 32 V
14 26 V
15 5 V
14 4 V
15 26 V
15 29 V
14 14 V
15 10 V
14 17 V
15 20 V
15 18 V
14 17 V
15 19 V
15 20 V
14 19 V
15 17 V
14 17 V
15 17 V
15 19 V
14 22 V
15 22 V
14 21 V
15 17 V
15 16 V
14 19 V
15 19 V
14 18 V
15 17 V
15 20 V
14 20 V
15 19 V
15 18 V
14 19 V
15 19 V
14 20 V
15 20 V
15 20 V
14 20 V
15 18 V
14 19 V
15 20 V
15 21 V
14 20 V
15 19 V
15 20 V
14 21 V
15 22 V
14 21 V
15 22 V
15 20 V
14 21 V
15 20 V
14 21 V
15 20 V
15 20 V
14 21 V
15 20 V
14 21 V
15 20 V
15 19 V
14 19 V
15 19 V
15 18 V
14 17 V
15 15 V
14 14 V
15 11 V
15 10 V
14 6 V
15 2 V
14 0 V
15 -5 V
15 -8 V
14 -12 V
15 -14 V
14 -16 V
15 -18 V
15 -20 V
14 -22 V
15 -22 V
15 -23 V
14 -23 V
15 -23 V
14 -23 V
15 -24 V
15 -24 V
14 -24 V
15 -25 V
14 -24 V
15 -24 V
15 -24 V
14 -23 V
15 -23 V
14 -23 V
15 -24 V
15 -23 V
14 -24 V
15 -23 V
15 -23 V
14 -23 V
15 -22 V
14 -23 V
15 -22 V
15 -22 V
14 -22 V
15 -22 V
14 -22 V
15 -22 V
15 -23 V
14 -24 V
15 -23 V
15 -19 V
14 -18 V
15 -18 V
14 -17 V
15 -17 V
15 -20 V
14 -26 V
15 -28 V
14 -22 V
15 -18 V
15 -20 V
14 -20 V
15 -18 V
14 -19 V
15 -21 V
15 -18 V
14 -12 V
15 -16 V
15 -25 V
14 -25 V
15 -14 V
14 -10 V
15 -17 V
15 -19 V
14 -16 V
15 -18 V
14 -24 V
15 -28 V
15 -25 V
14 -20 V
15 -12 V
14 -9 V
15 -13 V
15 -17 V
14 -23 V
15 -33 V
15 -46 V
14 -21 V
15 31 V
14 13 V
15 -46 V
15 -90 V
14 -48 V
15 5 V
14 -6 V
15 -17 V
15 -19 V
14 -41 V
4 -28 V
70 0 R
14 31 V
5 -31 V
LT1
2509 804 D
640 187 D
660 221 D
681 255 D
701 288 D
722 319 D
743 350 D
763 382 D
784 416 D
804 449 D
825 481 D
846 511 D
866 541 D
887 570 D
908 601 D
928 631 D
949 661 D
969 690 D
990 720 D
1011 748 D
1031 777 D
1052 806 D
1072 835 D
1093 864 D
1114 893 D
1134 921 D
1155 950 D
1176 980 D
1196 1009 D
1217 1038 D
1237 1067 D
1258 1096 D
1279 1125 D
1299 1154 D
1320 1183 D
1340 1212 D
1361 1241 D
1382 1271 D
1402 1300 D
1423 1329 D
1443 1359 D
1464 1389 D
1485 1419 D
1505 1448 D
1526 1478 D
1547 1508 D
1567 1539 D
1588 1569 D
1608 1599 D
1629 1629 D
1650 1660 D
1670 1690 D
1691 1720 D
1711 1750 D
1732 1780 D
1753 1810 D
1773 1839 D
1794 1867 D
1815 1894 D
1835 1920 D
1856 1943 D
1876 1964 D
1897 1981 D
1918 1994 D
1938 2037 D
1959 2040 D
1979 2000 D
2000 1990 D
2021 1975 D
2041 1956 D
2062 1932 D
2082 1904 D
2103 1875 D
2124 1844 D
2144 1813 D
2165 1780 D
2186 1747 D
2206 1714 D
2227 1681 D
2247 1648 D
2268 1614 D
2289 1581 D
2309 1548 D
2330 1515 D
2350 1483 D
2371 1450 D
2392 1418 D
2412 1386 D
2433 1354 D
2454 1322 D
2474 1291 D
2495 1260 D
2515 1229 D
2536 1198 D
2557 1167 D
2577 1137 D
2598 1106 D
2618 1076 D
2639 1046 D
2660 1016 D
2680 986 D
2701 956 D
2721 927 D
2742 897 D
2763 868 D
2783 839 D
2804 810 D
2825 780 D
2845 751 D
2866 722 D
2886 693 D
2907 664 D
2928 634 D
2948 604 D
2969 574 D
2989 544 D
3010 515 D
3031 488 D
3051 461 D
3072 433 D
3093 403 D
3113 370 D
3134 339 D
3154 311 D
3175 283 D
3196 250 D
3216 213 D
3237 178 D
3257 152 D
stroke
grestore
end
showpage
}
\put(2389,804){\makebox(0,0)[r]{$r=L/2^4$, Re=1.17E4}}
\put(3417,51){\makebox(0,0){10}}
\put(3123,51){\makebox(0,0){8}}
\put(2830,51){\makebox(0,0){6}}
\put(2536,51){\makebox(0,0){4}}
\put(2242,51){\makebox(0,0){2}}
\put(1949,51){\makebox(0,0){0}}
\put(1655,51){\makebox(0,0){-2}}
\put(1361,51){\makebox(0,0){-4}}
\put(1067,51){\makebox(0,0){-6}}
\put(774,51){\makebox(0,0){-8}}
\put(480,51){\makebox(0,0){-10}}
\put(420,2109){\makebox(0,0)[r]{0}}
\put(420,1783){\makebox(0,0)[r]{-1}}
\put(420,1456){\makebox(0,0)[r]{-2}}
\put(420,1130){\makebox(0,0)[r]{-3}}
\put(420,804){\makebox(0,0)[r]{-4}}
\put(420,477){\makebox(0,0)[r]{-5}}
\put(420,151){\makebox(0,0)[r]{-6}}
\end{picture}
\setlength{\unitlength}{0.1bp}
\special{!
/gnudict 40 dict def
gnudict begin
/Color false def
/Solid false def
/gnulinewidth 5.000 def
/vshift -33 def
/dl {10 mul} def
/hpt 31.5 def
/vpt 31.5 def
/M {moveto} bind def
/L {lineto} bind def
/R {rmoveto} bind def
/V {rlineto} bind def
/vpt2 vpt 2 mul def
/hpt2 hpt 2 mul def
/Lshow { currentpoint stroke M
  0 vshift R show } def
/Rshow { currentpoint stroke M
  dup stringwidth pop neg vshift R show } def
/Cshow { currentpoint stroke M
  dup stringwidth pop -2 div vshift R show } def
/DL { Color {setrgbcolor Solid {pop []} if 0 setdash }
 {pop pop pop Solid {pop []} if 0 setdash} ifelse } def
/BL { stroke gnulinewidth 2 mul setlinewidth } def
/AL { stroke gnulinewidth 2 div setlinewidth } def
/PL { stroke gnulinewidth setlinewidth } def
/LTb { BL [] 0 0 0 DL } def
/LTa { AL [1 dl 2 dl] 0 setdash 0 0 0 setrgbcolor } def
/LT0 { PL [] 0 1 0 DL } def
/LT1 { PL [4 dl 2 dl] 0 0 1 DL } def
/LT2 { PL [2 dl 3 dl] 1 0 0 DL } def
/LT3 { PL [1 dl 1.5 dl] 1 0 1 DL } def
/LT4 { PL [5 dl 2 dl 1 dl 2 dl] 0 1 1 DL } def
/LT5 { PL [4 dl 3 dl 1 dl 3 dl] 1 1 0 DL } def
/LT6 { PL [2 dl 2 dl 2 dl 4 dl] 0 0 0 DL } def
/LT7 { PL [2 dl 2 dl 2 dl 2 dl 2 dl 4 dl] 1 0.3 0 DL } def
/LT8 { PL [2 dl 2 dl 2 dl 2 dl 2 dl 2 dl 2 dl 4 dl] 0.5 0.5 0.5 DL } def
/P { stroke [] 0 setdash
  currentlinewidth 2 div sub M
  0 currentlinewidth V stroke } def
/D { stroke [] 0 setdash 2 copy vpt add M
  hpt neg vpt neg V hpt vpt neg V
  hpt vpt V hpt neg vpt V closepath stroke
  P } def
/A { stroke [] 0 setdash vpt sub M 0 vpt2 V
  currentpoint stroke M
  hpt neg vpt neg R hpt2 0 V stroke
  } def
/B { stroke [] 0 setdash 2 copy exch hpt sub exch vpt add M
  0 vpt2 neg V hpt2 0 V 0 vpt2 V
  hpt2 neg 0 V closepath stroke
  P } def
/C { stroke [] 0 setdash exch hpt sub exch vpt add M
  hpt2 vpt2 neg V currentpoint stroke M
  hpt2 neg 0 R hpt2 vpt2 V stroke } def
/T { stroke [] 0 setdash 2 copy vpt 1.12 mul add M
  hpt neg vpt -1.62 mul V
  hpt 2 mul 0 V
  hpt neg vpt 1.62 mul V closepath stroke
  P  } def
/S { 2 copy A C} def
end
}
\begin{picture}(3600,2160)(0,0)
\special{"
gnudict begin
gsave
50 50 translate
0.100 0.100 scale
0 setgray
/Helvetica findfont 100 scalefont setfont
newpath
-500.000000 -500.000000 translate
LTa
LTb
480 151 M
63 0 V
2874 0 R
-63 0 V
480 477 M
63 0 V
2874 0 R
-63 0 V
480 804 M
63 0 V
2874 0 R
-63 0 V
480 1130 M
63 0 V
2874 0 R
-63 0 V
480 1456 M
63 0 V
2874 0 R
-63 0 V
480 1783 M
63 0 V
2874 0 R
-63 0 V
480 2109 M
63 0 V
2874 0 R
-63 0 V
480 151 M
0 63 V
0 1895 R
0 -63 V
774 151 M
0 63 V
0 1895 R
0 -63 V
1067 151 M
0 63 V
0 1895 R
0 -63 V
1361 151 M
0 63 V
0 1895 R
0 -63 V
1655 151 M
0 63 V
0 1895 R
0 -63 V
1949 151 M
0 63 V
0 1895 R
0 -63 V
2242 151 M
0 63 V
0 1895 R
0 -63 V
2536 151 M
0 63 V
0 1895 R
0 -63 V
2830 151 M
0 63 V
0 1895 R
0 -63 V
3123 151 M
0 63 V
0 1895 R
0 -63 V
3417 151 M
0 63 V
0 1895 R
0 -63 V
480 151 M
2937 0 V
0 1958 V
-2937 0 V
480 151 L
LT0
538 430 M
15 12 V
15 19 V
14 24 V
15 23 V
15 14 V
14 0 V
15 0 V
14 12 V
15 14 V
15 8 V
14 14 V
15 26 V
14 29 V
15 21 V
15 10 V
14 -3 V
15 2 V
14 22 V
15 19 V
15 -3 V
14 -9 V
15 14 V
15 26 V
14 23 V
15 21 V
14 20 V
15 13 V
15 5 V
14 8 V
15 23 V
14 28 V
15 21 V
15 16 V
14 14 V
15 14 V
14 15 V
15 18 V
15 18 V
14 17 V
15 13 V
15 14 V
14 16 V
15 17 V
14 15 V
15 14 V
15 15 V
14 16 V
15 18 V
14 18 V
15 20 V
15 18 V
14 15 V
15 15 V
14 18 V
15 20 V
15 19 V
14 17 V
15 17 V
15 15 V
14 17 V
15 18 V
14 19 V
15 19 V
15 20 V
14 19 V
15 19 V
14 19 V
15 19 V
15 19 V
14 19 V
15 19 V
15 19 V
14 19 V
15 20 V
14 19 V
15 21 V
15 21 V
14 20 V
15 20 V
14 20 V
15 21 V
15 20 V
14 20 V
15 21 V
14 21 V
15 21 V
15 20 V
14 20 V
15 19 V
15 19 V
14 18 V
15 17 V
14 15 V
15 13 V
15 11 V
14 8 V
15 3 V
14 -1 V
15 -5 V
15 -10 V
14 -13 V
15 -16 V
14 -18 V
15 -20 V
15 -21 V
14 -22 V
15 -23 V
15 -24 V
14 -24 V
15 -23 V
14 -24 V
15 -24 V
15 -24 V
14 -24 V
15 -23 V
14 -24 V
15 -24 V
15 -24 V
14 -22 V
15 -23 V
14 -23 V
15 -22 V
15 -22 V
14 -22 V
15 -22 V
15 -23 V
14 -23 V
15 -23 V
14 -21 V
15 -20 V
15 -20 V
14 -20 V
15 -21 V
14 -20 V
15 -19 V
15 -21 V
14 -21 V
15 -21 V
15 -19 V
14 -19 V
15 -21 V
14 -21 V
15 -19 V
15 -16 V
14 -14 V
15 -18 V
14 -24 V
15 -26 V
15 -18 V
14 -14 V
15 -13 V
14 -12 V
15 -12 V
15 -16 V
14 -23 V
15 -23 V
15 -14 V
14 -12 V
15 -17 V
14 -19 V
15 -18 V
15 -17 V
14 -20 V
15 -22 V
14 -25 V
15 -18 V
15 -7 V
14 -7 V
15 -17 V
14 -17 V
15 -5 V
15 -13 V
14 -35 V
15 -37 V
15 -3 V
14 -6 V
15 -48 V
14 -45 V
15 33 V
15 27 V
14 -26 V
15 -57 V
14 -32 V
15 5 V
15 15 V
14 0 V
15 -32 V
14 -70 V
15 -113 V
12 -29 V
3 0 R
14 120 V
15 55 V
15 -34 V
14 -79 V
15 0 V
14 27 V
15 -31 V
LT1
2509 804 D
484 161 D
511 197 D
539 233 D
566 271 D
593 308 D
621 343 D
648 376 D
675 410 D
703 444 D
730 479 D
758 513 D
785 547 D
812 580 D
840 613 D
867 647 D
894 680 D
922 714 D
949 747 D
977 781 D
1004 815 D
1031 849 D
1059 883 D
1086 917 D
1113 951 D
1141 985 D
1168 1020 D
1196 1054 D
1223 1089 D
1250 1124 D
1278 1159 D
1305 1195 D
1332 1230 D
1360 1266 D
1387 1302 D
1415 1338 D
1442 1374 D
1469 1411 D
1497 1448 D
1524 1485 D
1552 1522 D
1579 1560 D
1606 1598 D
1634 1636 D
1661 1674 D
1688 1712 D
1716 1749 D
1743 1786 D
1771 1823 D
1798 1857 D
1825 1890 D
1853 1920 D
1880 1949 D
1907 1985 D
1935 2049 D
1962 2054 D
1990 1994 D
2017 1961 D
2044 1932 D
2072 1901 D
2099 1866 D
2126 1828 D
2154 1788 D
2181 1747 D
2209 1706 D
2236 1665 D
2263 1623 D
2291 1582 D
2318 1541 D
2345 1500 D
2373 1460 D
2400 1420 D
2428 1380 D
2455 1341 D
2482 1303 D
2510 1264 D
2537 1226 D
2565 1189 D
2592 1152 D
2619 1115 D
2647 1078 D
2674 1042 D
2701 1006 D
2729 970 D
2756 935 D
2784 900 D
2811 865 D
2838 830 D
2866 796 D
2893 762 D
2920 728 D
2948 694 D
2975 661 D
3003 627 D
3030 593 D
3057 560 D
3085 526 D
3112 493 D
3139 459 D
3167 425 D
3194 391 D
3222 358 D
3249 326 D
3276 295 D
3304 263 D
3331 228 D
3358 192 D
3386 159 D
stroke
grestore
end
showpage
}
\put(2389,804){\makebox(0,0)[r]{$r=L/2^5$, Re=1.17E4}}
\put(3417,51){\makebox(0,0){10}}
\put(3123,51){\makebox(0,0){8}}
\put(2830,51){\makebox(0,0){6}}
\put(2536,51){\makebox(0,0){4}}
\put(2242,51){\makebox(0,0){2}}
\put(1949,51){\makebox(0,0){0}}
\put(1655,51){\makebox(0,0){-2}}
\put(1361,51){\makebox(0,0){-4}}
\put(1067,51){\makebox(0,0){-6}}
\put(774,51){\makebox(0,0){-8}}
\put(480,51){\makebox(0,0){-10}}
\put(420,2109){\makebox(0,0)[r]{0}}
\put(420,1783){\makebox(0,0)[r]{-1}}
\put(420,1456){\makebox(0,0)[r]{-2}}
\put(420,1130){\makebox(0,0)[r]{-3}}
\put(420,804){\makebox(0,0)[r]{-4}}
\put(420,477){\makebox(0,0)[r]{-5}}
\put(420,151){\makebox(0,0)[r]{-6}}
\end{picture}
\newpage
\setlength{\unitlength}{0.1bp}
\special{!
/gnudict 40 dict def
gnudict begin
/Color false def
/Solid false def
/gnulinewidth 5.000 def
/vshift -33 def
/dl {10 mul} def
/hpt 31.5 def
/vpt 31.5 def
/M {moveto} bind def
/L {lineto} bind def
/R {rmoveto} bind def
/V {rlineto} bind def
/vpt2 vpt 2 mul def
/hpt2 hpt 2 mul def
/Lshow { currentpoint stroke M
  0 vshift R show } def
/Rshow { currentpoint stroke M
  dup stringwidth pop neg vshift R show } def
/Cshow { currentpoint stroke M
  dup stringwidth pop -2 div vshift R show } def
/DL { Color {setrgbcolor Solid {pop []} if 0 setdash }
 {pop pop pop Solid {pop []} if 0 setdash} ifelse } def
/BL { stroke gnulinewidth 2 mul setlinewidth } def
/AL { stroke gnulinewidth 2 div setlinewidth } def
/PL { stroke gnulinewidth setlinewidth } def
/LTb { BL [] 0 0 0 DL } def
/LTa { AL [1 dl 2 dl] 0 setdash 0 0 0 setrgbcolor } def
/LT0 { PL [] 0 1 0 DL } def
/LT1 { PL [4 dl 2 dl] 0 0 1 DL } def
/LT2 { PL [2 dl 3 dl] 1 0 0 DL } def
/LT3 { PL [1 dl 1.5 dl] 1 0 1 DL } def
/LT4 { PL [5 dl 2 dl 1 dl 2 dl] 0 1 1 DL } def
/LT5 { PL [4 dl 3 dl 1 dl 3 dl] 1 1 0 DL } def
/LT6 { PL [2 dl 2 dl 2 dl 4 dl] 0 0 0 DL } def
/LT7 { PL [2 dl 2 dl 2 dl 2 dl 2 dl 4 dl] 1 0.3 0 DL } def
/LT8 { PL [2 dl 2 dl 2 dl 2 dl 2 dl 2 dl 2 dl 4 dl] 0.5 0.5 0.5 DL } def
/P { stroke [] 0 setdash
  currentlinewidth 2 div sub M
  0 currentlinewidth V stroke } def
/D { stroke [] 0 setdash 2 copy vpt add M
  hpt neg vpt neg V hpt vpt neg V
  hpt vpt V hpt neg vpt V closepath stroke
  P } def
/A { stroke [] 0 setdash vpt sub M 0 vpt2 V
  currentpoint stroke M
  hpt neg vpt neg R hpt2 0 V stroke
  } def
/B { stroke [] 0 setdash 2 copy exch hpt sub exch vpt add M
  0 vpt2 neg V hpt2 0 V 0 vpt2 V
  hpt2 neg 0 V closepath stroke
  P } def
/C { stroke [] 0 setdash exch hpt sub exch vpt add M
  hpt2 vpt2 neg V currentpoint stroke M
  hpt2 neg 0 R hpt2 vpt2 V stroke } def
/T { stroke [] 0 setdash 2 copy vpt 1.12 mul add M
  hpt neg vpt -1.62 mul V
  hpt 2 mul 0 V
  hpt neg vpt 1.62 mul V closepath stroke
  P  } def
/S { 2 copy A C} def
end
}
\begin{picture}(3600,2160)(0,0)
\special{"
gnudict begin
gsave
50 50 translate
0.100 0.100 scale
0 setgray
/Helvetica findfont 100 scalefont setfont
newpath
-500.000000 -500.000000 translate
LTa
LTb
480 151 M
63 0 V
2874 0 R
-63 0 V
480 477 M
63 0 V
2874 0 R
-63 0 V
480 804 M
63 0 V
2874 0 R
-63 0 V
480 1130 M
63 0 V
2874 0 R
-63 0 V
480 1456 M
63 0 V
2874 0 R
-63 0 V
480 1783 M
63 0 V
2874 0 R
-63 0 V
480 2109 M
63 0 V
2874 0 R
-63 0 V
480 151 M
0 63 V
0 1895 R
0 -63 V
774 151 M
0 63 V
0 1895 R
0 -63 V
1067 151 M
0 63 V
0 1895 R
0 -63 V
1361 151 M
0 63 V
0 1895 R
0 -63 V
1655 151 M
0 63 V
0 1895 R
0 -63 V
1949 151 M
0 63 V
0 1895 R
0 -63 V
2242 151 M
0 63 V
0 1895 R
0 -63 V
2536 151 M
0 63 V
0 1895 R
0 -63 V
2830 151 M
0 63 V
0 1895 R
0 -63 V
3123 151 M
0 63 V
0 1895 R
0 -63 V
3417 151 M
0 63 V
0 1895 R
0 -63 V
480 151 M
2937 0 V
0 1958 V
-2937 0 V
480 151 L
LT0
538 535 M
15 -3 V
15 21 V
14 32 V
15 27 V
15 10 V
14 -12 V
15 -4 V
14 33 V
15 29 V
15 3 V
14 -3 V
15 13 V
14 16 V
15 4 V
15 6 V
14 21 V
15 19 V
14 5 V
15 8 V
15 26 V
14 24 V
15 7 V
15 0 V
14 11 V
15 18 V
14 18 V
15 15 V
15 8 V
14 7 V
15 12 V
14 17 V
15 20 V
15 18 V
14 11 V
15 8 V
14 12 V
15 14 V
15 15 V
14 15 V
15 13 V
15 13 V
14 14 V
15 17 V
14 19 V
15 18 V
15 13 V
14 10 V
15 13 V
14 15 V
15 16 V
15 16 V
14 15 V
15 15 V
14 15 V
15 16 V
15 16 V
14 17 V
15 16 V
15 17 V
14 16 V
15 18 V
14 17 V
15 17 V
15 18 V
14 17 V
15 17 V
14 17 V
15 19 V
15 18 V
14 18 V
15 19 V
15 18 V
14 19 V
15 19 V
14 19 V
15 19 V
15 20 V
14 20 V
15 20 V
14 20 V
15 19 V
15 21 V
14 20 V
15 21 V
14 21 V
15 20 V
15 21 V
14 21 V
15 20 V
15 21 V
14 20 V
15 19 V
14 17 V
15 16 V
15 14 V
14 9 V
15 5 V
14 -1 V
15 -7 V
15 -11 V
14 -16 V
15 -19 V
14 -21 V
15 -22 V
15 -23 V
14 -24 V
15 -24 V
15 -24 V
14 -24 V
15 -24 V
14 -25 V
15 -23 V
15 -24 V
14 -24 V
15 -24 V
14 -23 V
15 -22 V
15 -22 V
14 -22 V
15 -22 V
14 -23 V
15 -22 V
15 -21 V
14 -21 V
15 -21 V
15 -20 V
14 -20 V
15 -21 V
14 -20 V
15 -20 V
15 -19 V
14 -18 V
15 -21 V
14 -19 V
15 -19 V
15 -18 V
14 -19 V
15 -18 V
15 -15 V
14 -17 V
15 -21 V
14 -20 V
15 -15 V
15 -14 V
14 -14 V
15 -17 V
14 -19 V
15 -20 V
15 -17 V
14 -17 V
15 -18 V
14 -15 V
15 -9 V
15 -12 V
14 -20 V
15 -22 V
15 -14 V
14 -12 V
15 -13 V
14 -16 V
15 -17 V
15 -15 V
14 -11 V
15 -16 V
14 -24 V
15 -20 V
15 -1 V
14 -3 V
15 -23 V
14 -27 V
15 -10 V
15 -6 V
14 -20 V
15 -22 V
15 -6 V
14 -2 V
15 -12 V
14 -19 V
15 -20 V
15 -21 V
14 -21 V
15 -22 V
14 -21 V
15 -15 V
15 -4 V
14 10 V
15 14 V
14 -3 V
15 -38 V
15 -37 V
14 26 V
15 28 V
15 -13 V
14 -34 V
15 -22 V
14 -7 V
15 -5 V
LT1
2509 804 D
515 441 D
555 479 D
594 518 D
633 557 D
672 595 D
712 634 D
751 673 D
790 713 D
829 752 D
869 792 D
908 832 D
947 872 D
987 913 D
1026 953 D
1065 994 D
1104 1036 D
1144 1078 D
1183 1120 D
1222 1162 D
1261 1205 D
1301 1248 D
1340 1291 D
1379 1335 D
1418 1379 D
1458 1424 D
1497 1469 D
1536 1513 D
1575 1558 D
1615 1603 D
1654 1648 D
1693 1694 D
1733 1740 D
1772 1788 D
1811 1839 D
1850 1895 D
1890 1960 D
1929 2052 D
1968 2058 D
2007 1972 D
2047 1906 D
2086 1847 D
2125 1792 D
2164 1739 D
2204 1689 D
2243 1640 D
2282 1591 D
2322 1542 D
2361 1493 D
2400 1445 D
2439 1397 D
2479 1349 D
2518 1303 D
2557 1256 D
2596 1210 D
2636 1165 D
2675 1120 D
2714 1076 D
2753 1033 D
2793 990 D
2832 947 D
2871 905 D
2910 863 D
2950 822 D
2989 781 D
3028 740 D
3068 700 D
3107 660 D
3146 621 D
3185 581 D
3225 543 D
3264 504 D
3303 465 D
3342 427 D
3382 389 D
stroke
grestore
end
showpage
}
\put(2389,804){\makebox(0,0)[r]{$r=L/2^6$, Re=1.17E4}}
\put(3417,51){\makebox(0,0){10}}
\put(3123,51){\makebox(0,0){8}}
\put(2830,51){\makebox(0,0){6}}
\put(2536,51){\makebox(0,0){4}}
\put(2242,51){\makebox(0,0){2}}
\put(1949,51){\makebox(0,0){0}}
\put(1655,51){\makebox(0,0){-2}}
\put(1361,51){\makebox(0,0){-4}}
\put(1067,51){\makebox(0,0){-6}}
\put(774,51){\makebox(0,0){-8}}
\put(480,51){\makebox(0,0){-10}}
\put(420,2109){\makebox(0,0)[r]{0}}
\put(420,1783){\makebox(0,0)[r]{-1}}
\put(420,1456){\makebox(0,0)[r]{-2}}
\put(420,1130){\makebox(0,0)[r]{-3}}
\put(420,804){\makebox(0,0)[r]{-4}}
\put(420,477){\makebox(0,0)[r]{-5}}
\put(420,151){\makebox(0,0)[r]{-6}}
\end{picture}
\setlength{\unitlength}{0.1bp}
\special{!
/gnudict 40 dict def
gnudict begin
/Color false def
/Solid false def
/gnulinewidth 5.000 def
/vshift -33 def
/dl {10 mul} def
/hpt 31.5 def
/vpt 31.5 def
/M {moveto} bind def
/L {lineto} bind def
/R {rmoveto} bind def
/V {rlineto} bind def
/vpt2 vpt 2 mul def
/hpt2 hpt 2 mul def
/Lshow { currentpoint stroke M
  0 vshift R show } def
/Rshow { currentpoint stroke M
  dup stringwidth pop neg vshift R show } def
/Cshow { currentpoint stroke M
  dup stringwidth pop -2 div vshift R show } def
/DL { Color {setrgbcolor Solid {pop []} if 0 setdash }
 {pop pop pop Solid {pop []} if 0 setdash} ifelse } def
/BL { stroke gnulinewidth 2 mul setlinewidth } def
/AL { stroke gnulinewidth 2 div setlinewidth } def
/PL { stroke gnulinewidth setlinewidth } def
/LTb { BL [] 0 0 0 DL } def
/LTa { AL [1 dl 2 dl] 0 setdash 0 0 0 setrgbcolor } def
/LT0 { PL [] 0 1 0 DL } def
/LT1 { PL [4 dl 2 dl] 0 0 1 DL } def
/LT2 { PL [2 dl 3 dl] 1 0 0 DL } def
/LT3 { PL [1 dl 1.5 dl] 1 0 1 DL } def
/LT4 { PL [5 dl 2 dl 1 dl 2 dl] 0 1 1 DL } def
/LT5 { PL [4 dl 3 dl 1 dl 3 dl] 1 1 0 DL } def
/LT6 { PL [2 dl 2 dl 2 dl 4 dl] 0 0 0 DL } def
/LT7 { PL [2 dl 2 dl 2 dl 2 dl 2 dl 4 dl] 1 0.3 0 DL } def
/LT8 { PL [2 dl 2 dl 2 dl 2 dl 2 dl 2 dl 2 dl 4 dl] 0.5 0.5 0.5 DL } def
/P { stroke [] 0 setdash
  currentlinewidth 2 div sub M
  0 currentlinewidth V stroke } def
/D { stroke [] 0 setdash 2 copy vpt add M
  hpt neg vpt neg V hpt vpt neg V
  hpt vpt V hpt neg vpt V closepath stroke
  P } def
/A { stroke [] 0 setdash vpt sub M 0 vpt2 V
  currentpoint stroke M
  hpt neg vpt neg R hpt2 0 V stroke
  } def
/B { stroke [] 0 setdash 2 copy exch hpt sub exch vpt add M
  0 vpt2 neg V hpt2 0 V 0 vpt2 V
  hpt2 neg 0 V closepath stroke
  P } def
/C { stroke [] 0 setdash exch hpt sub exch vpt add M
  hpt2 vpt2 neg V currentpoint stroke M
  hpt2 neg 0 R hpt2 vpt2 V stroke } def
/T { stroke [] 0 setdash 2 copy vpt 1.12 mul add M
  hpt neg vpt -1.62 mul V
  hpt 2 mul 0 V
  hpt neg vpt 1.62 mul V closepath stroke
  P  } def
/S { 2 copy A C} def
end
}
\begin{picture}(3600,2160)(0,0)
\special{"
gnudict begin
gsave
50 50 translate
0.100 0.100 scale
0 setgray
/Helvetica findfont 100 scalefont setfont
newpath
-500.000000 -500.000000 translate
LTa
LTb
480 151 M
63 0 V
2874 0 R
-63 0 V
480 477 M
63 0 V
2874 0 R
-63 0 V
480 804 M
63 0 V
2874 0 R
-63 0 V
480 1130 M
63 0 V
2874 0 R
-63 0 V
480 1456 M
63 0 V
2874 0 R
-63 0 V
480 1783 M
63 0 V
2874 0 R
-63 0 V
480 2109 M
63 0 V
2874 0 R
-63 0 V
480 151 M
0 63 V
0 1895 R
0 -63 V
774 151 M
0 63 V
0 1895 R
0 -63 V
1067 151 M
0 63 V
0 1895 R
0 -63 V
1361 151 M
0 63 V
0 1895 R
0 -63 V
1655 151 M
0 63 V
0 1895 R
0 -63 V
1949 151 M
0 63 V
0 1895 R
0 -63 V
2242 151 M
0 63 V
0 1895 R
0 -63 V
2536 151 M
0 63 V
0 1895 R
0 -63 V
2830 151 M
0 63 V
0 1895 R
0 -63 V
3123 151 M
0 63 V
0 1895 R
0 -63 V
3417 151 M
0 63 V
0 1895 R
0 -63 V
480 151 M
2937 0 V
0 1958 V
-2937 0 V
480 151 L
LT0
538 636 M
15 13 V
15 12 V
14 7 V
15 3 V
15 8 V
14 18 V
15 15 V
14 3 V
15 4 V
15 16 V
14 15 V
15 7 V
14 10 V
15 22 V
15 19 V
14 5 V
15 6 V
14 18 V
15 17 V
15 3 V
14 -3 V
15 5 V
15 11 V
14 13 V
15 12 V
14 11 V
15 13 V
15 15 V
14 13 V
15 6 V
14 8 V
15 18 V
15 19 V
14 13 V
15 10 V
14 12 V
15 13 V
15 13 V
14 13 V
15 11 V
15 11 V
14 15 V
15 16 V
14 15 V
15 12 V
15 12 V
14 12 V
15 12 V
14 14 V
15 16 V
15 15 V
14 12 V
15 11 V
14 14 V
15 16 V
15 15 V
14 14 V
15 16 V
15 16 V
14 16 V
15 16 V
14 15 V
15 15 V
15 17 V
14 17 V
15 17 V
14 16 V
15 17 V
15 18 V
14 17 V
15 17 V
15 18 V
14 18 V
15 18 V
14 19 V
15 19 V
15 19 V
14 19 V
15 19 V
14 20 V
15 20 V
15 21 V
14 20 V
15 21 V
14 21 V
15 22 V
15 22 V
14 21 V
15 22 V
15 22 V
14 22 V
15 21 V
14 20 V
15 18 V
15 15 V
14 11 V
15 5 V
14 -1 V
15 -8 V
15 -14 V
14 -18 V
15 -20 V
14 -23 V
15 -25 V
15 -25 V
14 -25 V
15 -25 V
15 -25 V
14 -25 V
15 -24 V
14 -25 V
15 -24 V
15 -24 V
14 -23 V
15 -23 V
14 -23 V
15 -23 V
15 -22 V
14 -22 V
15 -22 V
14 -20 V
15 -20 V
15 -20 V
14 -21 V
15 -20 V
15 -20 V
14 -20 V
15 -19 V
14 -18 V
15 -18 V
15 -18 V
14 -19 V
15 -18 V
14 -17 V
15 -15 V
15 -16 V
14 -19 V
15 -18 V
15 -16 V
14 -15 V
15 -15 V
14 -17 V
15 -16 V
15 -15 V
14 -15 V
15 -14 V
14 -16 V
15 -17 V
15 -16 V
14 -13 V
15 -10 V
14 -13 V
15 -20 V
15 -20 V
14 -13 V
15 -11 V
15 -12 V
14 -14 V
15 -13 V
14 -12 V
15 -10 V
15 -13 V
14 -18 V
15 -21 V
14 -16 V
15 -8 V
15 -3 V
14 -8 V
15 -20 V
14 -24 V
15 -12 V
15 -3 V
14 0 V
15 -8 V
15 -23 V
14 -30 V
15 -20 V
14 -9 V
15 -2 V
15 -7 V
14 -20 V
15 -20 V
14 -4 V
15 -3 V
15 -20 V
14 -20 V
15 0 V
14 0 V
15 -21 V
15 -30 V
14 -15 V
15 0 V
15 4 V
14 4 V
15 2 V
14 -10 V
15 -35 V
LT1
2509 804 D
484 676 D
549 722 D
614 768 D
679 813 D
744 860 D
809 907 D
874 954 D
940 1001 D
1005 1049 D
1070 1098 D
1135 1148 D
1200 1198 D
1265 1250 D
1330 1304 D
1395 1359 D
1460 1417 D
1525 1479 D
1590 1545 D
1656 1616 D
1721 1695 D
1786 1784 D
1851 1888 D
1916 2024 D
1981 2034 D
2046 1897 D
2111 1787 D
2176 1691 D
2241 1607 D
2307 1530 D
2372 1459 D
2437 1393 D
2502 1331 D
2567 1273 D
2632 1216 D
2697 1162 D
2762 1109 D
2827 1057 D
2892 1006 D
2957 957 D
3023 908 D
3088 859 D
3153 812 D
3218 764 D
3283 717 D
3348 671 D
3413 625 D
stroke
grestore
end
showpage
}
\put(2389,804){\makebox(0,0)[r]{$r=L/2^7$, Re=1.17E4}}
\put(3417,51){\makebox(0,0){10}}
\put(3123,51){\makebox(0,0){8}}
\put(2830,51){\makebox(0,0){6}}
\put(2536,51){\makebox(0,0){4}}
\put(2242,51){\makebox(0,0){2}}
\put(1949,51){\makebox(0,0){0}}
\put(1655,51){\makebox(0,0){-2}}
\put(1361,51){\makebox(0,0){-4}}
\put(1067,51){\makebox(0,0){-6}}
\put(774,51){\makebox(0,0){-8}}
\put(480,51){\makebox(0,0){-10}}
\put(420,2109){\makebox(0,0)[r]{0}}
\put(420,1783){\makebox(0,0)[r]{-1}}
\put(420,1456){\makebox(0,0)[r]{-2}}
\put(420,1130){\makebox(0,0)[r]{-3}}
\put(420,804){\makebox(0,0)[r]{-4}}
\put(420,477){\makebox(0,0)[r]{-5}}
\put(420,151){\makebox(0,0)[r]{-6}}
\end{picture}
\label{fig-pdfevol-15V}
\end{figure}

\newpage

\begin{figure}[htb]
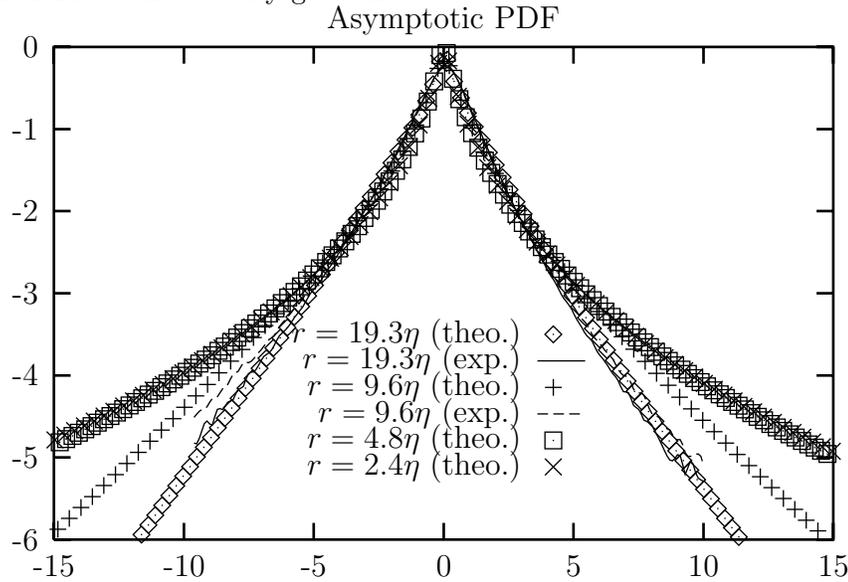

\caption{A comparison of theoretical  and experimental PDFs in the
viscous range. The dashed line corresponds to the scale of experimental
resolution. The last two curves show the asymptotic
form of the PDF of the velocity gradients.}
\input pictures/fig-pdfasym.tex
\label{fig-pdfasym}
\end{figure}

\begin{figure}[htb]
\caption{The theoretical prediction for the scale-dependent flatness $F(r)$ 
for different experimental runs. Note the sharp rise 
in the crossover regime, and the crossing of different curves in
the same regime.
  }
\setlength{\unitlength}{0.1bp}
\special{!
/gnudict 40 dict def
gnudict begin
/Color false def
/Solid false def
/gnulinewidth 5.000 def
/vshift -33 def
/dl {10 mul} def
/hpt 31.5 def
/vpt 31.5 def
/M {moveto} bind def
/L {lineto} bind def
/R {rmoveto} bind def
/V {rlineto} bind def
/vpt2 vpt 2 mul def
/hpt2 hpt 2 mul def
/Lshow { currentpoint stroke M
  0 vshift R show } def
/Rshow { currentpoint stroke M
  dup stringwidth pop neg vshift R show } def
/Cshow { currentpoint stroke M
  dup stringwidth pop -2 div vshift R show } def
/DL { Color {setrgbcolor Solid {pop []} if 0 setdash }
 {pop pop pop Solid {pop []} if 0 setdash} ifelse } def
/BL { stroke gnulinewidth 2 mul setlinewidth } def
/AL { stroke gnulinewidth 2 div setlinewidth } def
/PL { stroke gnulinewidth setlinewidth } def
/LTb { BL [] 0 0 0 DL } def
/LTa { AL [1 dl 2 dl] 0 setdash 0 0 0 setrgbcolor } def
/LT0 { PL [] 0 1 0 DL } def
/LT1 { PL [4 dl 2 dl] 0 0 1 DL } def
/LT2 { PL [2 dl 3 dl] 1 0 0 DL } def
/LT3 { PL [1 dl 1.5 dl] 1 0 1 DL } def
/LT4 { PL [5 dl 2 dl 1 dl 2 dl] 0 1 1 DL } def
/LT5 { PL [4 dl 3 dl 1 dl 3 dl] 1 1 0 DL } def
/LT6 { PL [2 dl 2 dl 2 dl 4 dl] 0 0 0 DL } def
/LT7 { PL [2 dl 2 dl 2 dl 2 dl 2 dl 4 dl] 1 0.3 0 DL } def
/LT8 { PL [2 dl 2 dl 2 dl 2 dl 2 dl 2 dl 2 dl 4 dl] 0.5 0.5 0.5 DL } def
/P { stroke [] 0 setdash
  currentlinewidth 2 div sub M
  0 currentlinewidth V stroke } def
/D { stroke [] 0 setdash 2 copy vpt add M
  hpt neg vpt neg V hpt vpt neg V
  hpt vpt V hpt neg vpt V closepath stroke
  P } def
/A { stroke [] 0 setdash vpt sub M 0 vpt2 V
  currentpoint stroke M
  hpt neg vpt neg R hpt2 0 V stroke
  } def
/B { stroke [] 0 setdash 2 copy exch hpt sub exch vpt add M
  0 vpt2 neg V hpt2 0 V 0 vpt2 V
  hpt2 neg 0 V closepath stroke
  P } def
/C { stroke [] 0 setdash exch hpt sub exch vpt add M
  hpt2 vpt2 neg V currentpoint stroke M
  hpt2 neg 0 R hpt2 vpt2 V stroke } def
/T { stroke [] 0 setdash 2 copy vpt 1.12 mul add M
  hpt neg vpt -1.62 mul V
  hpt 2 mul 0 V
  hpt neg vpt 1.62 mul V closepath stroke
  P  } def
/S { 2 copy A C} def
end
}
\begin{picture}(3600,2160)(0,0)
\special{"
gnudict begin
gsave
50 50 translate
0.100 0.100 scale
0 setgray
/Helvetica findfont 100 scalefont setfont
newpath
-500.000000 -500.000000 translate
LTa
LTb
480 151 M
63 0 V
2874 0 R
-63 0 V
480 431 M
31 0 V
2906 0 R
-31 0 V
480 594 M
31 0 V
2906 0 R
-31 0 V
480 710 M
31 0 V
2906 0 R
-31 0 V
480 800 M
31 0 V
2906 0 R
-31 0 V
480 874 M
31 0 V
2906 0 R
-31 0 V
480 936 M
31 0 V
2906 0 R
-31 0 V
480 990 M
31 0 V
2906 0 R
-31 0 V
480 1037 M
31 0 V
2906 0 R
-31 0 V
480 1080 M
63 0 V
2874 0 R
-63 0 V
480 1360 M
31 0 V
2906 0 R
-31 0 V
480 1523 M
31 0 V
2906 0 R
-31 0 V
480 1639 M
31 0 V
2906 0 R
-31 0 V
480 1729 M
31 0 V
2906 0 R
-31 0 V
480 1803 M
31 0 V
2906 0 R
-31 0 V
480 1865 M
31 0 V
2906 0 R
-31 0 V
480 1919 M
31 0 V
2906 0 R
-31 0 V
480 1966 M
31 0 V
2906 0 R
-31 0 V
480 2009 M
63 0 V
2874 0 R
-63 0 V
480 151 M
0 63 V
0 1795 R
0 -63 V
657 151 M
0 31 V
0 1827 R
0 -31 V
891 151 M
0 31 V
0 1827 R
0 -31 V
1010 151 M
0 31 V
0 1827 R
0 -31 V
1067 151 M
0 63 V
0 1795 R
0 -63 V
1244 151 M
0 31 V
0 1827 R
0 -31 V
1478 151 M
0 31 V
0 1827 R
0 -31 V
1598 151 M
0 31 V
0 1827 R
0 -31 V
1655 151 M
0 63 V
0 1795 R
0 -63 V
1832 151 M
0 31 V
0 1827 R
0 -31 V
2065 151 M
0 31 V
0 1827 R
0 -31 V
2185 151 M
0 31 V
0 1827 R
0 -31 V
2242 151 M
0 63 V
0 1795 R
0 -63 V
2419 151 M
0 31 V
0 1827 R
0 -31 V
2653 151 M
0 31 V
0 1827 R
0 -31 V
2773 151 M
0 31 V
0 1827 R
0 -31 V
2830 151 M
0 63 V
0 1795 R
0 -63 V
3006 151 M
0 31 V
0 1827 R
0 -31 V
3240 151 M
0 31 V
0 1827 R
0 -31 V
3360 151 M
0 31 V
0 1827 R
0 -31 V
3417 151 M
0 63 V
0 1795 R
0 -63 V
480 151 M
2937 0 V
0 1858 V
-2937 0 V
480 151 L
LT0
3174 1846 D
3220 696 D
3043 747 D
2866 845 D
2690 1024 D
2513 1195 D
2336 1260 D
2159 1276 D
1982 1282 D
1805 1285 D
1629 1286 D
1452 1287 D
1275 1287 D
1098 1288 D
921 1288 D
745 1288 D
LT1
3174 1746 A
3220 692 A
3043 732 A
2866 783 A
2690 874 A
2513 1051 A
2336 1241 A
2159 1325 A
1982 1346 A
1805 1353 A
1629 1356 A
1452 1358 A
1275 1359 A
1098 1360 A
921 1360 A
745 1360 A
LT2
3174 1646 B
3220 691 B
3043 728 B
2866 766 B
2690 808 B
2513 874 B
2336 1017 B
2159 1230 B
1982 1378 B
1805 1424 B
1629 1436 B
1452 1441 B
1275 1443 B
1098 1444 B
921 1445 B
745 1445 B
LT3
3174 1546 C
3220 691 C
3043 727 C
2866 765 C
2690 805 C
2513 863 C
2336 987 C
2159 1196 C
1982 1368 C
1805 1430 C
1629 1446 C
1452 1451 C
1275 1454 C
1098 1455 C
921 1456 C
745 1457 C
LT4
3174 1446 T
3220 691 T
3043 727 T
2866 764 T
2690 802 T
2513 852 T
2336 955 T
2159 1150 T
1982 1348 T
1805 1435 T
1629 1457 T
1452 1464 T
1275 1467 T
1098 1469 T
921 1470 T
745 1471 T
stroke
grestore
end
showpage
}
\put(3054,1446){\makebox(0,0)[r]{Re=2.88E4}}
\put(3054,1546){\makebox(0,0)[r]{Re=1.17E4}}
\put(3054,1646){\makebox(0,0)[r]{Re=1.07E4}}
\put(3054,1746){\makebox(0,0)[r]{Re=1.86E3}}
\put(3054,1846){\makebox(0,0)[r]{Re=1.36E3}}
\put(1948,2109){\makebox(0,0){Flatness (Theory)}}
\put(3417,51){\makebox(0,0){1}}
\put(2830,51){\makebox(0,0){0.1}}
\put(2242,51){\makebox(0,0){0.01}}
\put(1655,51){\makebox(0,0){0.001}}
\put(1067,51){\makebox(0,0){0.0001}}
\put(480,51){\makebox(0,0){1e-05}}
\put(420,2009){\makebox(0,0)[r]{100}}
\put(420,1080){\makebox(0,0)[r]{10}}
\put(420,151){\makebox(0,0)[r]{1}}
\end{picture}
\label{fig-flat-theo}
\end{figure}
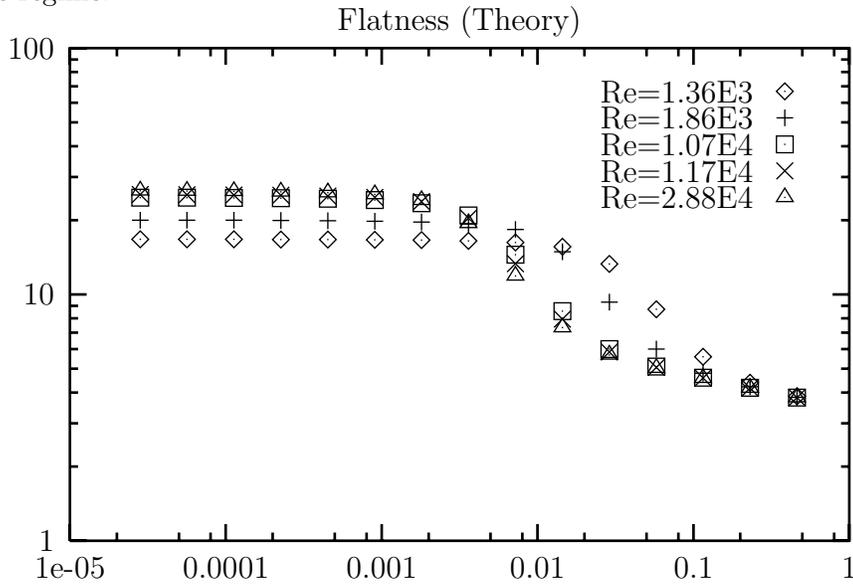

\newpage

\begin{figure}[htb]
\caption{A comparison of the experimental data for $F(r)$ with the 
theoretical prediction for $Re = 1.17E4$.
  }
\setlength{\unitlength}{0.1bp}
\special{!
/gnudict 40 dict def
gnudict begin
/Color false def
/Solid false def
/gnulinewidth 5.000 def
/vshift -33 def
/dl {10 mul} def
/hpt 31.5 def
/vpt 31.5 def
/M {moveto} bind def
/L {lineto} bind def
/R {rmoveto} bind def
/V {rlineto} bind def
/vpt2 vpt 2 mul def
/hpt2 hpt 2 mul def
/Lshow { currentpoint stroke M
  0 vshift R show } def
/Rshow { currentpoint stroke M
  dup stringwidth pop neg vshift R show } def
/Cshow { currentpoint stroke M
  dup stringwidth pop -2 div vshift R show } def
/DL { Color {setrgbcolor Solid {pop []} if 0 setdash }
 {pop pop pop Solid {pop []} if 0 setdash} ifelse } def
/BL { stroke gnulinewidth 2 mul setlinewidth } def
/AL { stroke gnulinewidth 2 div setlinewidth } def
/PL { stroke gnulinewidth setlinewidth } def
/LTb { BL [] 0 0 0 DL } def
/LTa { AL [1 dl 2 dl] 0 setdash 0 0 0 setrgbcolor } def
/LT0 { PL [] 0 1 0 DL } def
/LT1 { PL [4 dl 2 dl] 0 0 1 DL } def
/LT2 { PL [2 dl 3 dl] 1 0 0 DL } def
/LT3 { PL [1 dl 1.5 dl] 1 0 1 DL } def
/LT4 { PL [5 dl 2 dl 1 dl 2 dl] 0 1 1 DL } def
/LT5 { PL [4 dl 3 dl 1 dl 3 dl] 1 1 0 DL } def
/LT6 { PL [2 dl 2 dl 2 dl 4 dl] 0 0 0 DL } def
/LT7 { PL [2 dl 2 dl 2 dl 2 dl 2 dl 4 dl] 1 0.3 0 DL } def
/LT8 { PL [2 dl 2 dl 2 dl 2 dl 2 dl 2 dl 2 dl 4 dl] 0.5 0.5 0.5 DL } def
/P { stroke [] 0 setdash
  currentlinewidth 2 div sub M
  0 currentlinewidth V stroke } def
/D { stroke [] 0 setdash 2 copy vpt add M
  hpt neg vpt neg V hpt vpt neg V
  hpt vpt V hpt neg vpt V closepath stroke
  P } def
/A { stroke [] 0 setdash vpt sub M 0 vpt2 V
  currentpoint stroke M
  hpt neg vpt neg R hpt2 0 V stroke
  } def
/B { stroke [] 0 setdash 2 copy exch hpt sub exch vpt add M
  0 vpt2 neg V hpt2 0 V 0 vpt2 V
  hpt2 neg 0 V closepath stroke
  P } def
/C { stroke [] 0 setdash exch hpt sub exch vpt add M
  hpt2 vpt2 neg V currentpoint stroke M
  hpt2 neg 0 R hpt2 vpt2 V stroke } def
/T { stroke [] 0 setdash 2 copy vpt 1.12 mul add M
  hpt neg vpt -1.62 mul V
  hpt 2 mul 0 V
  hpt neg vpt 1.62 mul V closepath stroke
  P  } def
/S { 2 copy A C} def
end
}
\begin{picture}(3600,2160)(0,0)
\special{"
gnudict begin
gsave
50 50 translate
0.100 0.100 scale
0 setgray
/Helvetica findfont 100 scalefont setfont
newpath
-500.000000 -500.000000 translate
LTa
LTb
480 151 M
63 0 V
2874 0 R
-63 0 V
480 431 M
31 0 V
2906 0 R
-31 0 V
480 594 M
31 0 V
2906 0 R
-31 0 V
480 710 M
31 0 V
2906 0 R
-31 0 V
480 800 M
31 0 V
2906 0 R
-31 0 V
480 874 M
31 0 V
2906 0 R
-31 0 V
480 936 M
31 0 V
2906 0 R
-31 0 V
480 990 M
31 0 V
2906 0 R
-31 0 V
480 1037 M
31 0 V
2906 0 R
-31 0 V
480 1080 M
63 0 V
2874 0 R
-63 0 V
480 1360 M
31 0 V
2906 0 R
-31 0 V
480 1523 M
31 0 V
2906 0 R
-31 0 V
480 1639 M
31 0 V
2906 0 R
-31 0 V
480 1729 M
31 0 V
2906 0 R
-31 0 V
480 1803 M
31 0 V
2906 0 R
-31 0 V
480 1865 M
31 0 V
2906 0 R
-31 0 V
480 1919 M
31 0 V
2906 0 R
-31 0 V
480 1966 M
31 0 V
2906 0 R
-31 0 V
480 2009 M
63 0 V
2874 0 R
-63 0 V
480 151 M
0 63 V
0 1795 R
0 -63 V
657 151 M
0 31 V
0 1827 R
0 -31 V
891 151 M
0 31 V
0 1827 R
0 -31 V
1010 151 M
0 31 V
0 1827 R
0 -31 V
1067 151 M
0 63 V
0 1795 R
0 -63 V
1244 151 M
0 31 V
0 1827 R
0 -31 V
1478 151 M
0 31 V
0 1827 R
0 -31 V
1598 151 M
0 31 V
0 1827 R
0 -31 V
1655 151 M
0 63 V
0 1795 R
0 -63 V
1832 151 M
0 31 V
0 1827 R
0 -31 V
2065 151 M
0 31 V
0 1827 R
0 -31 V
2185 151 M
0 31 V
0 1827 R
0 -31 V
2242 151 M
0 63 V
0 1795 R
0 -63 V
2419 151 M
0 31 V
0 1827 R
0 -31 V
2653 151 M
0 31 V
0 1827 R
0 -31 V
2773 151 M
0 31 V
0 1827 R
0 -31 V
2830 151 M
0 63 V
0 1795 R
0 -63 V
3006 151 M
0 31 V
0 1827 R
0 -31 V
3240 151 M
0 31 V
0 1827 R
0 -31 V
3360 151 M
0 31 V
0 1827 R
0 -31 V
3417 151 M
0 63 V
0 1795 R
0 -63 V
480 151 M
2937 0 V
0 1858 V
-2937 0 V
480 151 L
LT0
3174 1846 D
3220 691 D
3043 727 D
2866 765 D
2690 805 D
2513 863 D
2336 987 D
2159 1196 D
1982 1368 D
1805 1430 D
1629 1446 D
1452 1451 D
1275 1454 D
1098 1455 D
921 1456 D
745 1457 D
LT1
3174 1746 A
3376 705 A
3199 740 A
3022 776 A
2845 808 A
2668 842 A
2492 889 A
2315 951 A
2138 1013 A
1961 1043 A
stroke
grestore
end
showpage
}
\put(3054,1746){\makebox(0,0)[r]{Exp., Re=1.17E4}}
\put(3054,1846){\makebox(0,0)[r]{Theo., Re=1.17E4}}
\put(1948,2109){\makebox(0,0){Flatness (Theory vs. Experiment)}}
\put(3417,51){\makebox(0,0){1}}
\put(2830,51){\makebox(0,0){0.1}}
\put(2242,51){\makebox(0,0){0.01}}
\put(1655,51){\makebox(0,0){0.001}}
\put(1067,51){\makebox(0,0){0.0001}}
\put(480,51){\makebox(0,0){1e-05}}
\put(420,2009){\makebox(0,0)[r]{100}}
\put(420,1080){\makebox(0,0)[r]{10}}
\put(420,151){\makebox(0,0)[r]{1}}
\end{picture}
\label{fig-flat-compr.tex}
\end{figure}
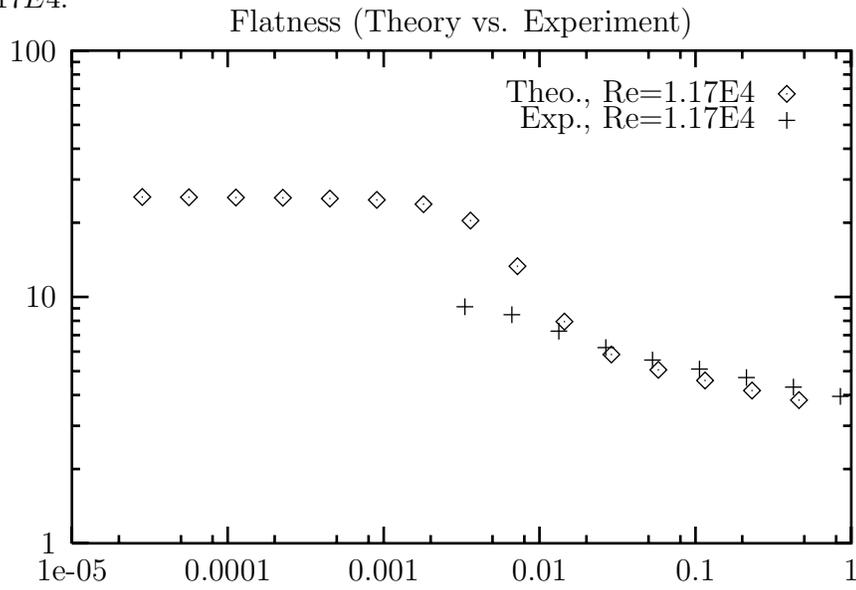

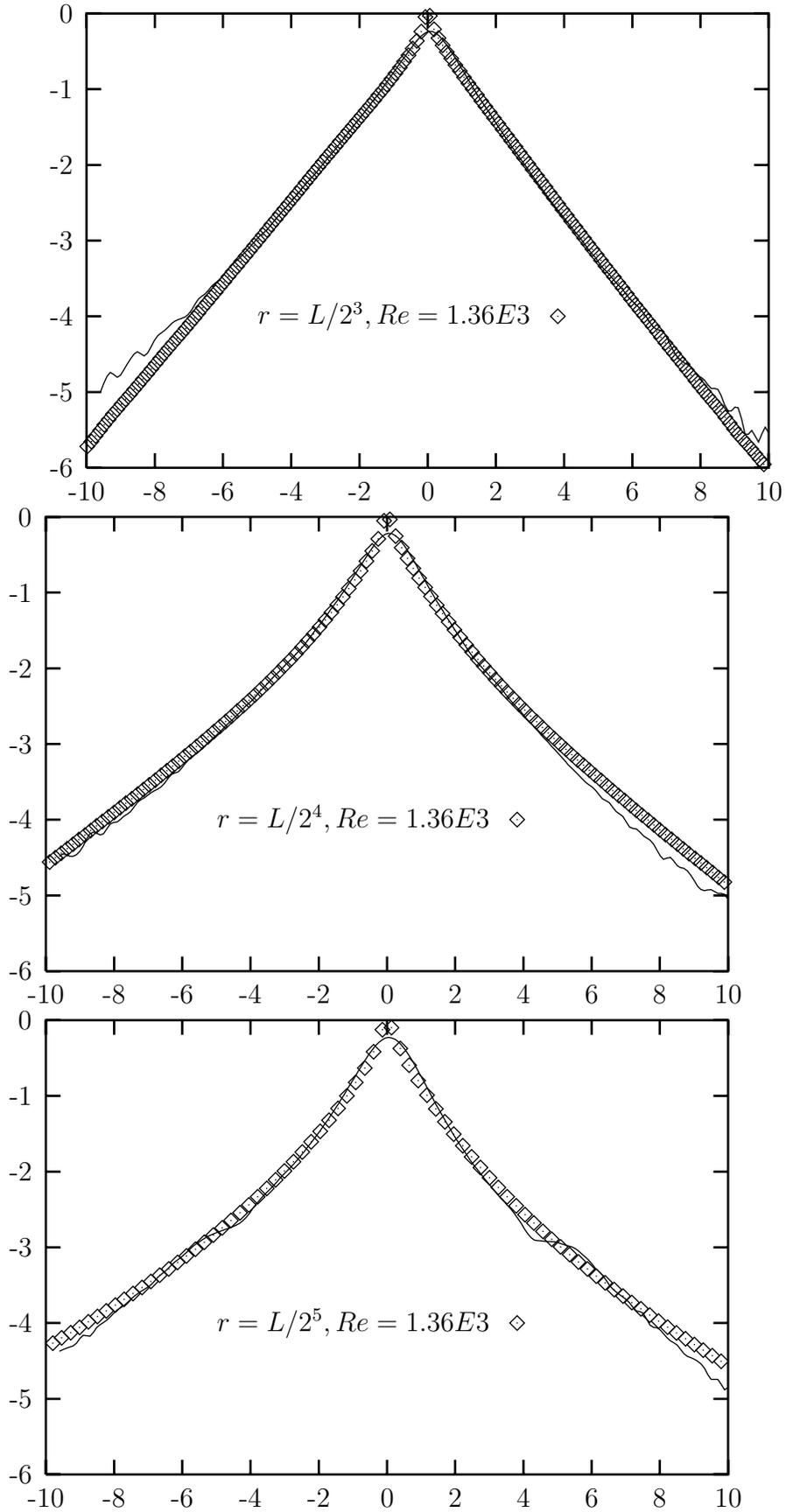
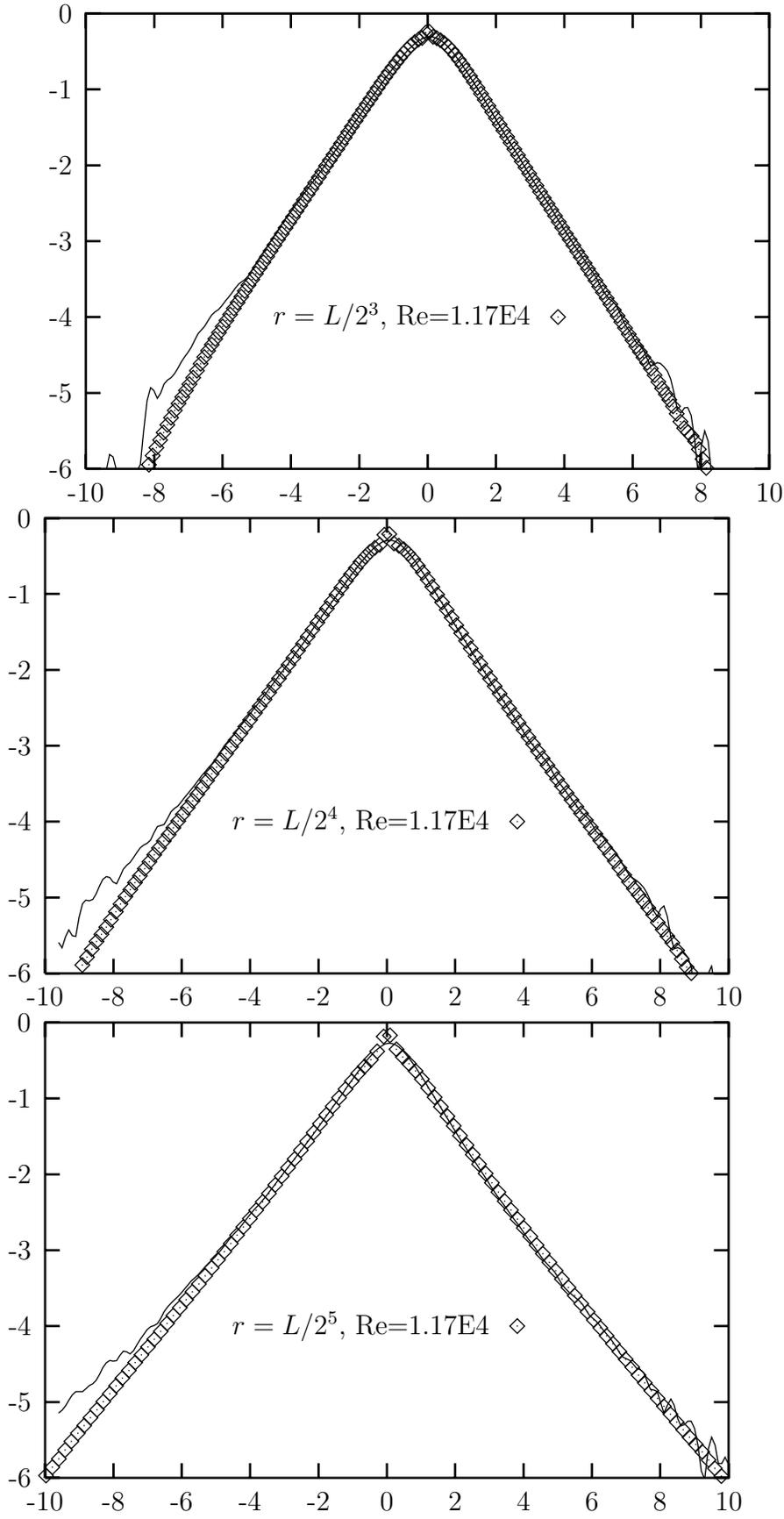
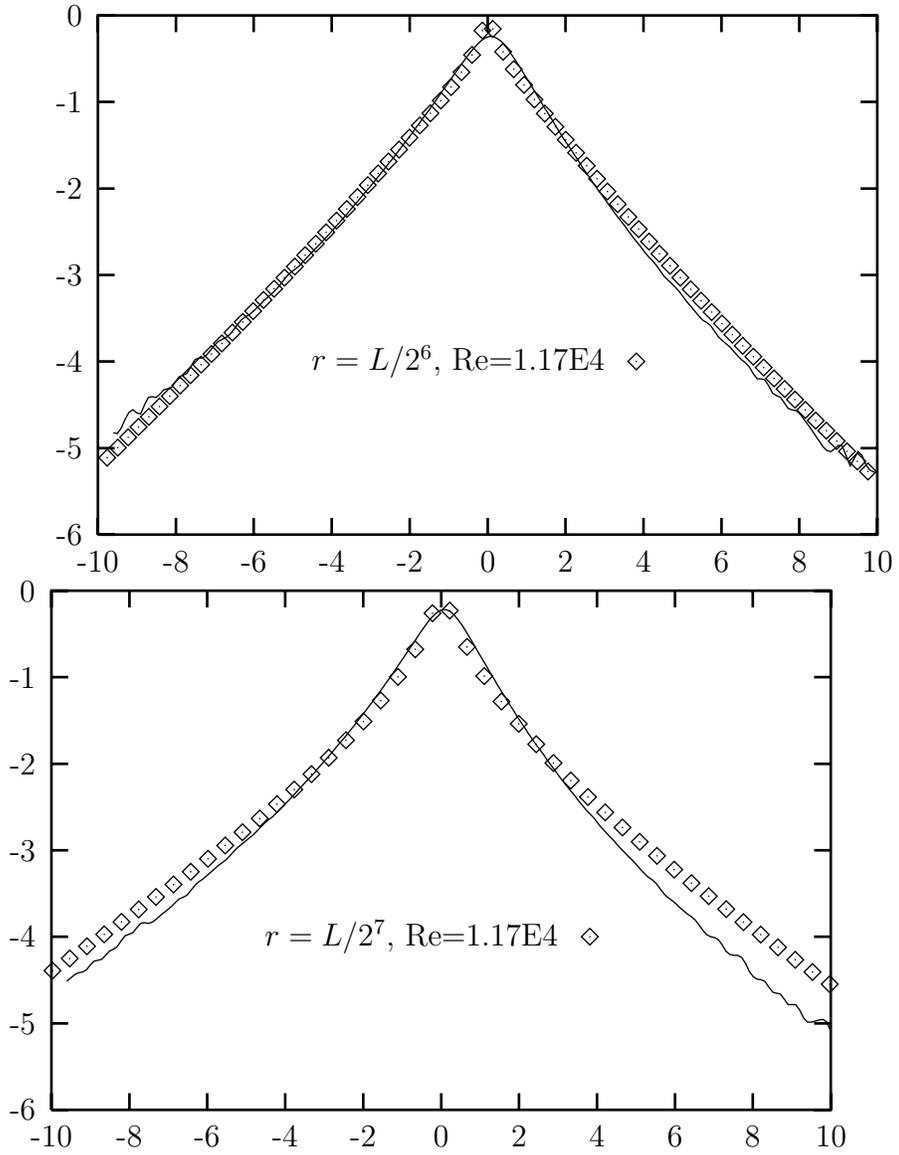
\end{document}